\documentclass[twocolumn]{aastex63}
\usepackage{xcolor}
\usepackage{color}
\usepackage{graphicx}
\usepackage{dcolumn}
\usepackage{bm}
\usepackage{hyperref}
\usepackage{amsmath}
\usepackage{multirow}
\usepackage{cancel}

\newcommand{\Ms}{{\rm ~M}_\odot}

\newcommand{\gw}{{\rm GW}}
\newcommand{\der}{{\rm d}}
\newcommand{\bbh}{{\rm BBH}}

\received{}
\revised{}
\accepted{}
\submitjournal{ApJ}

\shorttitle{Formation of GW190521 and IMBH mergers in YMCs}
\shortauthors{Arca Sedda et al.}
\graphicspath{{./}{}}

\begin{document}

\title{Breaching the limit: formation of GW190521-like and IMBH mergers in young massive clusters}

\correspondingauthor{Manuel Arca Sedda}
\email{m.arcasedda@gmail.com}

\author[0000-0002-3987-0519]{Manuel Arca Sedda}
\affil{Astronomisches Rechen-Institut, Zentrum f\"{u}r Astronomie der Universit\"{a}t  Heidelberg, M\"onchhofstr. 12-14, D-69120 Heidelberg, Germany}
\author{Francesco Paolo Rizzuto}
\affil{Max Planck Institute for Astrophysics, Karl-Schwarzschild-Str. 1, D-85740, Garching, Germany}
\author{Thorsten Naab}
\affil{Max Planck Institute for Astrophysics, Karl-Schwarzschild-Str. 1, D-85740, Garching, Germany}
\author{Jeremiah Ostriker} 
\affil{Department of Astronomy, Columbia University, 550 W, 120th Street, New York, NY 10027, USA}
\affil{Department of Astrophysical Sciences, Princeton University, Peyton Hall, Princeton, NJ 08544, USA}
\author{Mirek Giersz}
\affil{Nicolaus Copernicus Astronomical Centre, Polish Academy of Sciences, ul. Bartycka 18, 00-716 Warsaw, Poland}
\author{Rainer Spurzem}
\affil{Astronomisches Rechen-Institut, Zentrum f\"{u}r Astronomie der Universit\"{a}t  Heidelberg, M\"onchhofstr. 12-14, D-69120 Heidelberg, Germany}
\affil{National Astronomical Observatories and Key Laboratory of Computational Astrophysics, Chinese Academy of Sciences, 20A Datun Rd.,Chaoyang District, 100101, Beijing, China}
\affil{Kavli Institute for Astronomy and Astrophysics, Beijing University, Yiheyuan Lu 5, Haidian Qu, 100871, Beijing, China}

\begin{abstract}
The LIGO-Virgo-Kagra collaboration (LVC) discovered recently GW190521, a gravitational wave (GW) source associated with the merger between two black holes (BHs) with mass $66$ M$_\odot$ and $>85$ M$_\odot$. GW190521 represents the first BH binary (BBH) merger with a primary mass falling in the "upper mass-gap" and the first leaving behind a $\sim 150$ M$_\odot$ remnant. So far, the LVC reported the discovery of four further mergers having a total mass $>100$ M$_\odot$, i.e. in the intermediate-mass black holes (IMBH) mass range. Here, we discuss results from a series of 80 $N$-body simulations of young massive clusters (YMCs) that implement relativistic corrections to follow compact object mergers. We discover the development of a GW190521-like system as the result of a 3rd-generation merger, and four IMBH-BH mergers with total mass $~(300-350)$ M$_\odot$. We show that these IMBH-BH mergers are low-frequency GW sources detectable with LISA and DECIGO out to redshift $z=0.01-0.1$ and $z>100$, and we discuss how their detection could help unravelling IMBH natal spins. For the GW190521 test case, we show that the 3rd-generation merger remnant has a spin and effective spin parameter that matches the $90\%$ credible interval measured for GW190521 better than a simpler double merger and comparably to a single merger. Due to GW recoil kicks, we show that retaining the products of these mergers require birth-sites with escape velocities $\gtrsim 50-100$ km s$^{-1}$, values typically attained in galactic nuclei and massive clusters with steep density profiles.
\end{abstract}

\section{Introduction}
\label{sec:intro} 

The third observational campaign ran by the LIGO-Virgo-Kagra collaboration (LVC) led to the discovery of four black hole binary (BBH) mergers with a total mass $>100\Ms$, i.e. in the mass range of intermediate-mass black holes (IMBHs) \citep{gwtc2}. All these mergers are characterised by a primary having a mass $>64\Ms$, in a range of masses where stellar evolution theories predict the absence of such compact objects. Massive stars developing a Helium burning core in the range of $\sim 64-135\Ms$ end their lives in disruptive pair instability supernovae (PISNe) leaving no BH remnant behind \citep{fowler64,woosley07}. If the Helium core falls in the $32-64\Ms$ mass range, the star develops rapid pulses that enhance mass loss before the star goes supernova, a process called pulsational pair instability supernova (PPISN) \citep{woosley17}. These processes result in the so called ``upper mass-gap'' with no stellar BHs in the mass range of $\sim 64 - 120 \Ms$  \citep[see e.g.][]{spera17,woosley17}. 

The lower edge of the gap is highly uncertain and can be shifted down to $40\Ms$ or the gap might even not exist, depending on the metallicity, the rotation of the progenitor star, the rates of Carbon-Oxygen nuclear reactions \citep[see e.g.][]{farmer19,stevenson19,mapelli20,costa21,woosley21}. The physical mechanisms that govern core overshooting and stellar winds can make the picture even more complex, especially for metal-poor stars with masses $> 90\Ms$ which could collapse to BHs with masses right in the upper mass-gap \citep[e.g.][]{vink21,tanikawa21} 

At least one of the four LVC gravitational wave (GW) sources, named GW190521, have a primary mass $>65\Ms$ at $99\%$ credibility and a companion with mass $66_{-18}^{+17}$ \citep{gw190521b}, thus potentially both the GW190521 components fall in the upper mass-gap. 

There are several potential formation channels for the components involved in such type of mergers, like formation from population III stars \citep{liu20,kinugawa21,tanikawa21}, binary evolution \citep{belczynski21}, stellar collisions \citep{mapelli16,dicarlo19, kremer20, rizzuto20,renzo20} and hierarchical mergers \citep{gerosa17,rodriguez19,antonini19,arcasedda20b,fragione20,doctor20,mapelli20,arcasedda20d,belczynski20c} in young (YMCs), globular (GCs), and nuclear clusters (NCs), or in active galactic nuclei \citep{mckernan18,yang19,graham20,secunda20, tagawa20}.

Dense star clusters represent an ideal environment to nurture mass-gap BHs via the dynamical assembly mechanism, either through stellar collisions or hierarchical mergers. However, the modelling of such environments relied so far on either direct $N$-body simulations, tailored to low-mass clusters, i.e. lighter than $5\times 10^4\Ms$ \citep[e.g.][]{banerjee18,dicarlo19}, or adopting outdated stellar evolution prescriptions \citep{wang16,anagnostou20a}, on Monte Carlo methods \citep[e.g.][]{giersz15,giersz19,rodriguez19,kremer20}, or on semi-analytic approaches \citep[e.g.][]{gerosa17,fragione20,arcasedda20d}. 
In this work, we present and discuss the properties of BH binary mergers with total masses in the range $150-350\Ms$ that we identify in a suite of 80 direct $N$-body simulations modelling compact clusters composed of $110,000$ stars \citep{rizzuto20}, which implement relativistic corrections to enable a reliable follow-up of the coalescence of compact objects. 
One of the mergers discussed here is actually the byproduct of a series of four BH mergers that lead to the formation of a final merger with component masses $68\Ms$ and $70\Ms$, serendipitously compatible with GW190521 measured properties. 
We use a statistical approach to discuss the impact of BH natal spins and recoil kick received upon GW emission in determining the likelihood for this type of mergers to take place in young and massive clusters and similar environments.

The paper is organised as follows: in Section \ref{sec:2} we briefly describe the main features of the $N$-body models; in Section \ref{sec:3} we focus on the evolution of mergers beyond the upper mass-gap, i.e. involving IMBHs; Section \ref{sec:4} is devoted to mergers that contribute to the formation of BHs in the mass-gap; in Section \ref{sec:5} we discuss the impact of GW kicks; in Section \ref{sec:6} we compare our results with similar works based on $N$-body models; finally, Section \ref{sec:7} summarizes our main conclusions.

\section{Numerical simulation of young compact clusters}
\label{sec:2}
In the following, we analyse the database of 80 direct $N$-body simulations performed with an improved version of the \texttt{NBODY6++GPU} code \citep{wang15}, presented in \cite{rizzuto20}. These models represent young massive star clusters evolved up to $\sim 300-500$ Myr composed of $110,000$ stars, including $10\%$ of primordial binaries, with stellar masses in the range $(0.08-100)\Ms$, and typical central densities in the range $10^5-3\times 10^7\Ms$ pc$^{-3}$. The key features of these models include stellar evolution of single and binary stars \citep{hurley00,hurley02,belczynski02}, natal kicks for neutron stars \citep{hobbs}, a fallback prescription to compute BH masses \citep{belczynski02}, a dedicated treatment for tight binary evolution \citep{mikkola98}, and general relativistic corrections \citep{mikkola08} to the equation of motions of compact stellar objects\footnote{We refer the reader to \cite{rizzuto20} for further details about the simulations}. 

Although our models do not include PISNe and PPISNe, we note that the initial mass function adopted in our models, in the range $0.08 - 100 \Ms$, should be not severely affected by such mechanisms, as they should act on stars heavier than $100-120 \Ms$ for the metallicity value adopted here \citep[see][]{spera17}.

The stellar evolution prescriptions in these models is slightly outdated, but they represent among the first fully self-consistent $N$-body simulations in which all the relevant ingredients listed above are accounted for simultaneously. Similar works are either based on smaller \citep[e.g.][]{dicarlo19,dicarlo20}, or sparser clusters \citep[e.g.][]{banerjee16,banerjee20}. An improved version of these simulations that include state-of-the-art stellar evolution prescriptions (Kamlah et al in prep) is currently underway (Arca Sedda et al in prep.).

Among all the simulations we identify 2 single mergers involving one IMBH and a stellar BH, 1 merger in which one IMBH undergoes two subsequent mergers with stellar BHs, and 1 case in which a sequence of three consecutive BH mergers builds-up an IMBH with total mass $\sim 140\Ms$.
In all the cases presented here, the binary merger is triggered by repeated interactions with passing-by stars and BHs and, in some cases, to the temporary formation of hierarchical triples that trigger an increase of the binary eccentricity and facilitate the merger \citep[cfr. Fig. 6-7 in][]{rizzuto20}. 

Hereafter, we label BH mergers consistently with the runs naming adopted in \cite{rizzuto20}. Models' names consist of three parts: the first part identifies the initial value of the cluster half-mass radius in pc, the second part identifies the value adopted for the central value of the adimensional potential well according to \cite{King} models, whereas the third refers to the number of the simulation. We add a further label to identify the number of previous mergers that the BH underwent, e.g. Gen-1(Gen-2) refers to a first(second) generation merger. For clarity's sake, we use letter $a$ to identify the heavier component of the binary and $b$ to identify the companion. 

In the following, we refer to mergers beyond the mass-gap as those involving at least one BH with a mass $>150\Ms$, whereas we define IMBHs all BHs with a mass $>100\Ms$. Similarly, we refer to  mergers falling in the mass-gap as those involving at least one component in the mass range $65-120\Ms$.

In the next section, we study separately mergers with BHs beyond or inside the upper mass-gap.

For all the mergers we record the semimajor axis and eccentricity the first time that the pericentre of the BH binary orbit falls below 100 times the sum of BHs Schwarzschild radii. We follow the final inspiral phase integrating the \cite{peters64} equations:
\begin{eqnarray}
\frac{\der a}{\der t} &=& -\frac{64}{5}\frac{G^3}{c^5}\beta(M_{1},M_{2})\frac{F(e)}{a^3},\\
\frac{\der e}{\der t} &=& -\frac{304}{15}\frac{G^3}{c^5}\beta(M_{1},M_{2})\frac{e G(e)}{a^4},
\label{peters2}
\end{eqnarray}
with
\begin{eqnarray}
F(e)          &=& (1 - e^2)^{-7/2}\left(1 +\frac{73}{24} e^2 + \frac{37}{96}e_f^4\right);\\
\beta(M_{1},M_{2})  &=&  M_{1}M_{2}(M_{1}+M_{2});\\
G(e)          &=& (1-e^2)^{-5/2}\left(1+\frac{121}{304}e^2\right).
\end{eqnarray}
This procedure enable us to determine the orbital properties of the binary during the last stages prior to the merger, a phase that is not captured from the direct $N-$body simulations. Along with the orbital evolution we calculate the associated GW strain and frequency, which are key ingredients to determine whether these mergers can be seen with GW detectors. 

As detailed in Appendix \ref{app:Spins}, we exploit numerical relativity fitting formula to estimate the remnant BH mass, spin, effective spin parameter \citep{jimenez17} and the post-merger GW recoil kick \citep{campanelli07,lousto08,lousto12}. 

Varying the distribution of stellar BH natal spins and for IMBHs, we place constraints on: i) the spin distribution of remnants, and ii) the likelihood for a merger remnant to be retained in the parent cluster and possibly undergo a further merger.

In the following, we study separately the mergers involving BHs beyond the mass-gap and the one case falling right inside the mass-gap, calculating for each merger the time evolution of the binary semimajor axis and eccentricity, the associated GW strain, the mass, spin, and effective spin parameter of the BH remnant, and the GW recoil kick.

\section{Black hole mergers beyond the mass-gap}
\label{sec:3}

Among the 80 simulations, we find a handful of interesting examples of BH binary mergers that led to a final BH with a total mass $>100\Ms$, thus in the IMBHs mass range. In two models, labelled as R06W9sim3 and R06W6sim6, the primary BH reaches a mass $M_a = 285-328\Ms$ through the accretion of a massive star onto a stellar mass BH, and after 11.3 and 113 Myr, respectively, form a binary with a stellar BH that undergoes coalescence in a few yr. In another model (named R06W9sim7), the primary BH grows up to a mass $M_a = 307\Ms$ and undergoes two subsequent merger events, first with a BH with mass $M_b = 22\Ms$ (R06W9sim7Gen-1) and later with another BH with mass $M_b = 26\Ms$ (R06W9sim7Gen-2).

In these cases, the heavier BH in the binary forms from the collision between a stellar mass BH and a very massive star (VMS) that assembled via a series of collisions in the cluster core among massive main sequence stars (mass $>50\Ms$) over a short timescale, $<10$ Myr. 

The amount of mass that is actually accreted onto a BH during a BH-star collision or accretion event is highly uncertain \citep[see e.g.][]{shiokawa15,metzger16,law19}. In \cite{rizzuto20}, this quantity is regulated through an accretion parameter $f_{\rm acc}$ that represents the fraction of the stellar mass fed to the BH. The whole set of simulations explores the impact of $f_{\rm acc} = 0.1-0.5-1.0$ values\footnote{for further details about the treatment of BH - star collision and mass loss during MS-MS mergers we refer the reader to \cite{rizzuto20}}. 
The actual mass of an IMBH candidate in this simulation is thus given by $M_{\rm IMBH} = M_{\rm BH,pro} + f_{\rm acc} M_{\rm VMS}$, where $M_{\rm BH,pro}$ and $M_{\rm VMS}$ are the masses of the accretor stellar BH and the donor VMS, respectively.
All the mergers discussed in this work take place in simulations with $f_{\rm acc} = 1$, thus in which the VMS-BH accretion is maximized.

Whether such high-efficiency accretion can be attained is unclear. In the majority of our models, typically the VMS-BH binary tightens via three-body interactions with fly-by stars up to the point at which the pericentral distance falls below the VMS typical radius. The BH will thus penetrate the outer stellar layer and rapidly sink to its centre, accreting mass in its course and after settling in the star centre. 

Many factors can contribute to determine the actual amount of mass accreted onto a compact object, like the type of the accretor (a NS or a BH), the evolutionary stage of the star, the size of the VMS, the structure of the VMS envelope, and the inclusion or not of General Relativistic effects \citep[see e.g.][]{macleod15, osorio20}. 

Simulations of the common envelope phase in tight binaries composed of a BH and a more massive star show that the accretion of the core of the companion
onto the BH and the expulsion of the envelope can trigger explosive phenomena that resemble SNe-like transients \citep{schroder20}. A VMS formed out of repeated collisions among main sequence stars is expected to have a compact core containing almost all the VMS mass and a tenuous envelope with densities as low as $10^{-10}$ g cm$^{-3}$ \citep{glebbeek09}. 

Therefore, a value $f_{\rm acc} = 1$ seems reasonable under the assumption that most of the VMS mass is concentrated in its core and that the core is completely accreted onto the BH.

In such a picture, the accretion process would spin-up the BH up to nearly extremal values regardless the initial BH spin \citep{schroder20}, thus suggesting that IMBHs formed out of BH-VMS interactions could be characterised by spins close to unity.

As summarized in Table \ref{tab:t1}, at formation the semimajor axis ($a$) and eccentricity ($e$) of our IMBH-BH binaries span a wide range, with two of the cases exhibiting an extremely large eccentricity of $e = 0.995 - 0.997$. 
Whether the large eccentricity is preserved or not when the binary enters the frequency window accessible to GW detectors depends mainly on the binary orbital properties. 
The associated GW signal emitted by eccentric sources is characterized by a broad frequency spectrum, with the frequency of the dominant harmonic given by 
\citep{oleary09}
\begin{align}
f_\gw =& \frac{1.2 ~{\rm mHz}}{1+z}\left(\frac{M_a+M_b}{300 \Ms}\right)^{1/2}
\left(\frac{a}{2 {\rm ~R}_\odot}\right)^{-3/2}\times \nonumber\\
&\times {\rm ceil}\left[1.15\frac{(1+e)^{1/2}}{(1-e)^{3/2}}\right]
\left(1-e\right)^{-3/2},
\end{align}
note that around $90\%$ of the GW emitted power is associated with harmonics at frequencies $0.2<f/f_\gw<3$ \citep{oleary09,kocsis12}. 

\begin{figure}
\includegraphics[width=\columnwidth]{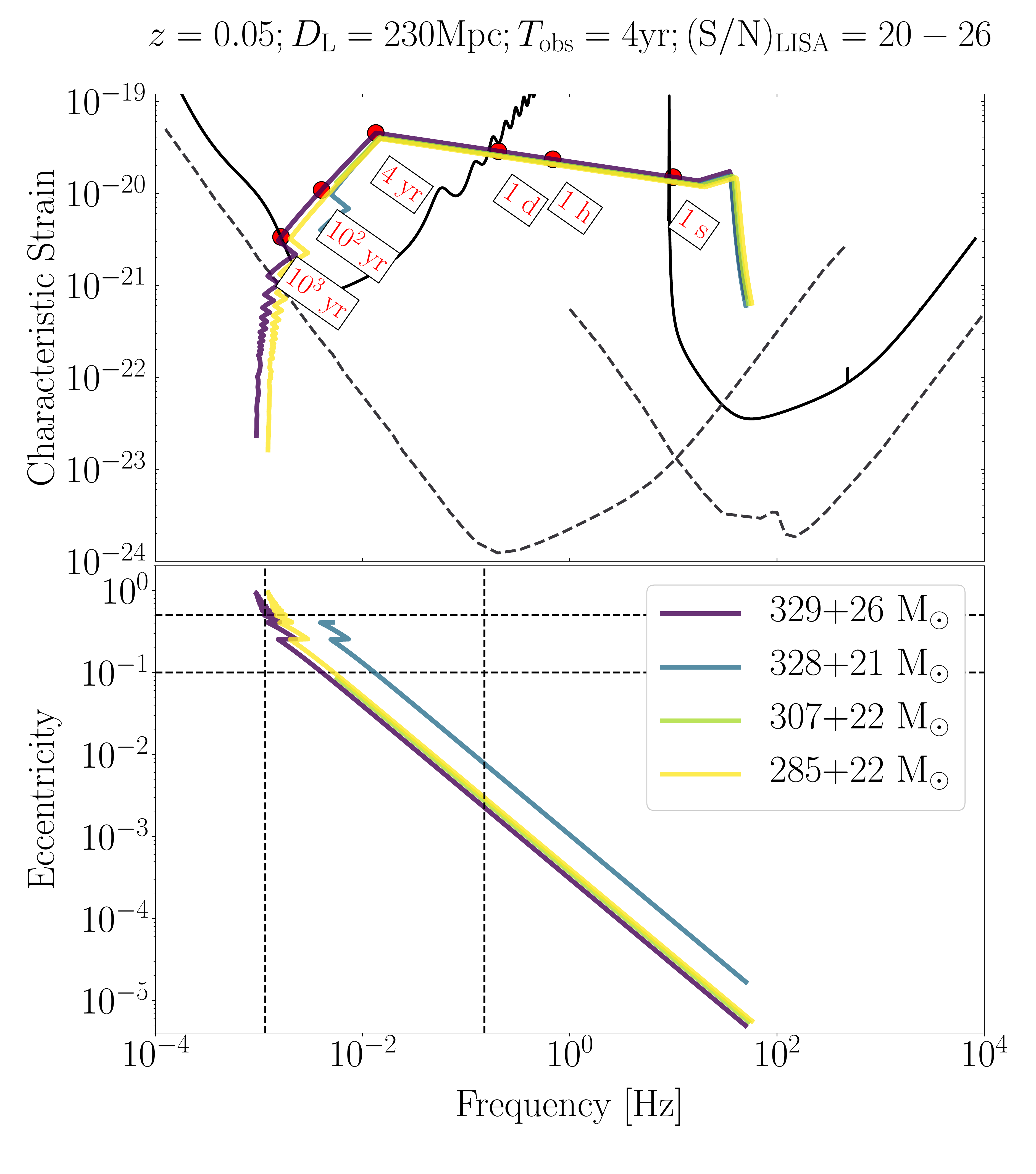}
\caption{Top panel: GW strain evolution as a function of the frequency for the mergers beyond the mass-gap (coloured lines). Simulated tracks are overlapped to the sensitivity curves of LISA and LIGO (solid black lines), and DECIGO and Einstein Telescope (dashed black lines). The white boxes indicate the time to merger for the heaviest merger. Bottom panel: eccentricity evolution as a function of the frequency. The vertical lines identify the moment in which the IMBH-BH mergers enter and exit the LISA sensitivity window, whereas horizontal lines identify the eccentricity values $e=0.1, 0.5$. All the mergers are assumed to happen at a redshift $z=0.05$, corresponding to a luminosity distance $D_{\rm L} = 230$ Mpc. The figure title report the typical signal-to-noise ratio $({\rm S}/{\rm N}) \sim 20-26$, assuming for LISA a $T_{\rm obs} = 4$ yr long mission.}
\label{fig:freq}
\end{figure}

At formation, the dominant frequency for all the IMBH-BH mergers described above spans the $f_\gw \sim 10^{-3} - 10^{-2}$ Hz frequency range, thus falling inside the LISA sensitive window. 
Figure \ref{fig:freq} shows the evolution of the GW strain, the frequency, and eccentricity for all IMBH-BH mergers in our model, assuming that they are located at a redshift $z = 0.05$. 
As shown in the figure, at such distance the GW strain of these IMBH-BH mergers will move inside the LISA sensitivity window from $10^3$ yr to $\sim 1$ day prior to the merger. Within this timespan, the IMBH-BH eccentricity decreases from $e\simeq 0.5$ to $e < 0.01$ when the time to merger is $4$ yr, and reaches values below $e<10^{-4}$ one second before the merger. 

To characterize the typical eccentricity of the binary as it sweeps through the LISA frequency window, we proceeds in two ways. On the one hand, we calculate the average eccentricity weighted with the time to the merger, which provides a measure of the probability to observe the binary in a given orbital configuration. On the other hand, we measure the binary eccentricity 10 yr and 5 yr prior to the merger, which is a period of time during which LISA can likely follow the binary inspiral.  

As summarized in Table \ref{tab:t1}, we infer an average eccentricity of $\langle e \rangle \sim 0.302$ for model with $(M_1+M_2) = (328 + 21)\Ms$, 
$\langle e \rangle \sim 0.288$ for model with $(M_1+M_2) = (329 + 26)\Ms$, and $\langle e \rangle \sim 0.296$ for model with $(M_1+M_2) = (285 + 22)\Ms$. The eccentricity 10 yr and 5 yr prior to the merger, instead, ranges in between $e_{5,10} = 0.04 - 0.13$. Even measuring such a relatively small $e$ values would provide a unique proof of a dynamical origin for this class of mergers, as the typical eccentricities for mergers in field binaries falling in the LISA band are expected to be much smaller, i.e. $10^{-6}<e<10^{-4}$ \citep{kowalska11,nishizawa16}.

Detecting BH mergers with such masses using LISA depends on several parameters, like the redshift at which the merger take place, the signal-to-noise ratio (SNR, $(S/N)$) required to obtain a clear GW signal, the observation time. Assuming that the mergers occur closer than $\sim 44$ Mpc, corresponding to a redshift $z=0.01$, and adopting an observation time of $4$ yr, we find that LISA could detect the last phase of these IMBH-BH mergers with a $(S/N) = 100-150$, well above the minimum threshold required for detection, i.e. $(S/N)_{\rm min} = 15$ \citep{amaro17lisa}. More in general, we derive the LISA horizon redshift, i.e. the maximum distance that LISA can reach for IMBHs with masses in the range $100-400\Ms$ merging with stellar BHs (mass $5-50\Ms$), as detailed in Appendix \ref{app:GWstr}, in the limit of nearly circular sources. This choice is motivated by the fact that, as mentioned above, our IMBH-BH mergers are expected to be circularized in the LISA band.
Assuming a minimum $(S/N)_{\rm min}=15$ and a 4 yr observation time, we find that LISA could detect these mergers out to a redshift $z = 0.01 - 0.1$, depending on the masses of the binary components. For clarity's sake, the GW strain-frequency evolution shown in Figure \ref{fig:freq} assumes that the sources are located at a redshift $z = 0.05$, well within the boundaries imposed by the horizon redshift.
We note that taking into account the small residual eccentricity in the calculation of the SNR affects the horizon redshift by less than $10\%$. Our analysis suggests that mergers occurring in the Milky Way (MW) and its surrounding could be extremely bright low-frequency GW sources, making LISA an unique tool to unravel IMBH formation routes in young dense clusters.

 \begin{table*}
    \centering
    \begin{tabular}{cccccccccc|ccc}
    \hline
    \hline
    Simulation ID & Merger    & $\sigma_{YC}$ & $M_a$ & $M_b$ & $q$ & $a$ & $e$ & $ t_{\rm form} $&$t_{\rm gw}$ & & &   \\ 
                  & generation& km s$^{-1}$   & $\Ms$ & $\Ms$ & & R$_\odot$ & & Myr &yr & & & \\
    \hline
                \multicolumn{10}{c|}{IMBH-BH mergers} & \multicolumn{3}{c}{$f = 10^{-4}-10^{-2}$ Hz} \\
                \multicolumn{10}{c|}{} & $\langle e\rangle$ &  $e_{\rm 5 yr}$ &  $e_{\rm 10 yr}$\\
    \hline
    R06W9sim3     & Gen-1&  $7$ & $328$ &  $21$ & $0.064 $ & $1.21 $& $0.41   $&$ 113$ &$ 44 $   &$ 0.32$&$ 0.098$&$ 0.128$  \\ 
    R06W9sim7     & Gen-1&  $6$ & $307$ &  $22$ & $0.072 $ & $0.7  $& $0.085  $&$ 348$ &$ 16$    &$ 0.06$&$ 0.035$&$ 0.046$  \\ 
    R06W9sim7     & Gen-2&  $6$ & $329$ &  $26$ & $0.079 $ & $697  $& $0.997  $&$ 369$ &$ 30,000$&$ 0.76$&$ 0.031$&$ 0.041$  \\ 
    R06W6sim6     & Gen-1& $10$ & $285$ &  $22$ & $0.077 $ & $36.7 $& $0.955  $&$ 11.2$&$ 8500 $ &$ 0.71$&$ 0.037$&$ 0.049$  \\
    \hline
                \multicolumn{10}{c|}{Mass-gap mergers}  & \multicolumn{3}{c}{$f>10$ Hz}\\
                \multicolumn{10}{c|}{}  & $e_{\rm 10s}$ & $e_{\rm 1s}$ & $e_{\rm 0.1s}$ \\
    \hline
    R06W6sim1     & Gen-1&  $9$ &  $28$ &  $17$ & $0.607 $ & $337  $& $0.999  $&$ 38.6$&$ 36.7$&$1.1\times 10^{-4}$&$ 4.7\times 10^{-5}$&$ 2.2\times10^{-5}$\\ 
    R06W6sim1     & Gen-2&  $9$ &  $45$ &  $25$ & $0.556 $ & $83.4 $& $0.99978$&$ 46.9$&$ 10 $ &$ 0.02$&$ 0.008$&$ 0.004$\\ 
    R06W6sim1     & Gen-3&  $8$ &  $70$ &  $68$ & $0.971 $ & $1.08 $& $0.8927 $&$ 83.9$&$ 10 $ &$ 6.4\times10^{-4} $&$ 3.6\times10^{-4} $&$ 3.0\times10^{-4}$\\ 
    \hline
    \end{tabular}
    \caption{Col. 1: simulation ID. Col. 2: number of previous mergers involving one of the components. Col. 3: cluster velocity dispersion at the time of merger. Col. 4-5: mass of the primary (subscript ``$a$'') and companion (``$b$''), respectively. Col. 6: binary mass ratio. Col. 7-8: semimajor axis and eccentricity at the last snapshot before the merger. Col. 9: time at which the binary formed. Col. 10: merger time since the last available snapshot. Col. 11-13: eccentricity of the binary, depending on the merger mass. For IMBH-BH mergers columns identify the eccentricity averaged over the merger time and the eccentricity measured 5 yr and 10 yr before merger. For stellar BH mergers, columns identify the eccentricity measured 10, 1, 0.1 second before merger, respectively.}
    \label{tab:t1}
 \end{table*}

Another important information that can be extracted from the GW signal is the spin of the merging BHs, which in turn provides insights on the BH formation mechanisms. The current picture about the natal spin distribution of stellar BHs is poorly constrained, and the impact of binary evolution, stellar rotation, or supernova explosion on the natal BH spin is unclear \citep{demink13,seoane16,Belczynski17,Qin18,Qin19}. The detection of GWs helped in placing some constraints on the spin distribution of merging BH binaries \citep{gwtc2b}, although there are a number of potential correlations between the spin and the binary formation history that are difficult to disentangle \citep[e.g.][]{arcasedda19b, bavera20, arcasedda20b,zevin20b}. For IMBHs, the picture is even more complex, as the spin will be inevitably connected with their formation history. 
For instance, matter accretion onto a stellar BH from a much heavier companion could efficiently spin-up the BH and lead to a final IMBH with a spin close to unity \citep{schroder20}. On the other hand, it is unclear whether the accretion of material from a ``normal'' stellar companion can efficiently spin-up the growing BH \citep{Qin18,Qin19}. 

\begin{figure*}
    \centering
    \includegraphics[width=1.5\columnwidth]{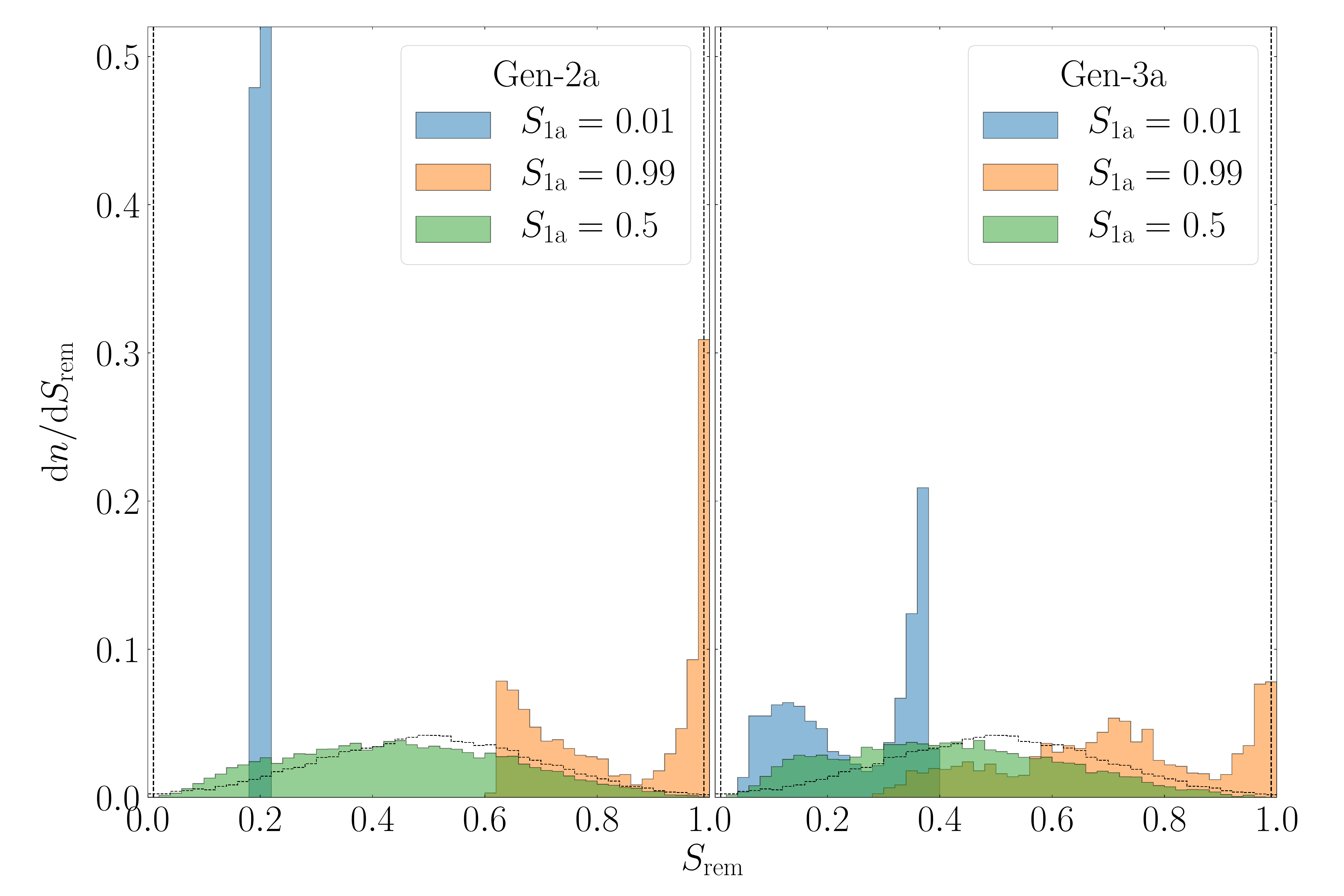}
    \caption{Spin distribution of the remnant IMBH in model R06W9sim7 after the first (Gen-1, left panel) and the second merger (Gen-2, right panel). We assume that, prior to the merger, the IMBH spin is either $S_{\rm 1a} = 0.01$ (blue steps), $S_{\rm 1a} = 0.99$ (orange steps), or it is drawn from the same distribution adopted for stellar BHs (green steps). The stellar BH spin is draw from a Gaussian distribution with average value $S_{\rm 1b}=0.5$  and dispersion $\sigma_S = 0.2$. The black dotted lines identify the different distributions adopted for the IMBH initial spin.}
    \label{fig:IBHsp}
\end{figure*}

Therefore, we follow a conservative approach to get insights on the IMBH and BH spin distribution and evolution. For each IMBH-BH merger, we assume that either the IMBH is almost non-spinning ($S_{\rm 1a} = 0.01$), nearly extremal ($S_{\rm 1a} = 0.99$), or it follows the same spin distribution of stellar BHs. For stellar BH spins we test either a flat distribution in the range $S_{\rm 1b} = 0-1$\footnote{We indicate all stellar BHs with letter ``b'' as they are the lightest component in the IMBH-BH binary, regardless of the merger generation.}  or a gaussian distribution peaked at $S_{\rm 1b} = 0.2-0.5-0.7$ with a dispersion of $\sigma_S = 0.2$. Upon these hypothesis, we sample 2,000 values of the stellar BH spin for each merger and each value of $S_{\rm 1a}$, calculating the remnant mass and spin via numerical relativity fitting formulae \citep{jimenez17}. Figure \ref{fig:IBHsp} shows the spin distribution of the IMBH remnant for model R06W9sim7 after the first- and the second-generation merger, assuming a Gaussian peaked at $s_{\rm 1b} = 0.5$ for the stellar BH spin distribution.  

We find that the spin of the remnant depends critically on the pre-merging IMBH spin, whereas it is weakly dependent on the stellar BH spin distribution. The remnant IMBH spin distribution narrowly peaks around $S_{\rm 2a} = 0.2$, whereas for a nearly extremal IMBH the remnant spin distribution is widely distributed and exhibits two peaks at $S_{\rm 2a} = 0.6$ and $0.99$. The second merger event changes the IMBH spin further, leading to a double peak distribution with peaks at 0.15 and 0.3 if the IMBH was originally non-spinning, while leading to a wide three-peaks distribution in the case of an initially nearly-extremal IMBH. If the IMBH spin distribution follows the same distribution of stellar BHs, e.g. a Gaussian, we found that the distribution gets skewed toward smaller spin values. The distribution peak shifts toward smaller values at every subsequent merger event, moving from $S_{1_a,{\rm peak}} = 0.5$ for Gen-1a, to $S_{2_a,{\rm peak}} = 0.4$ for Gen-2a, and $S_{3_a,{\rm peak}}=0.3$. Even in a single merger event, our analysis outlines that the natal IMBH spin has a crucial impact on the remnant spin. Spin measurements in IMBH-BH mergers via GW detections have thus the potential to unravel IMBH natal spin distribution and, thus, to help constraining IMBH formation routes.

Multiple-generation mergers can carry further insights on the IMBH formation and evolution, but their development can be prevented by recoil kicks imparted to the merger remnant due to anysotropic GW emission, which can be as large as $10^3$ km s$^{-1}$ \citep{lousto08} and thus can kick the IMBH out of the cluster after the first merger \citep[e.g.][]{bockelmann08}.

In clusters with sufficiently large escape velocities, repeated mergers can lead the IMBH spin to slowly decrease and attain final values $<0.1-0.3$, regardless of the IMBH natal spin \citep[see e.g.][]{arcasedda20c}. 
Owing to this, the distribution of IMBH spins formed in high escape velocity clusters should contain little information about the IMBH natal spin, which insted will dominate the spin distribution of IMBHs developed in low escape velocity clusters.

Unfortunately, GW recoil are not implemented in our $N$-body models, so we resort to post-processing of the data to explore how GW recoil kick would impact the retention of the merger products in dense clusters and how this would impact the spin distribution of merged IMBHs, as detailed in Section \ref{sec:5}.  

\section{Black hole mergers in the mass-gap}
\label{sec:4}
\begin{figure*}
    \centering
    \includegraphics[width=0.9\textwidth]{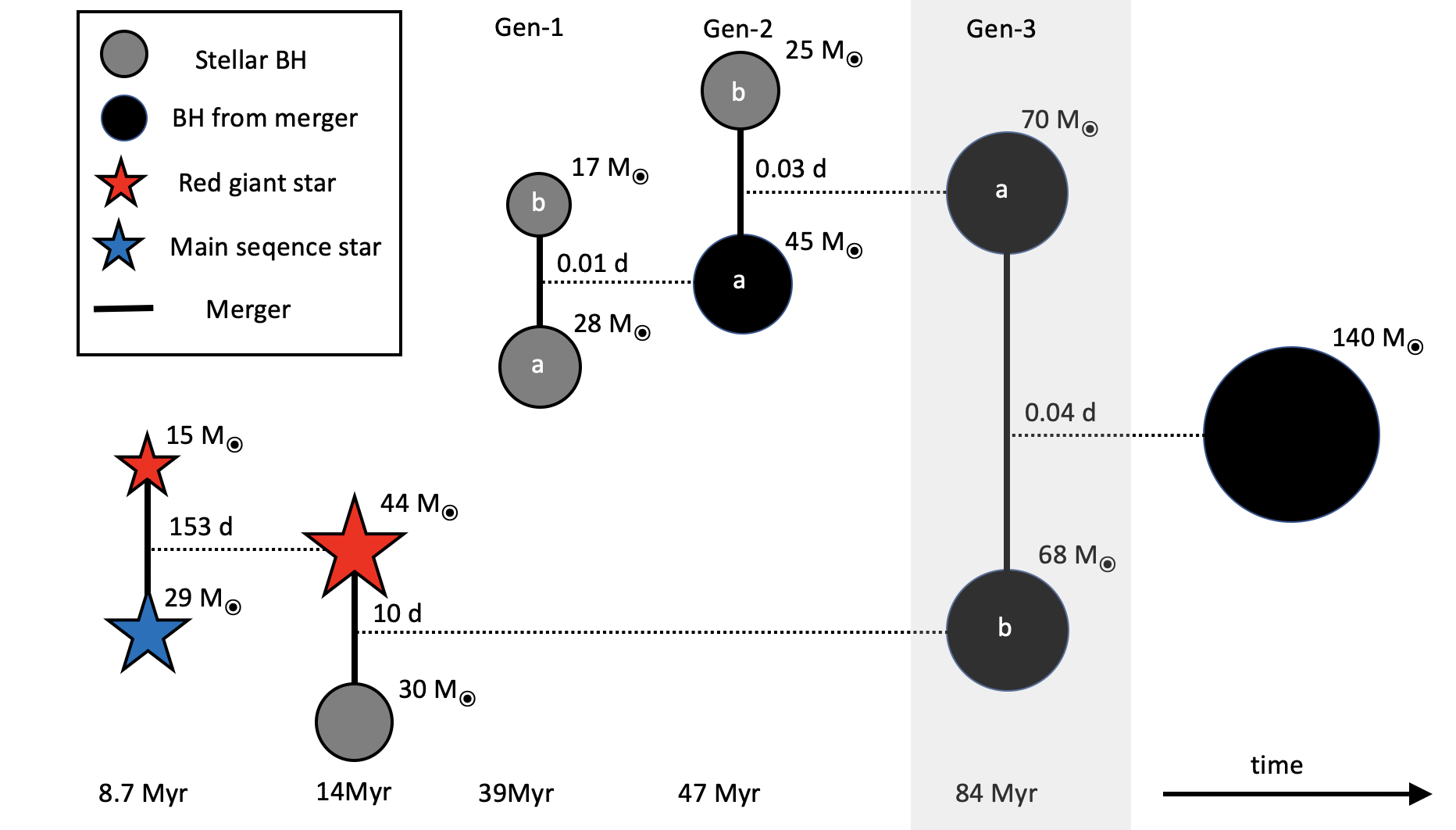}
    \caption{Visualization of the formation path towards a ``mass gap'' merger (grey region, Gen-4) of two black holes with mass $70 M_\odot$ (Gen-3a) and $68 M_\odot$ (Gen-3b) developed in a high-resolution $N$-body simulation of a dense star cluster \cite{rizzuto20}. The more massive BH (Gen-3a) grows by two preceding BH mergers (Gen-1, Gen-2). The lower mass BH (Gen-3b) builds up in a stellar merger of a red giant with a main sequence star followed by the collision with a stellar mass BH. The masses of the components (in solar masses) and the orbital periods (in days) are indicated at the respective times of the merger (black horizontal lines) after the start of the simulation. \label{fig0}}
\end{figure*}

In another model (R06W6sim1), we found a binary BH merger with components in the upper mass-gap and interestingly similar to GW190521. This merger was formed via a peculiar channel, namely a series of three subsequent merger events, as sketched in Figure \ref{fig0}. 

The sequence of mergers that built-up our ``GW190521-like'' merger starts with a first generation merger of two stellar mass BHs with masses $28\Ms$ (hereafter Gen-1a) and $17\Ms$ (Gen-1b), after $39$ million years (Myr). The Gen-1a merger product, which we name Gen-2a, captures a companion, Gen-2b, with mass $ 25\Ms$ and merges 10 Myr later, leaving behind a product, Gen-3a, with a total mass of $ 70\Ms$. Gen-3a forms a binary with Gen-3b, a BH with mass $68\Ms$ formed out of the accretion of a red giant star with mass $\sim 50\Ms$ onto a stellar BH with mass $\sim 30\Ms$. The binary Gen-3ab undergoes coalescence within $84$ Myr, assemblying Gen-4, a 4th generation BH with total mass $138\Ms$\footnote{This is the mass added from all components and does not account for matter loss due to GW emission}. Note that Gen-3a(b) represents the primary(secondary) in the final merger of the sequence, in analogy with GW190521 components naming. 

About 200 years prior to the merger, the Gen-3ab binary has a semi-major axis of $864$ solar radii and eccentricity 0.99982. However, under the assumption that the system is isolated, GW emission induces the circularization of the binary and leads its eccentricity to drop below $10^{-3}$ by the time the binary enters the LIGO sensitivity band, at frequencies $>10$ Hz. Measuring the eccentricity for GW sources in the LIGO band is quite hard, as the merger spends a short time in band. So far, the analysis of LIGO-Virgo data seems to disfavor a significant eccentricity at merger \citep{gw190521b}, although a more robust assessment about the binary eccentricity is complicated by a partial degeneracy between high eccentricity and precession \citep{gw190521b,romero20}.

Assuming a redshift $z=0.8$, i.e. the value estimated for GW190521, we found that LISA would not be able to detect any of the three mergers. When the last merger enters the LIGO sensitive band, i.e. frequency range $f=10-1000$ Hz, the binary eccentricity is already $e<10^{-4}$. This is due to the fact that at the last snapshot our binary has a relatively wide orbit, with a semimajor axis of ${a}\sim 4$ AU. For instance, reducing the ${a}$ by a factor $10(50)$ would lead the eccentricity measured at 10 Hz to increase up to $0.02(0.06)$, yet smaller than the maximum eccentricity inferred for GW190521 \citep{romero20}. An eccentricity above 0.1 could be reached if the ${a} = 0.4$ AU and the corresponding eccentricity is $e = 0.99995$. Note that the typical semimajor axis of long-lived (hard) binaries in star clusters scales with the cluster velocity dispersion as ${a}_{\rm hard}\propto \sigma^2$, thus statistically speaking, forming binaries ten times tighter would require a host cluster with a velocity dispersion $\simeq \sqrt{10}$ times larger.

Following the same scheme adopted for IMBH mergers, we calculate how the remnant mass and spin vary after every new merger event and upon different assumptions of the BH natal spin distribution. For the BH natal spin distribution we adopt the same assumption as in the previous section. We assume that the natal spin of Gen-3b is not affected substantially by the accretion from the previous merger with a red giant companion \citep{valsecchi10,wong12,Qin19}. 

\begin{figure*}
    \centering
    \includegraphics[width=0.9\textwidth]{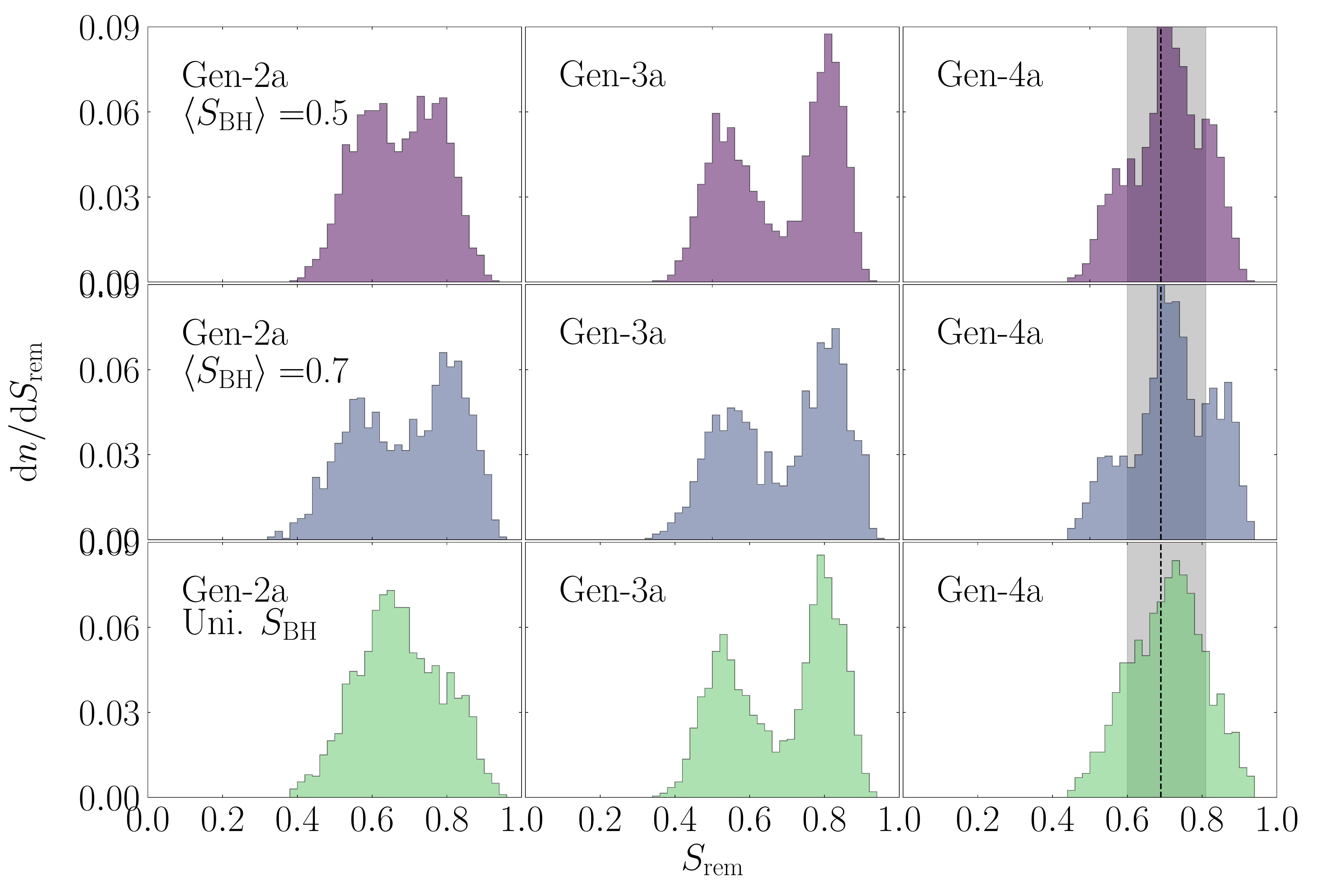}
    \caption{Spin distribution of the BH remnant in model R06W6sim1 after the first (BH remnant name Gen-2a, left panels), second (Gen-3a, central panels), and third (Gen-4a, right panels) merger. From top to bottom, panel rows refer to different assumptions on the natal BH spin distribution, namely a gaussian peaked either on $0.5$ (upper row) or $0.7$ (central row), or a uniform distribution (lower row).}
    \label{fig:my_label}
\end{figure*}

As shown in Figure \ref{fig:my_label}, we find that a chain of three-mergers as the one described here leads to final BH spin and effective spin parameter $\chi_{\rm eff}$ distributions that are fully compatible with the $90\%$ credible level measured for GW190521. In Appendix \ref{app:cmp} we compare the distribution for $S$ and $\chi_{\rm eff}$ assuming that GW190521 formed through of one, two, or three consecutive mergers. We show that a triple merger provides a better match to observations than a double merger, depending on the subtle BH natal spin distribution.

However, the successful development of a multiple generation merger depends critically on the ratio between two quantities, namely the cluster escape velocity and the GW recoil kick imparted onto the remnant BH after each merger event. In the next section we quantify the probability for such merger chain to occur in our simulated cluster or in denser environments.

\subsection{Degeneracy in multiple merger scenarios behind the GW190521 origin}
\label{app:cmp}
Several works in the recent literature pointed out that GW190521 could be the byproduct of a hierarchical merger, i.e. a series of mergers that built-up the GW190521 primary first and later gave rise to the detected merger. In our simulations we find one such case in which a series of three mergers led to a final merger similar to GW190521. 

\begin{figure*}
    \centering
    \includegraphics[width=0.95\columnwidth]{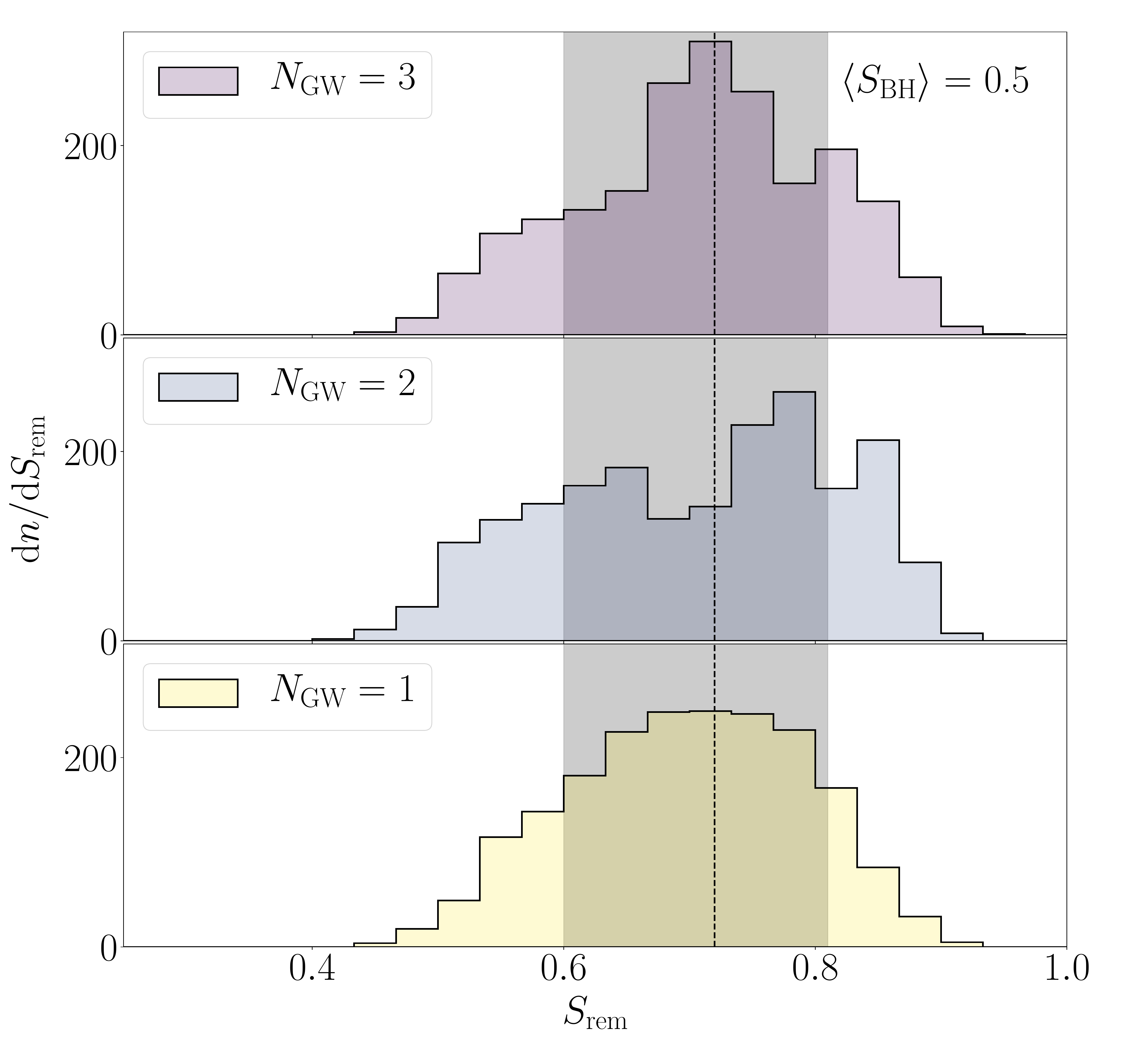}
    \includegraphics[width=0.95\columnwidth]{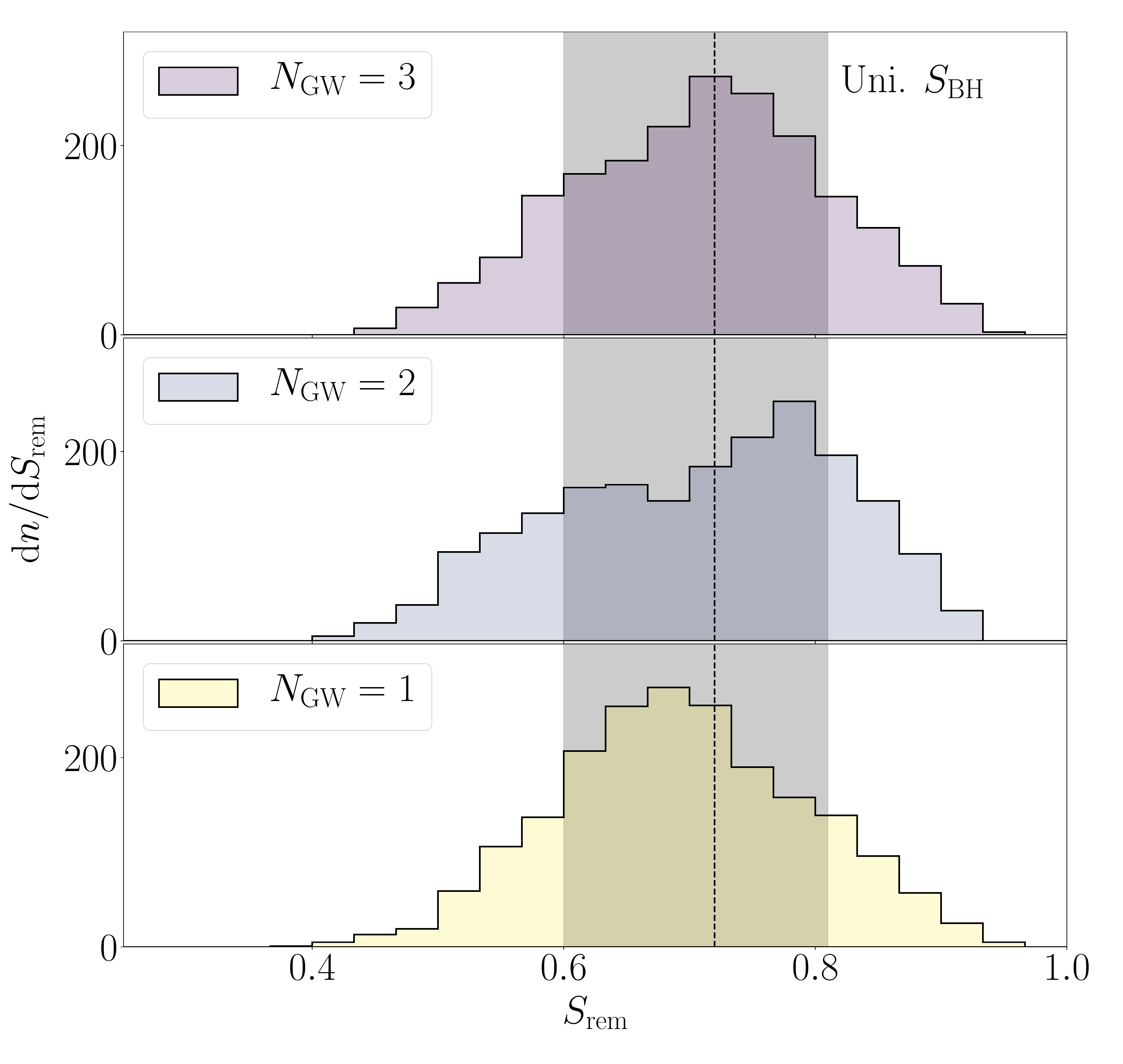}
    \includegraphics[width=0.95\columnwidth]{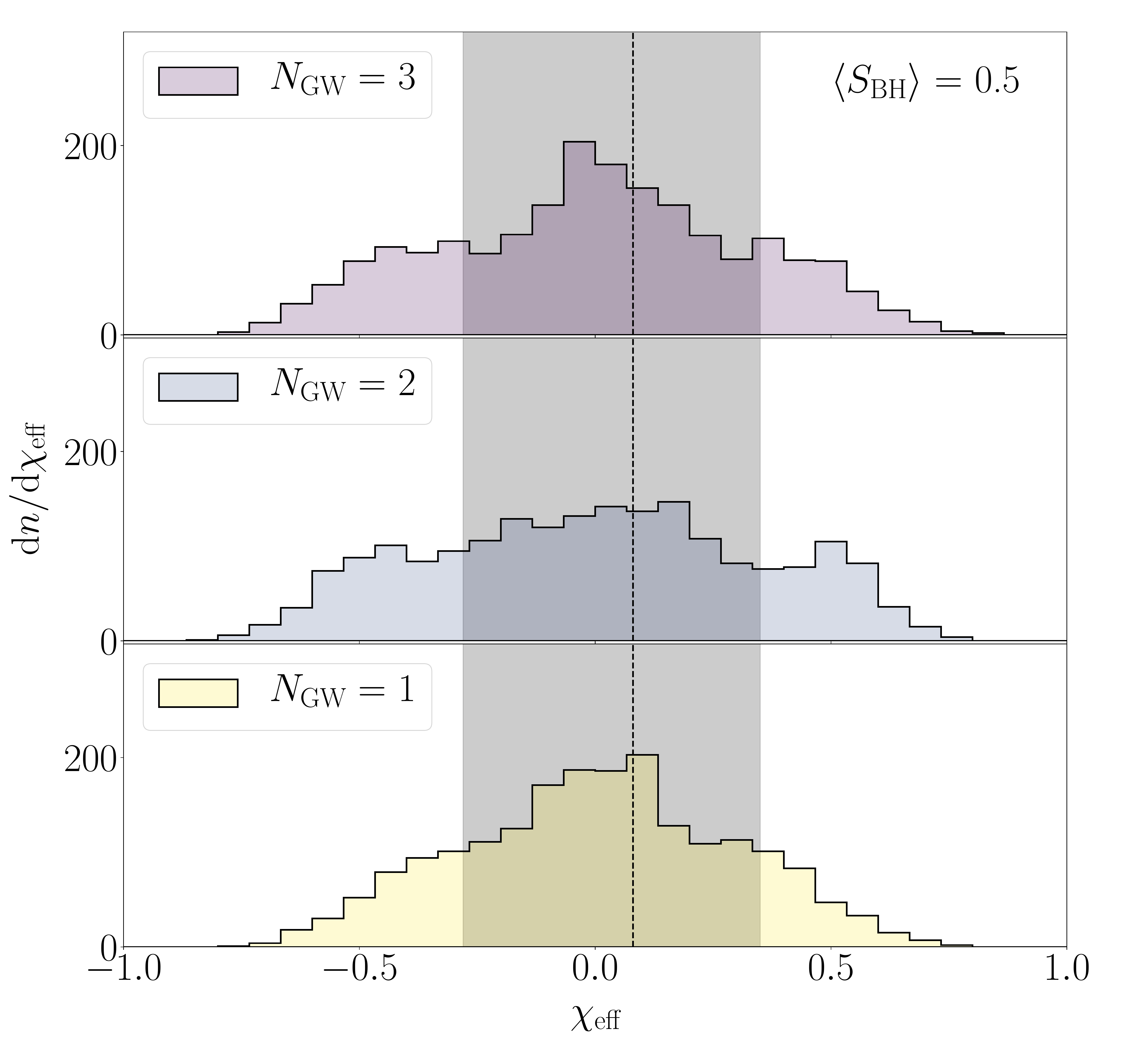}
    \includegraphics[width=0.95\columnwidth]{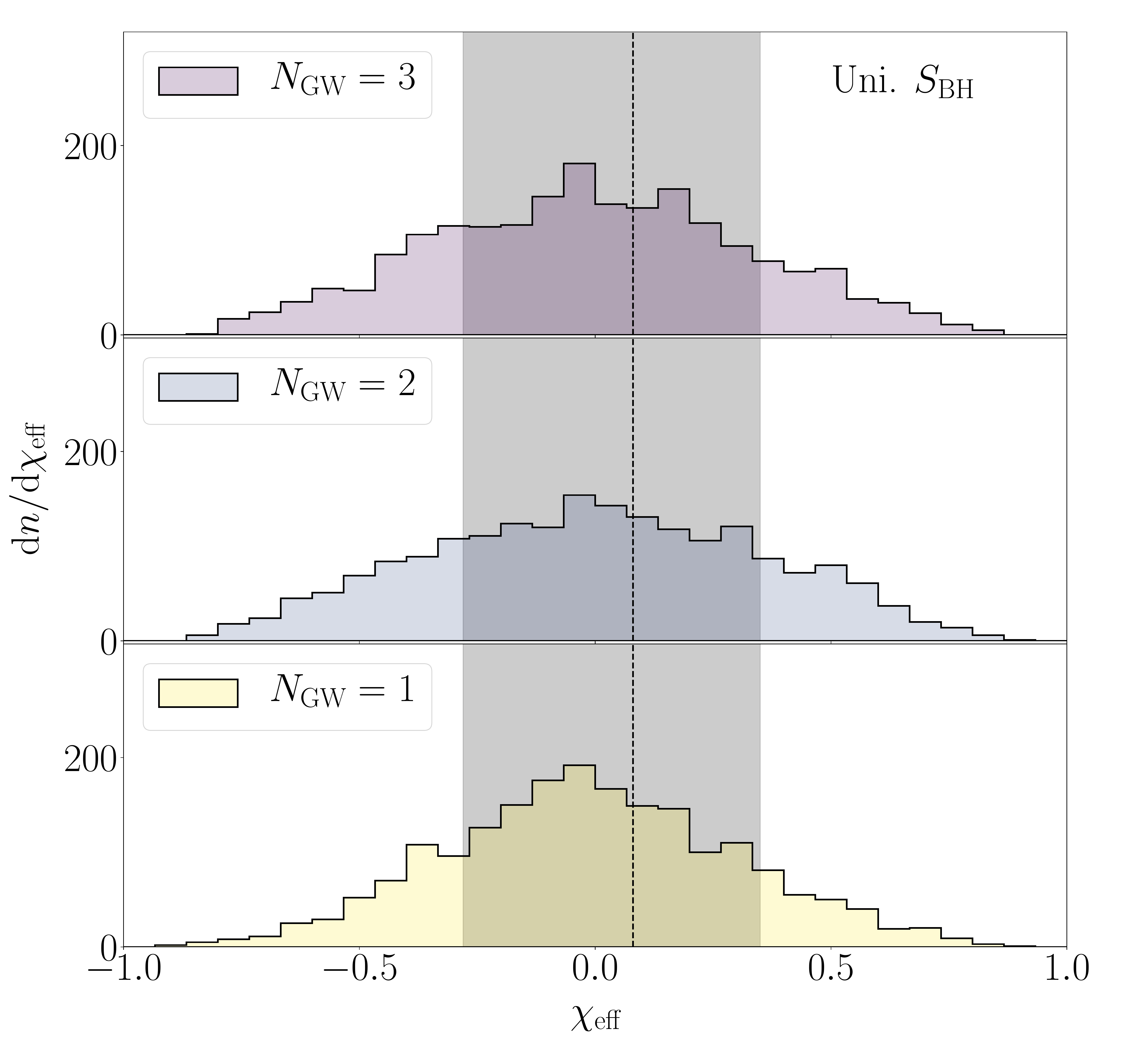}
    \caption{Top row: remnant spin distribution for our GW190521-like merger assuming that the final remnant is the product of three (top panel in each sub-figure), two (central panel) or one merger (bottom panel). Bottom row: as above, but here we calculate the effective spin parameter. Left plots correspond to the model in which natal BH spins are drawn from a Gaussian peaked at 0.5, whereas right plots refer to a uniform BH spin distribution.}
    \label{fig:cmpspin}
\end{figure*}

In this section, we compare a triple merger channel for the origin of GW190521 with simpler scenarios, namely a single merger, although this is challenged by stellar evolution theory of both single and binary stars, and a double merger. Figure \ref{fig:cmpspin} compares the remnant spin and effective spin distribution for these three channels (single, double, and triple merger) assuming that the natal BH spin distribution is either Gaussian with a peak at 0.5 or uniform. Regardless of the BH natal spins, the plot highlights that a double merger scenario seems less favored over a single or a triple merger scenarios, as they both produce a remnant spin distribution that nicely fits the $90\%$ credible level measured for GW190521. Quantitatively speaking, the fraction of merger remnants with a remnant spin compatible with GW190521 ranges between $\sim 0.66-0.72$ for the single and triple merger channels, while it is limited to $0.58-0.6$ for the single channel. Similarly, a single or triple merger is more likely to produce a remnant with $\chi_{\rm eff}$ compatible with that measured for GW190521. 
Our analysis suggests that measuring spin and effective spin parameters of massive BBH mergers can help unravelling the BH natal spin distribution and constraining further the BH natal mass distribution. For instance, the single merger channel would hint at a scenario in which the GW190521 components grew their mass via either stellar accretion or stellar collision, or would imply that some processes in single and binary stellar evolution are yet not fully understood. As we show in the next section, the triple merger channel would instead have crucial implications on the properties of the host cluster. 
Table \ref{tab:app} summarizes the fraction of models compatible with GW190521 remnant spin for the different models explored. 

\begin{table}
    \centering
    \begin{tabular}{c|ccc|ccc}
        \hline
        \hline
        BH spin        & \multicolumn{3}{c|}{$S_{\rm rem}$} & \multicolumn{3}{c}{$\chi_{\rm eff}$} \\
     distribution      &    1   &   2   &   3   & 1   &   2    &   3  \\
        \hline
        Uniform        &  0.716 & 0.579 & 0.666 & 0.679 & 0.587 & 0.624 \\
        Gaussian 0.2   &  0.929 & 0.618 & 0.782 & 0.936 & 0.616 & 0.707 \\
        Gaussian 0.5   &  0.716 & 0.579 & 0.666 & 0.691 & 0.569 & 0.614 \\
        Gaussian 0.7   &  0.582 & 0.582 & 0.597 & 0.582 & 0.558 & 0.581 \\ 
        \hline
    \end{tabular}
    \caption{Occurrence of GW190521-like merger remnant spin and effective spin parameter for single (first sub-column in main columns 2 and 3), double (second sub-column), and triple (third sub-column) mergers.}
    \label{tab:app}
\end{table}

\section{Gravitational wave recoil}
\label{sec:5}

In this section we calculate the GW recoil kick imparted onto our IMBH and BH mergers and the implication for IMBH retention and the development of multiple mergers in YMCs. As detailed in Appendix \ref{app:GWstr}, for each merger we calculate the GW recoil via numerical relativity fitting formulae \citep{campanelli07,lousto08,lousto12} and compare it with the cluster velocity dispersion. 

\subsection{Mergers beyond the mass-gap}
\begin{figure*}
    \centering
    \includegraphics[width=0.8\textwidth]{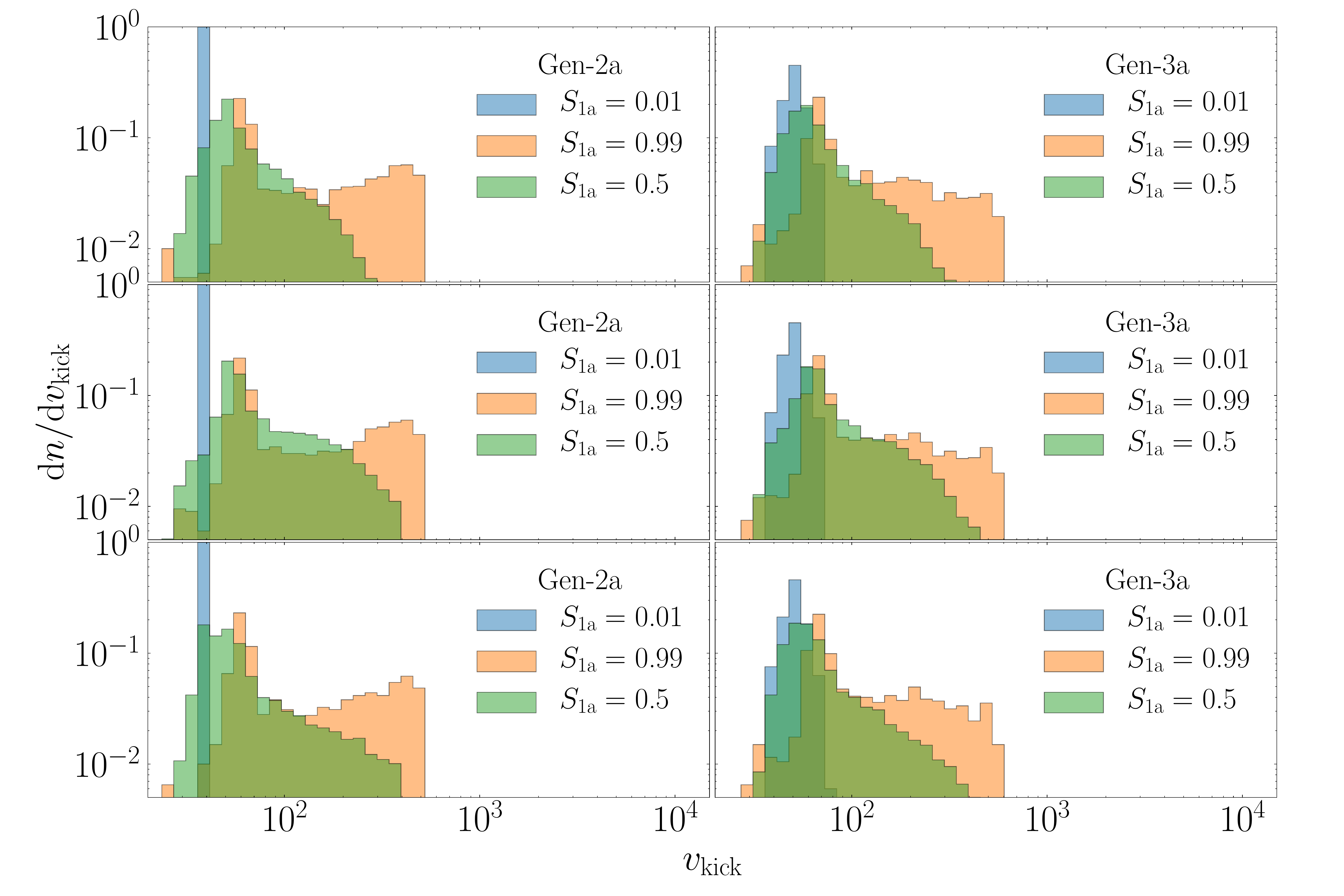}
    \caption{Gravitational wave recoil imparted to the IMBH in model R06W6sim7 after the first (left panels) and the second merger (right panels). In each panel we differentiate between the case of an initially slowly spinning (blue steps), a nearly extremal (orange steps) IMBHs, or assuming that IMBHs and BHs follow the same spin distribution. From top to bottom, we assume that the stellar BH spin distribution is a Gaussian peaked on 0.5 (top) or 0.7 (central), or is described by a Uniform distribution (bottom).}
    \label{fig:IBHkick}
\end{figure*}

We focus on IMBH-BH mergers first, particularly on the double merger found in model R06W6sim7. Figure \ref{fig:IBHkick} shows the kick distribution imparted on the post-merged IMBH in the case in which the IMBH is initially either non-rotating, nearly extremal, or has a spin drawn from the same distribution adopted for BHs. For stellar BHs, the spin distribution is either a Gaussian peaked on $0.5,~0.7$, or a uniform distribution. If the IMBH is initially slowly spinning, the kick received after both the first and second merger attains values around $v_{\rm kick} = 30$ km s$^{-1}$, and in general smaller than $80$ km s$^{-1}$. Larger IMBH spins would instead lead the GW kick to increase considerably, reaching values up to $500$ km s$^{-1}$. Nonetheless, even for spinning IMBHs the GW kick distribution shows a peak at $30-40$ km s$^{-1}$. Our simulated cluster has a total mass of $M_c = 7.4\times10^4\Ms$ and half-mass radius of $r_h = 1$ pc. If we assume that the central part of such cluster can be described by a power-law with slope $\gamma$, it is possible to express the central escape velocity as \citep{Deh93}
\begin{equation}
    v_{\rm esc}^2  = \frac{2GM_c}{(2-\gamma)r_h}\left(2^{1/(3-\gamma)}-1\right),
    \label{eq:vesc}
\end{equation}
which returns $v_{\rm esc} \simeq 45.8-51.6$ km s$^{-1}$ assuming $\gamma = 0-1$. We note that the estimate obtained via the Equation above is around twice the escape velocity value calculated through the gravitational potential calculated directly from the simulation. 

In the case of a steeper density profile, the escape velocity for a \cite{Deh93} model diverges in the cluster centre. If we assume that the merger takes place in the cluster inner part, e.g. $r<0.1r_h$, the escape velocity rises to up to $60-112$ km s$^{-1}$ for $\gamma = 1.5-2$.

Assuming $v_{\rm esc} = 40$ km s$^{-1}$, i.e. a value compatible with the one measured from the $N$-body simulation, we find that the remnant of a first generation IMBH-BH merger has a retention probability of $P_{\rm Gen-2a} = 96.9 - 99.1\%$ if the IMBH initial spin is almost null, whilst the second generation merger remnant is retained with a $P_{\rm Gen-3a} = 1.8-2.1\%$ probability. This probability rises up to $P_{\rm Gen-3a} = 50\%$ assuming $v_{\rm esc} = 50$ km s$^{-1}$. 
If the initial IMBH spin is $\sim 1$, instead, the retention probability decreases down to $P_{\rm Gen-2a} = 3(0.1)\%$ for the first(second) generation merger product 

Therefore, measuring the spin in an IMBH-BH merger could provide crucial insights on the possible host environment in which the merger developed.

\subsection{Mergers in the mass-gap}

To place constraints on the GW recoil kick imparted to BH remnants in the triple merger channel that leads to the GW190521-like merger in one of our simulations, we assign to the merger components a spin extracted from either a Gaussian, centered on either $S_{\rm} = 0.2-0.5-0.7$, or a Uniform distribution. For each merger event, we sample 10,000 values of the spin for each component and calculate the GW recoil imparted on the remnant. We find that in all the three mergers the remnant has more than $90\%$ of probability to receive a kick larger than $80$ km s$^{-1}$, regardless of the BH natal spin distribution.

\begin{figure}
\centering
\includegraphics[width=\columnwidth]{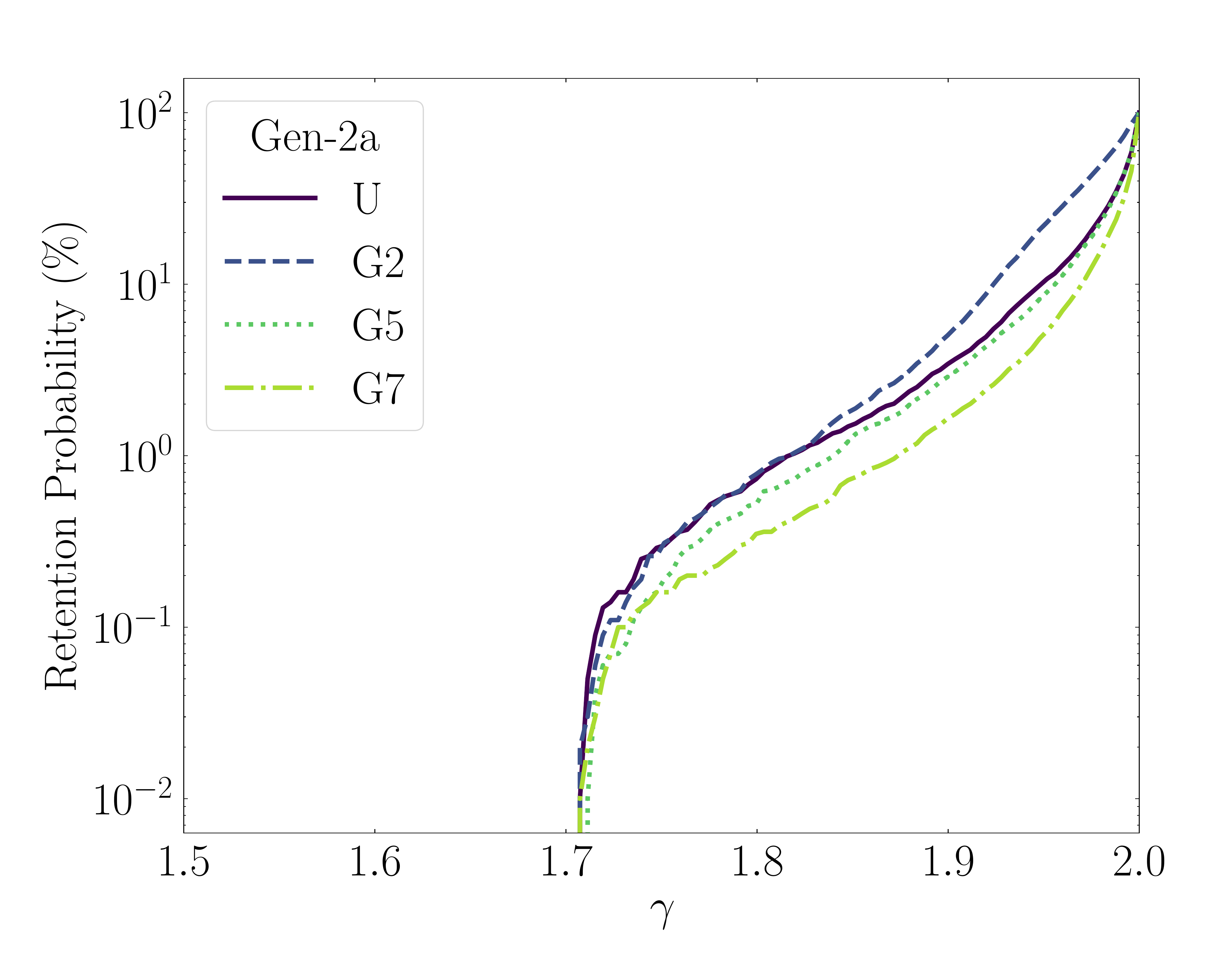}
\caption{ Retention probability (in percentage points) for Gen-2a merger remnant as a function of the cluster density slope $\gamma$. Different curves correspond to a different choice for the stellar BH spin distribution.}
\label{vdehret}
\end{figure}

From Equation \ref{eq:vesc} we thus can conclude that the retention of remnants receiving such large kicks is maximized in clusters with a steep density distribution. Using the values that characterise the $N$-body model in which the triple merger develops, i.e. a cluster mass of $7.4\times10^4\Ms$ and a half-mass radius of $0.6$ pc, we compute the Gen-2a retention probability, namely the percentage of models for which $v_{\rm kick} < v_{\rm esc}$, at varying the cluster density slope $\gamma$. Figure \ref{vdehret} shows how the retention probability for Gen-2a merger varies at increasing the slope of the density profile. Our analysis suggests that a slope larger than $\gamma > 1.8-2$ would ensure a retention probability $>0.3-80\%$, depending on the underlying BH natal spin distribution. 

\begin{figure*}
    \centering
    \includegraphics[width=0.7\textwidth]{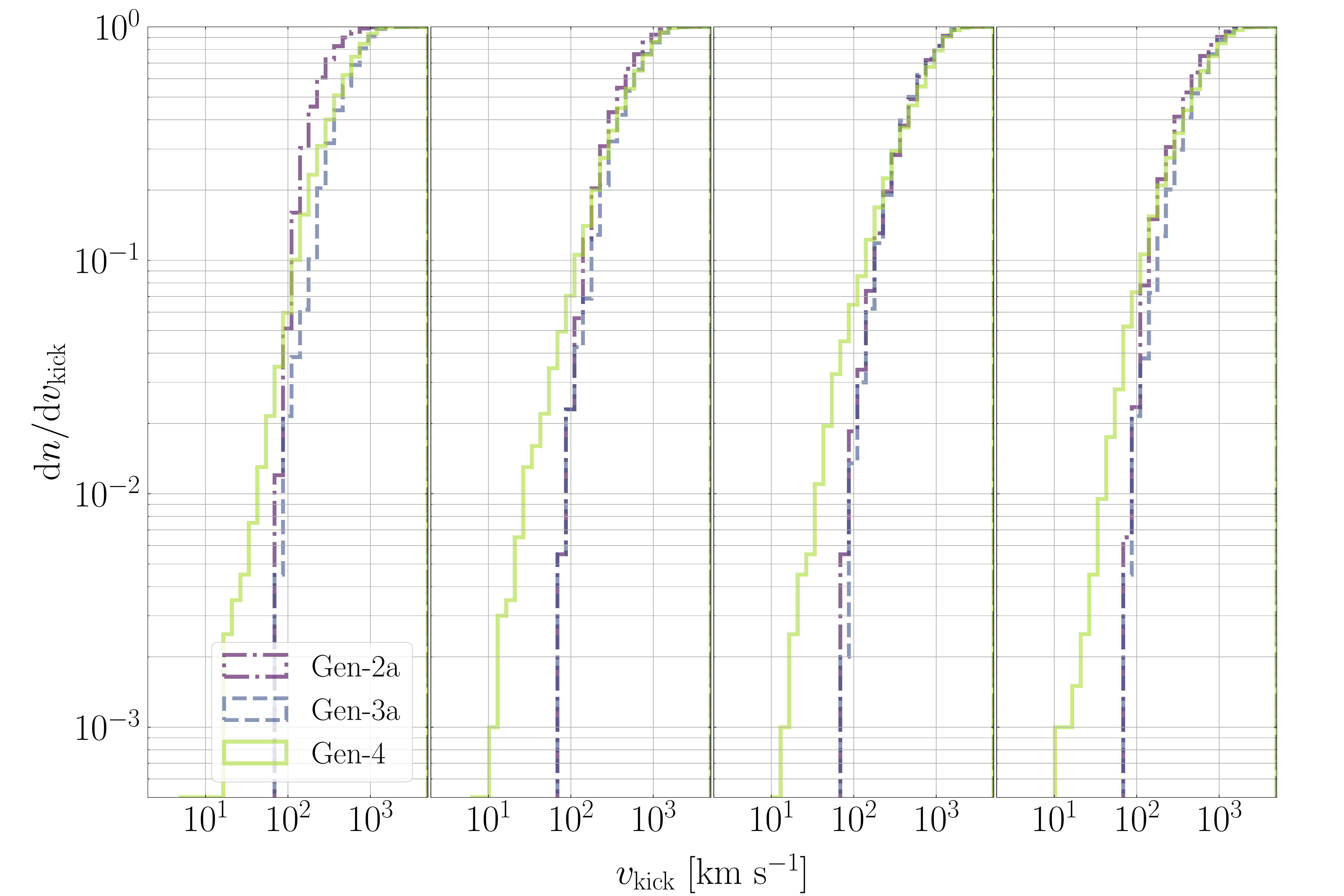}\\
    \includegraphics[width=0.7\textwidth]{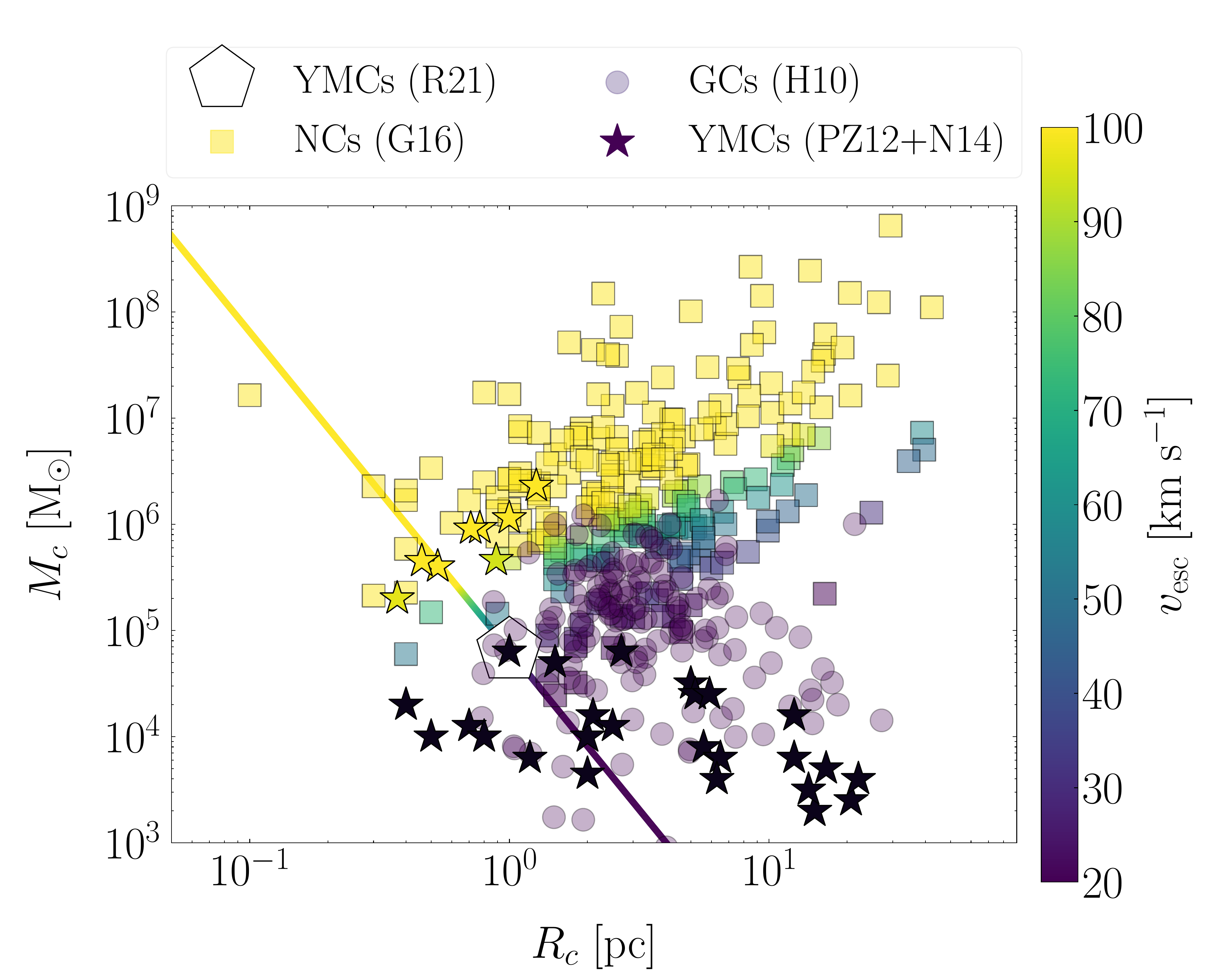}
    \caption{Top: cumulative distribution of recoil kick imparted to the remnant after the first, second, and third merger in our model R06W6Sim1. From left to right, panels refer to the adopted BH natal spin distribution: Gaussian peaked at 0.2, 0.5, 0.7, or uniform, respectively.
    Bottom: Cluster mass (y-axis) and half-mass radius (x-axis) for local NCs \citep[][G16, squares]{georgiev16}, Galactic GCs \citep[][H10, circles]{harris10}, local YMCs \citet[][PZ10, stars]{pz10}, and super star clusters (SSCs) detected in the Henize 2-10 starburst galaxy \citet[][N10, yellow stars]{ngu14}. The color-coding identify the cluster central escape velocity assuming that the inner regions of all the clusters shown can be described by a power-law with slope $\gamma = 0.3$. The coloured straight line identifies the loci of points for which the cluster relaxation time equals the value inferred for our $N$-body model, which is identified as a white star \citep[][R21]{rizzuto20}. }
    \label{fig:esc}
\end{figure*}

This implies that is nearly impossible to develop a long chain of mergers in a YMC similar to the simulated one unless the cluster is characterised by a quite steep density profile. In principle, in order for the R6W6sim1 triple merger to develop, our analysis suggests that the cluster escape velocity should increase by a factor of at least 3-5, a threshold that could be achieved with a heavier or denser cluster. However, the development of the triple merger chain is intrinsically locked to the cluster evolution. For instance, the mass-segregation timescale regulates the time over which the BHs reach the cluster core and start interacting with each other. In our $N$-body models the segregation of the most massive stars takes place within the first 1-2 Myr. Since both the collision rate and the mass-segregation timescale are tightly linked to the cluster relaxation time \citep{zwart02}, the minimum requirement to attempt an extrapolation from our models to heavier and denser clusters is that the mass-segregation time, which scales with the cluster relaxation time $t_{\rm rel}\propto M_c^{1/2} r_h^{3/2}$ \citep{bt}, remains constant. Note that this roughly corresponds to the requirement that the mass-segregation time in the cluster remains constant as well \citep[see e.g.][]{AS16}. At the same time, we need that the cluster escape velocity, which for Dehnen models scales with the $M_c/r_h$ ratio as shown in Equation \ref{eq:vesc} above, increases above the threshold imposed by $v_{\rm kick}$. Thus, by requiring that the relaxation time is the same as in the $N$-body model and that $v_{\rm esc} > v_{\rm kick}$ enables us to find the  $(M_c,r_h)$ values that describe heavier clusters that should be equivalent to our $N$-body models from the dynamical point of view. Clearly, this is an oversimplification that neglects the impact of several processes, e.g. a substantial primordial mass segregation or fraction of binaries.
Figure \ref{fig:esc} shows the locus of points with constant relaxation time compared to the typical mass and half-mass radius of NCs in the local Universe \citep[taken from][]{georgiev16}, MW GCs \citep{harris10}, YMCs in the MW and MW satellites \citep{pz10}, and the 11 YMCs detected in the starburst galaxy Henize 2-10 \citep{ngu14,ASCD15He}, often referred to as super star clusters (SSCs).

From the plot it is apparent that only clusters with a mass $M_c>3\times10^5\Ms$ and half-mass radius $r_h < 0.6$ pc have a central escape velocity $v_{\rm esc}\gtrsim 90$ km s$^{-1}$ and thus could retain the remnant of the first merger in our triple-merger channel. The comparison with observed clusters suggest that a fraction of NCs ($\sim 5\%$ in \cite{georgiev16} sample of low redshift NCs) and SSCs similar to the one found in Henize 2-10 are sufficiently dense to fulfill these two conditions and potentially harbour a multiple merger chain. The fact that 
the triple merger chain occurs very early in our models ($<100$ Myr) seems to favour the SSC scenario, as these class of clusters are usually very young -- the inferred age of the SSCs in Henize 2-10 is $\sim 5$ Myr -- massive, and dense \citep[e.g.][]{ngu14}. Alternatively, a young nuclear cluster formed out of SSC collisions \citep{ASCD15He} could represent a potential host for our triple merger chain.

Note that among all the models explored, we find that the kick imparted to the first merger remnant exceeds $100$ km s$^{-1}$ with a probability of $>90\%$. This would suggest that even a double merger scenario seems unlikely for GW190521, unless the host cluster was sufficiently dense and massive as shown in Figure \ref{fig:esc}. 

\subsection{Merger rates} 

Although our analysis is based on a handful models, the few IMBH-BH mergers developed in our simulations can be used to infer a rough estimate of the merger rate for this type of GW sources. 
In the following, we infer the merger rate for $1$st-generation IMBH-BH mergers. Note that the development of these mergers arise from dynamics and is not affected at all by the lack of a treatment of GW kicks and spins.

One possible way to calculate the merger rate is to assume that the formation of young clusters 
proceeds at a fraction $\alpha<1$ of the cosmic star formation rate (SFR) $\psi(z)$, which can be expressed as \citep{madau17}:
\begin{equation}
\psi(z) = 0.01 \frac{(1.+z)^{2.6}}{[1.+((1.+z)/3.2)^{6.2}]}{\rm ~\Ms ~yr^{-1} ~Mpc^{-3}}.
\end{equation}
In the following, we adopt $\alpha = 0.01$ \citep[e.g.][]{belczinski17b}.

From our models, we can define a merger efficiency as the ratio between the number of merger $N_{\rm mer}=3$ and the total number of simulations $N_{\rm sim} = 80$, divided by the cluster mass $M_c = 7.4\times 10^4\Ms$. 

The fraction of star clusters with a mass similar to the simulated one within a given $\epsilon_M$ range can be obtained as:
\begin{equation}
\eta_M = \frac{\int_{M_1}^{M_2} f(M)dM}{\int_{M_{\rm min}}^{M_{\rm max}} f(M)dM}=\frac{M_2^{1-s} - M_1^{1-s}}{M_{\rm max}^{1-s} - M_{\rm min}^{1-s}},
\end{equation}
where $M_{1,2} = M_c (1 \mp \epsilon_M)$ and $M_{\rm min, max} = 10^{4-7}\Ms$. In the latter equality of the equation above, we assume that the cluster mass function is well described by a power-law $f(M)\propto M^{-s}$ with slope $s=2$ \citep[e.g.][]{gieles09}. If we assume an $\epsilon_M = 0.3$ value, thus implying that only clusters with a mass deviating less than $\sim 30\%$ from the simulated one can develop the IMBH-BH mergers discussed here, we get $\eta_M \simeq 0.09$.

\begin{figure}
\centering
\includegraphics[width=\columnwidth]{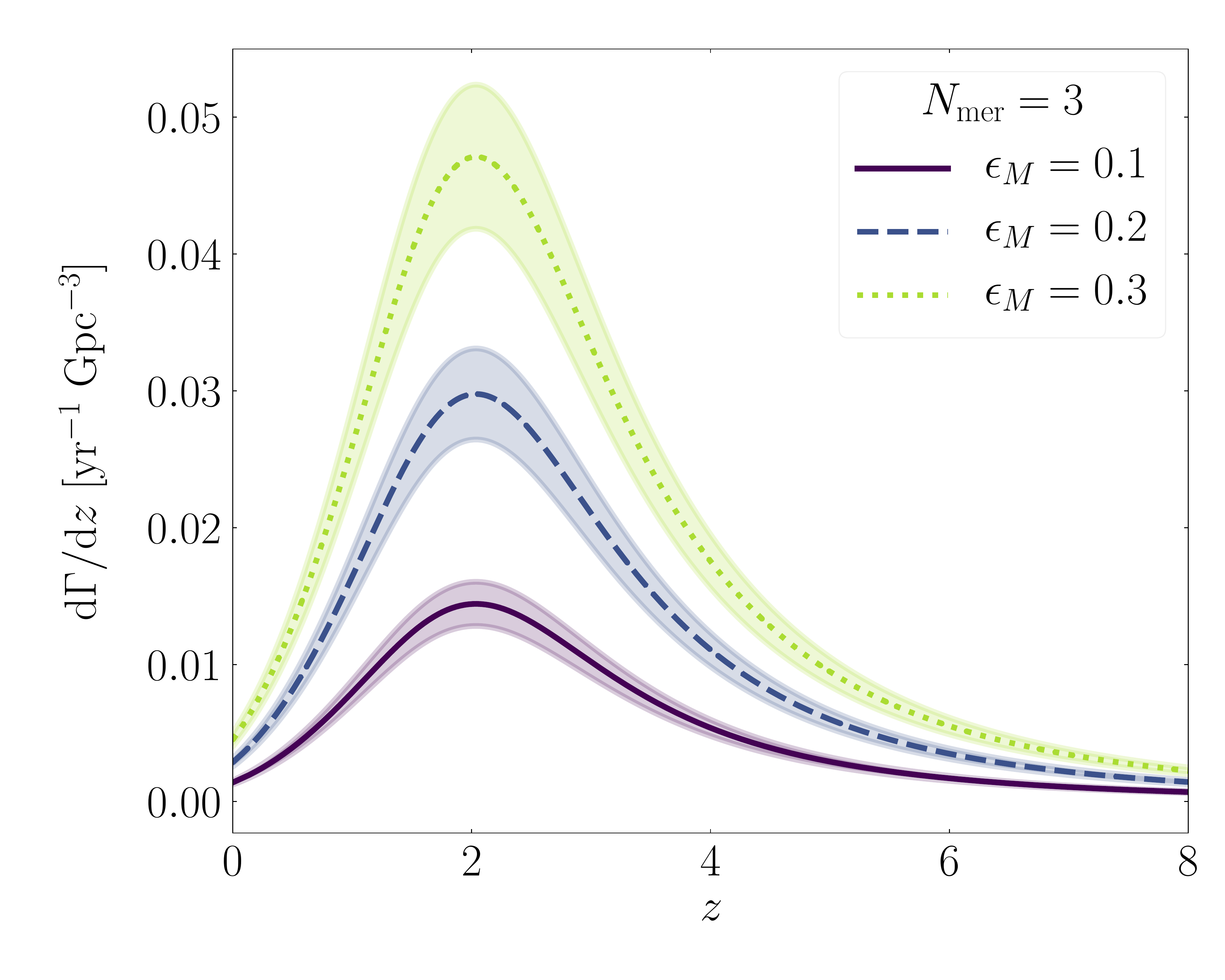}
\includegraphics[width=\columnwidth]{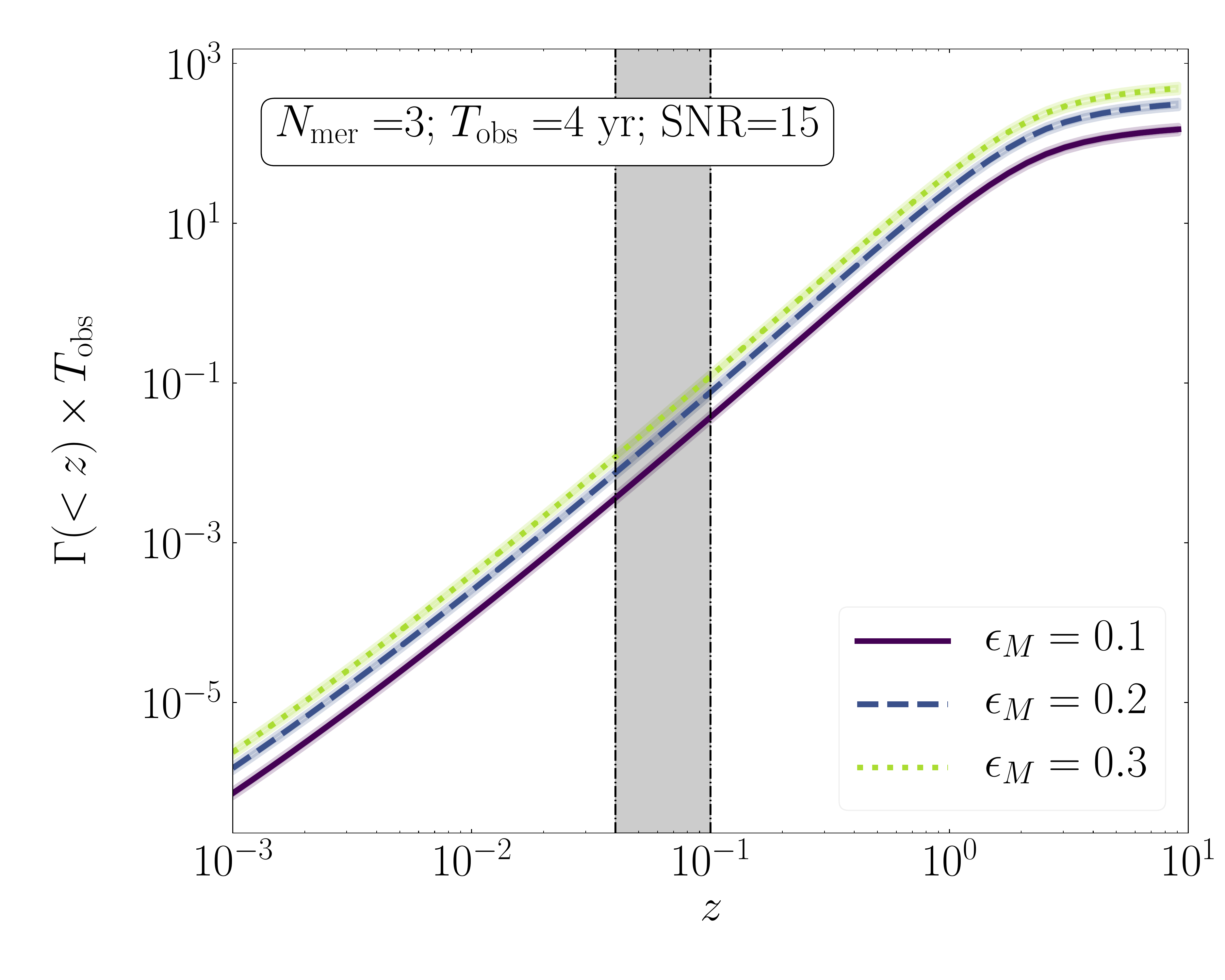}
\caption{Top panel: merger rate density as a function of the redshift for three values of $\epsilon_M = 0.1,~0.2,~0.3$ and assuming three IMBH-BH mergers out of 80 simulations, i.e. $N_{\rm mer} = 1$. Bottom panel: as above, but here we show the cumulative merger rate within a given redshift. The shaded area between the dot dashed black lines mark the limiting values of LISA horizon redshift (see Appendix \ref{app:GWstr}).}
\label{gwrate}
\end{figure}

Combining together the considerations above we can infer the merger rate as a function of redshift as:
\begin{equation}
\frac{{\rm d}\Gamma}{{\rm d}z} = \frac{N_{\rm mer}}{N_{\rm sim} M_c} \eta_M \alpha\psi(z),
\end{equation}
whose behaviour as a function of redshift is shown in the top panel of Figure \ref{gwrate}. We note that the peak at redshift $z=2$ is intrinsically inherited from the adopted SFR. 

The overall merger rate within a redshift $z$ could be thus obtained by integrating the equation above over the cosmological volume:
\begin{equation}
\Gamma(<z) = \int_0^{z_{\rm max}} \frac{{\rm d}\Gamma}{{\rm d}z} \frac{{\rm d}V}{{\rm d}z}\frac{1}{1+z}{\rm d}z ,
\end{equation}
where $T_{\rm obs}$ is the mission duration time, ${\rm d}V/{\rm d}z$ is the comoving cosmological volume element and $(1+z)^{-1}$ is a factor accounting for the dilation time.
The bottom panel of Figure \ref{gwrate} shows the cumulative merger rate $\Gamma(<z)$ as a function of the redshift. Given the horizon redshift for LISA, and assuming an observation time of $T_{\rm obs} = 4$ yr and a minimum SNR of 15, we infer a total number of detections $\Gamma T_{\rm obs} = 10^{-3} - 10^{-1}$. 

We note that in order to reach a detection rate of $\Gamma = 0.2-2.5$ yr$^{-1}$ with the same observation time, LISA horizon should expand beyond redshift $z = 0.3-1$. Reaching such redshift would require an increase in the LISA sensitivity curve especially in the $0.01-1$ Hz frequency range.

Assuming the most recent sensitivity curve \citep{amaro17lisa,robson19}, we find that increasing LISA's sensitivity by a factor 2 would already permit to detect IMBH-BH mergers out to a redshift $z = 0.2$, thus implying $\Gamma \sim 0.2$ yr$^{-1}$.

Whilst detecting these sources with LISA might be tricky, future detectors sensitive to decihertz GWs could clearly pitch their signals. For instance, in the same range of masses of our interest, DECIGO could reach a redshift $z > 10^2$, thus enabling the possibility to probe IMBH-BH mergers up to the time of the formation of the first stars. A mid-range detector could fill the gap between LISA and DECIGO. For instance, adopting the sensitivity curve proposed for the conservative decihertz observatory concept design discussed in \cite{arcasedda20cqg}, we find an horizon redshift $z > 4$, leading to a detection rate $\Gamma \sim 10-100$ yr$^{-1}$.

\section{$N$-body simulations and comparison with similar works}
\label{sec:6}
The suite of direct $N$-body simulations of young dense clusters in which we find the GW190521-like merger is carried out with an improved version of the  NBODY6++GPU direct $N$-body code \citep{wang15}. The key features of these models include stellar evolution of single and binary stars \citep{hurley00,hurley02}, natal kicks for neutron stars \citep{hobbs}, a fallback prescription to compute BH masses \citep{belczynski02}, a dedicated treatment for tight binary evolution \citep{mikkola98}, and general relativistic corrections to the equation of motions of compact stellar objects \citep{rizzuto20}. The current implementation of stellar evolution does not include the latest recipes for supernovae mechanisms (e.g. pair instability and pulsational pair instability supernova).

We note that the mass function adopted in this work, which is truncated at $m_{\rm max} = 100 \Ms$,  and the metallicity value assumed ($Z = 0.0002$), make PISN and PPISN mechanisms inefficient in our models, as they should affect mostly stars with initial masses in the range $100-200\Ms$ \citep[e.g. cfr. with Figure 2 in][]{spera17}.

In this regard, it should also be noted that the actual extension of the upper-mass gap is largely unknown and a number of recent stellar evolution models showed that it is even possible to populate the mass-gap with BHs formed through ``ordinary'' stellar evolution \citep[see e.g.][]{woosley21,vink21}.

Although a handful stars in some of our models can exceed the $100\Ms$ mass limit imposed by the IMF, e.g. through stellar mergers, it must be noted that the remnant BH mass is limited to $30-40\Ms$, owing to the adopted stellar evolution recipes \citep[see Fig. 2 in][]{rizzuto20}. The few BHs formed this way will have masses overlapping the overall BH mass spectrum, and given their small number (order of a few in 110,000 stars) are not expected to affect significantly the overall cluster evolution.

Nonetheless, these are among the first $N$-body models that take into account simultaneously both general relativistic effects, stellar evolution for single and binary stars, and a substantial fraction of primordial binaries. 
The creation of a new simulations database is currently underway and will feature up-to-date stellar evolution prescriptions, a larger number of stars, and a more abundant population of primordial binaries.

Our models simulate the evolution of young and dense star clusters with typical central densities in the range $10^5-3\times 10^7$ solar mass per cubic parsec ($\Ms$ pc$^{-3}$)\footnote{We refer to \cite{rizzuto20} for more details about the simulations}.
 
In comparison to other works in the literature \citep[e.g.][]{dicarlo19,dicarlo20,banerjee16,banerjee20} our models represent on average heavier clusters, 2-10 times compared to \cite{dicarlo19,dicarlo20} and 1-10 times compared to \cite{banerjee16}, and denser, since our clusters have a half-mass radius $0.6-1$ pc. 

Moreover, \cite{rizzuto20} explores the impact of the primordial binary fraction adopting a value $f_b = 0.1$, thus corresponding to the maximum value explored by \cite{banerjee16} and being definitely smaller than the value used in \citep{dicarlo19}, who assumed $f_b = 40\%$. 

The models presented here treats stellar feeding onto BHs through a parameter, $f_{\rm acc}$, which regulate the amount of mass that is accreted onto the BH during an accretion or collision event. \cite{rizzuto20} explores the range $f_{\rm acc} = 0.1,~0.5,~1.0$. In comparison to our models, \cite{dicarlo19} adopted the more conservative assumption that the stellar material is completely lost and the BH does not accrete at all, whereas \cite{banerjee20} explored the range $f_{\rm acc} = 0.7,~0.9$ in their latest models.

Another element of difference is in the maximum mass adopted for the initial mass function: \cite{rizzuto20} adopts a $m_{\rm max} = 100 \Ms$ value, whereas \cite{banerjee16} uses 
a cluster mass dependent relation \citep{kroupa13} that leads to $m_{\rm max}<100\Ms$ in the cluster mass range discussed here, and \cite{dicarlo19} assumes $m_{\rm max} = 150 \Ms$.

Finally, we note that only \cite{banerjee20} took into account GW recoil kick self-consistently in the simulations in the recent literature of direct $N$-body models. However, the models discussed in \cite{banerjee20} cover a different portion of the phase-space in terms of cluster half-mass radius, density profile, metallicity and initial binary fraction. Out of the 65 models described in \cite{banerjee20}, only one has structural properties compatible with \cite{rizzuto20} models, e.g. a similar cluster mass and metallicity, although crucial differences remain, i.e. a different metallicity, fraction of binaries and stellar evolution recipes. 

We also note that the approach presented in this paper is one of the first adopting a post-processing procedure to quantify the impact of such mechanism \citep[see also][]{anagnostou20b}. This represents a fast technique to assess statistically the impact of GW kicks and spins without the need of running the exact same simulation many times, and thus saving considerable computational and human time. In these regards, it should be noted that, despite the limitations of our $N$-body models, all the $1$st-generation mergers presented in this paper involve a stellar BH that did never experience previous mergers and an IMBH formed out of stellar accretion onto a stellar BH, and thus they can be considered a reliable outcome of stellar dynamics and evolution.

The discussion above highlights how our models occupy a portion of the phase space not fully covered by previous works, providing a substantially new exploration of young massive clusters dynamics and evolution.

\section{Conclusions}
\label{sec:7}

We analysed the outputs from a suite of 80 $N$-body models of young massive clusters with $N=110,000$ stars, $10\%$ of which initially paired in binaries. Our  $N$-body simulations are among the first in which the number of stars is $>10^5$ and the motion of compact remnants takes into account general relativistic corrections, thus ensuring a reliable description of possible merger events.
Our main results can be summarized as follows:

\begin{itemize}
    \item we found 3 mergers out of 80 models involving a BH beyond the upper mass-gap and 1 model in which a merger similar to GW190521 develops through a triple generation merger chain, all taking place over timescales $t\sim 10-300$ Myr, at an early stage of the  host cluster evolution;
    \item in mergers labelled as ``beyond the mass-gap'', the primary is an IMBH with a mass in the range $M_1 = 205-329\Ms$, whereas the companion is a stellar BH with $M_2 \simeq 20\Ms$. In one of these cases, the IMBH undergoes two subsequent mergers. These mergers fall into the class of intermediate-mass ratio inspiral (IMRI) GW sources;
    \item we show that this type of mergers can be detected with LISA up to a redshift $z=0.01-0.1$ assuming a mission duration of 4 yr and a minimum signal-to-noise ratio $({\rm S/N})_{{\rm min}}=15$;
    \item when the IMBH-BH binary decouples from the cluster dynamics the eccentricity can be very large, $e = 0.085-0.997$. However, 10(5) year prior to the merger it reduces to $e=0.041-0.128(0.031-0.098)$, due to angular momentum loss via GW emission. This is likely due to the large separation at formation of the IMBH-BH mergers, which in turn owes to the relatively low cluster mass. Despite small, measuring such eccentricity values with LISA would uniquely probe the dynamical channel, since isolated binaries are expected to have $10^{-6}<e<10^{-4}$ in the LISA band;
    \item we find that the IMBH-BH remnant carries insights on the IMBH ``natal'' spin. The spin distribution for IMBH remnant in an IMBH-BH merger is well defined: if the IMBH is initially non-rotating, the remnant spin distributin would be narrowly peaked around $S_{\rm rem} = 0.2$, whereas for nearly extremal IMBHs the potential remnant spin values are broadly distributed between $0.6-1$. The different distribution is preserved upon a double merger, thus suggesting that detecting IMBH mergers beyond the mass gap is crucial to assess the IMBH natal spin and thus IMBHs' formation processes;
    \item in 1 simulation out of 80, we found a triple merger chain event that lead to a merger between BHs likely in the upper mass-gap, $M_{1,2}=(70+68)\Ms$, quite similar to the LVC source named GW190521. Regardless the BH natal spin distribution adopted, we find that such triple merger chain lead to a remnant with a final spin that matches well the value inferred for GW190521. 
    \item at formation, the mass-gap merger has an eccentricity of $e=0.897$, reducing to $e\leq 3\times 10^{-4}$ when the binary enters the GW frequency window of 10 Hz (the LVC window);
    \item Comparing models and the detected final spin and effective spin parameter, we show that a triple merger scenario has a larger probability ($59.7-78.2\%$) to produce a remnant with a spin and effective spin parameter matching GW190521 than a double merger scenario ($57.9-61.8\%$). The probability is similar to the case in which GW190521 developed in a single merger event ($58.2-92.9\%$);
    \item we estimate the recoil kick imparted to the mrger remnant in the mergers both beyond and inside the mass-gap. In the case of IMBH-BH mergers, we show that the remnant receives a recoil in the range $v_{\rm kick} = 30-50$ km s$^{-1}$ for Schwarzschild IMBHs, whereas $v_{\rm kick}$ is as large as 500 km s$^{-1}$ for Kerr IMBHs. Therefore, finding signatures of IMBHs in young massive clusters can tell us more about IMBH natal spins. Statistically speaking, detecting signatures of an IMBH in a YMC could imply three possibilities: 1) the IMBH natal spin is very small, thus the recoil kick is small, 2) the IMBH did not undergo mergers with other BHs, 3) the birthsite is an extremely dense YMC; 
    \item we derive a rough estimate for the merger rate and detection rate of IMBH-BH mergers for LISA and mid-range observatories like DECIGO. In the case of LISA, we find a detection rate of $\Gamma = 0.1$ yr$^{-1}$, assuming a 4 yr long mission and a signal-to-noise threshold value of $(S/N)=15$. We show that detectors sensitive to decihertz have the potential to shed a light on IMBH formation mechanisms, reaching redshift of up to $z > 10$, with inferred detection rates of $10-100$ yr$^{-1}$ for DECIGO-like detectors.
    \item regarding the merger-gap merger, we show that the kick imparted to the merger remnant of the first event in the merger chain is generally larger than $70-100$ km s$^{-1}$, thus making unlikely for such channel to develop in ``normal'' YMCs. Nonetheless, retention can be ensured at a $80\%$ level if the YMC density distribution is particularly steep, i.e. with a slope $\gamma > 1.7$. We show that, under optimistic assumptions, sufficiently dense and young nuclear clusters could be the nursery of such triple merger channel. 
    
\end{itemize}

\section*{Acknowledgements} 
The authors acknowledge the anonymous referees for their constructive reports. The authors are grateful to Ataru Tanikawa, Jorick Vink, and Chris Belczynski for useful comments and discussions. MAS acknowledges support from the Alexander von Humboldt Foundation and the Federal Ministry for Education and Research for the research project "Black Holes at all the scales". RS acknowledges the Strategic Priority Research Program (Pilot B) “Multi-wavelength gravitational wave universe” of the Chinese Academy of Sciences (No. XDB23040100) and National Science Foundation of China grant No. 11673032. MG was partially supported by the Polish National Science Center (NCN) through the grant UMO-2016/23/B/ST9/02732. This work benefited of support from the Volkswagen Foundation Trilateral Partnership project No. I/97778 ``Dynamical Mechanisms of Accretion in Galactic Nuclei'', the Deutsche Forschungsgemeinschaft (DFG, German Research Foundation) -- Project-ID 138713538 -- SFB 881 (``The Milky Way System''), and the COST Action CA16104. 

\appendix

\section{Detecting IMBH-BH mergers in young massive clusters with LISA}
\label{app:GWstr}

\begin{figure}
    \centering
    \includegraphics[width=0.65\textwidth]{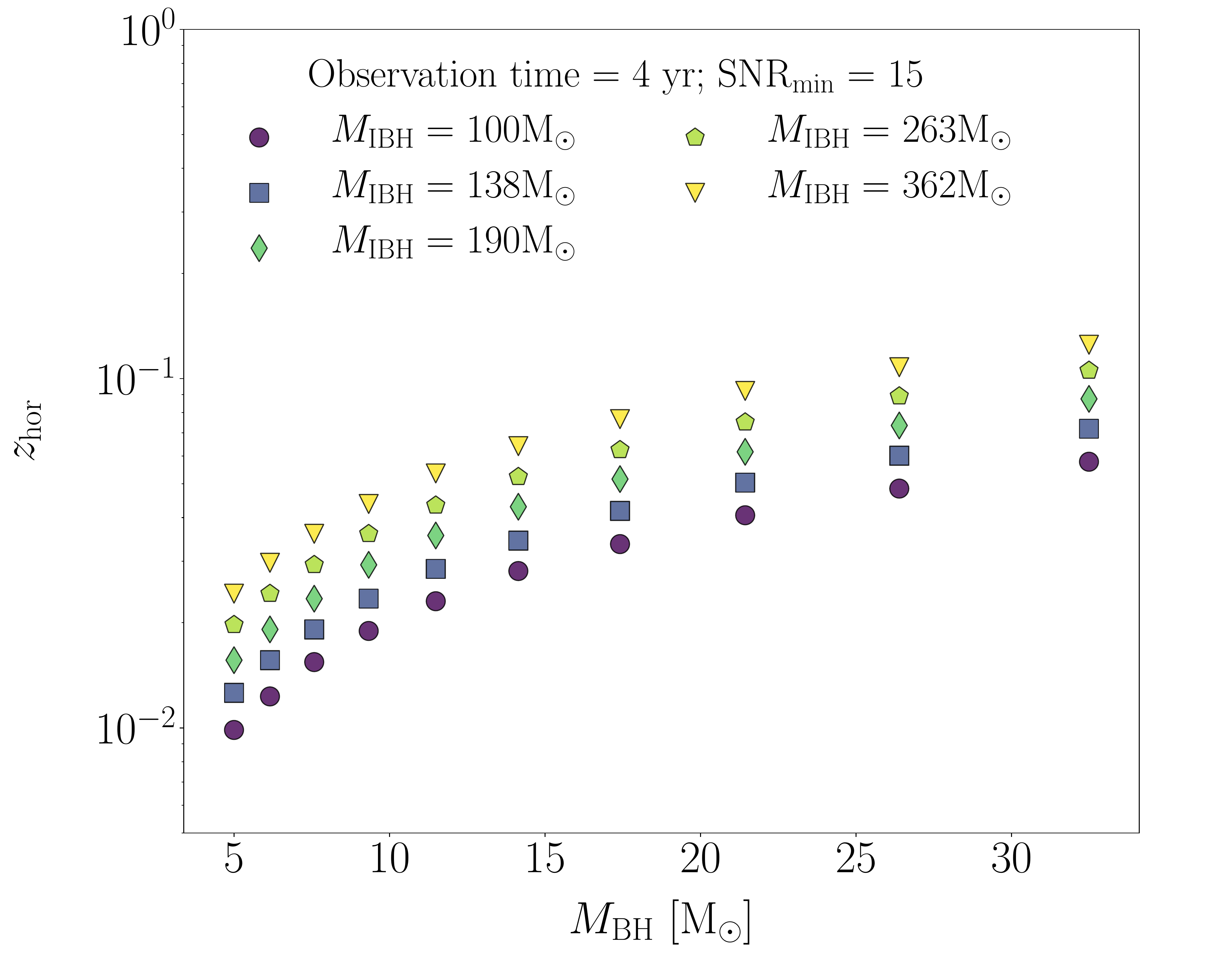}
    \caption{Horizon redshift for circular IMBH-BH mergers merging in the LISA band, assuming a mission lifetime of 4 yr.}
    \label{fig:LISAHor}
\end{figure}

To make predictions on the detectability of our IMBH-BH mergers by LISA, we need to infer the GW source {\it horizon}, which determines the maximum distance in space, or the redshift $z_{\rm hor}$, at which the source signal is detected with a threshold signal-to-noise ratio (SNR), namely:
\begin{equation}
(S/N)^2 = \int_{f_1}^{f_2} \displaystyle{\frac{h_c^2(f,z_{\rm hor})}{S_n^2(f)}} {\rm d}f,
\end{equation}
with $f_{1,2}$ the initial and final frequency of the GW signal, $h_c(f,z)$ its characteristic strain, and $S_n(f)$ is the detector sensitivity. To determine $z_{\rm hor}$ we calculate $f_1$ and $f_2$ by integrating the final stage of the IMBH-BH merger signal assuming an observation time of 4 yr -- i.e. the nominal duration time of the LISA mission -- and adopting $(S/N)=15$. We assume the set of cosmological parameters measured by the Planck mission, namely $H_0 = 67.74$ km/s/Mpc$^{3}$, $\Omega_m = 0.3089$, $\Omega_\Lambda = 0.6911$ \citep{planck15}. Figure \ref{fig:LISAHor} shows how the horizon redshift changes for LISA\footnote{\url{https://www.elisascience.org/}} at varying the IMBH mass in the range $M_{\rm IBH} = 100-400\Ms$ and the stellar BH companion mass in the range $M_{\rm BH}= 5-30 \Ms$.

We find that circular IMBH-BH mergers similar to the one found in our simulations can be observed out to a redshift $z_{\rm hor}= 0.01-0.15$, corresponding to a luminosity distance of $D_L \simeq 44 - 500$ Mpc. We note however that this represents a lower limit to $z_{\rm hor}$, as the GW power emitted by the higher order harmonics in the case of eccentric orbits sums up to the 0th-order harmonic, making the source brighter and, thus, observable from slightly farther away.

\section{Modelling merged black holes: remnant mass, spin, and gravitational wave recoil}
\label{app:Spins}

The $N$-body simulations presented here do not implement a treatment for BH spins and GW recoil, and do not account for GW energy loss in the evaluation of the remnant mass, which is calculated as the mass of the merging BHs.

To include these important ingredients we post-process the simulated data creating a set of mergers with BH masses given by the $N$-body model and spins drawn by different BH spin distributions. 

In the case of IMBHs (mass $>200\Ms$) we explore two limiting cases, whether the IMBH is slowly ($S_{\rm IBH} = 0.01$) or highly ($S_{\rm IBH} = 0.99$) spinning. For stellar BHs the spin should be somehow connected with the BH formation process, but currently the picture around BH natal spins is poorly constrained. Given the lack of a general consensus around BH natal spin, we explore four different cases: either the BH natal spin distribution is a Gaussian peaked at either 0.2, 0.5, or 0.7, or it is uniform in the 0-1 range of values.   

We create 10,000 versions of each merger event listed in Table \ref{tab:t1} and for each spin distribution adopted. 
The mass and spin of the remnant BH are calculated through numerical fitting formulae derived by \cite{jimenez17}. In the case of multiple generation mergers we use the remnant mass and spin from the previous generation. For each merger we also calculate the chirp mass $\mathcal{M}=(m_1m_2)^{3/5}/(m_1+m_2)^{1/5}$ and the effective spin parameters 
\begin{equation}
    \chi = \frac{M_a S_a \cos\theta_1 + M_b S_b \cos\theta_2}{M_a+M_b},
\end{equation}
where $\cos\theta_i$ represents the angle between the $i$-th component spin vector and the binary angular momentum. 
Note that these two quantities are particularly important for stellar BH mergers, as they are better constrained in high-frequency GW detectors than total mass and remnant spin.

The GW recoil velocity received by a remnant after a merger can be calculated as \cite{campanelli07,lousto08,lousto12}
\begin{eqnarray}
\vec{v}_\gw   =& v_m\hat{e}_{\bot,1} + v_\bot(\cos \xi \hat{e}_{\bot,1} + \sin \xi \hat{e}_{\bot,2}) + v_\parallel \hat{e}_\parallel, \label{eqKick1}\\
v_m         =& A\eta^2 \sqrt{1-4\eta} (1+B\eta), \\
v_\bot      =& \displaystyle{\frac{H\eta^2}{1+q_\bbh}}\left(S_{2,\parallel} - q_\bbh S_{1,\parallel} \right), \\
v_\parallel =& \displaystyle{\frac{16\eta^2}{1+q_\bbh}}\left[ V_{11} + V_A \Xi_\parallel + V_B \Xi_\parallel^2 + V_C \Xi_\parallel^3 \right] \times \nonumber \\
             & \times \left| \vec{S}_{2,\bot} - q_\bbh\vec{S}_{1,\bot} \right| \cos(\phi_\Delta - \phi_1).  \label{eqKick2}
\end{eqnarray}
In the equation above, $\eta \equiv q_\bbh/(1+q_\bbh)^2$ is the symmetric mass ratio, $\vec{\Xi} \equiv 2(\vec{S}_2 + q_\bbh^2 \vec{S}_1) / (1 + q_\bbh)^2$, and subscripts $\bot$ and $\parallel$ mark directions (perpendicular and parallel) of the BH spin vector with respect to the direction of the binary angular momentum. 
The orthonormal basis defined by the unit vectors ($\hat{e}_\parallel, \hat{e}_{\bot,1}, \hat{e}_{\bot,2}$) has one component directed perpendicular to ($\hat{e}_\parallel$) and two components lying in the binary orbital plane. We set $A = 1.2 \times 10^4$ km s$^{-1}$, $B = -0.93$, $H = 6.9\times 10^3$ km s$^{-1}$, and $\xi = 145^\circ$ \cite{gonzalez07,lousto08}, and $V_{A,B,C} = (2.481, 1.793, 1.507)\times 10^3$ km s$^{-1}$ are scaling constants\cite{lousto12}. $\phi_\Delta$ represents the angle between the direction of the infall at merger (which we randomly draw in the binary orbital plane) and the in-plane component of $\vec{\Delta} \equiv (M_a+M_b)^2 (\vec{S}_b - q_\bbh \vec{S}_a)/(1+q_\bbh)$, while $\phi_1 = 0-2\pi$ is the phase of the binary, extracted randomly between the two limiting values. 

\begin{figure}
    \centering
    \includegraphics[width=\columnwidth]{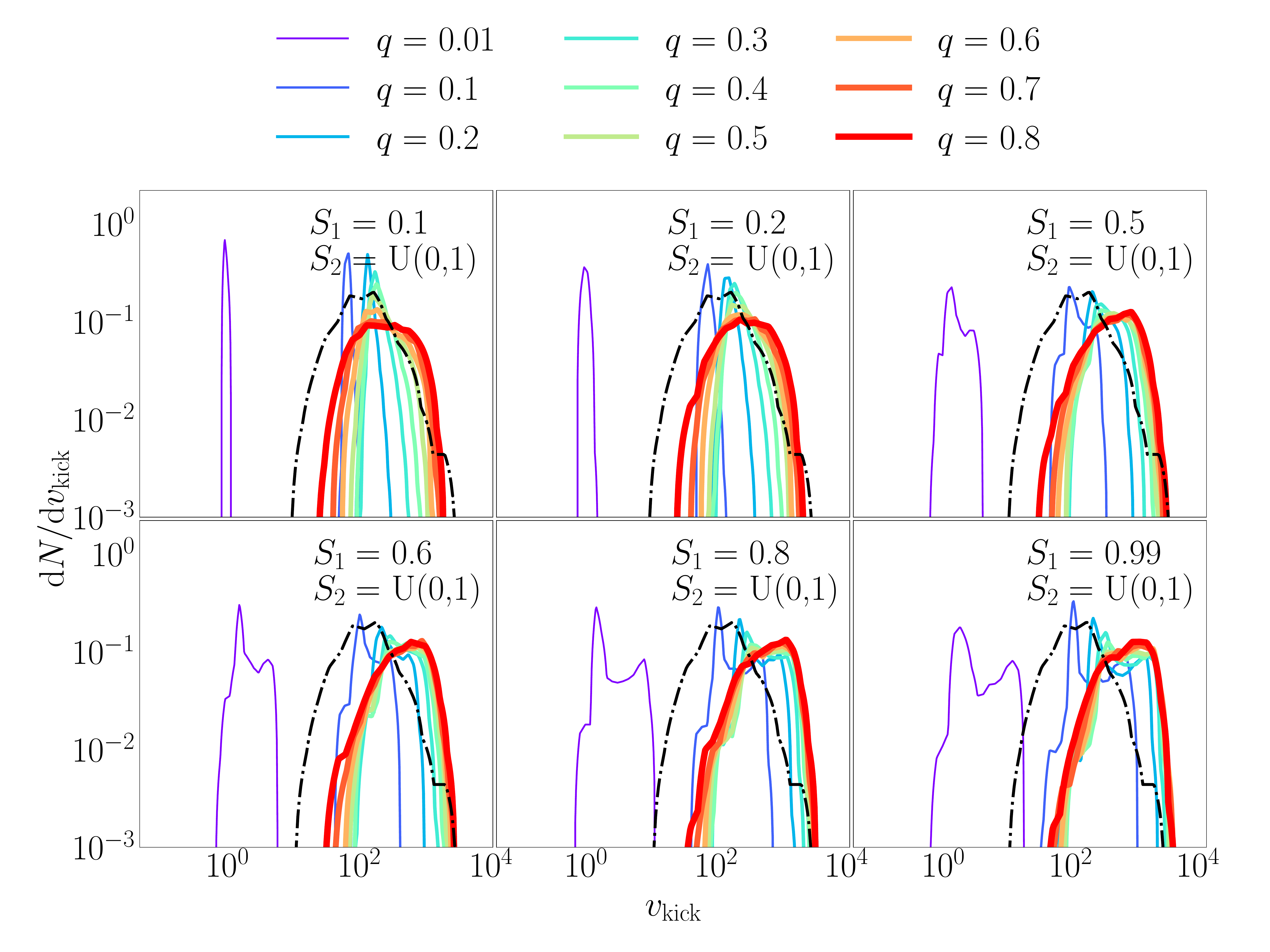}
    \caption{Recoil kick distribution for merging BH binaries with mass ratio in the range $q=0.01-0.9$, assuming that the primary BH has a given value of the spin while the secondary BH spin is drawn from a uniform distribution in the 0-1 spin range. The black dot-dashed line represent the central escape velocity of NCs in the local universe \citep{georgiev16} assuming a density profile $\propto r^{-1}$.}
    \label{fig:kick}
\end{figure}

The amplitude of the kick depends critically on the spins of the two components and the binary mass ratio.
Figure \ref{fig:kick} shows the kick distribution for BH mergers with mass ratio between $0.01-0.9$. As shown in the plot, the recoil is modest ($<100$ km s$^{-1}$) only for highly-asymmetric mergers (i.e. with $q<0.1$). As shown in Table \ref{tab:t1}, the mass ratio is of the order of $q \sim 0.07$ for mergers involving an IMBH, and $q\simeq 0.6-0.9$ for stellar BH mergers, thus implying kicks $\sim 10$ km s${-1}$ and $>100$ km s$^{-1}$, respectively.

\clearpage
\bibliography{apssamp}{}

\newcommand{\noop}[1]{}
\begin{thebibliography}{}
\expandafter\ifx\csname natexlab\endcsname\relax\def\natexlab#1{#1}\fi
\providecommand{\url}[1]{\href{#1}{#1}}
\providecommand{\dodoi}[1]{doi:~\href{http://doi.org/#1}{\nolinkurl{#1}}}
\providecommand{\doeprint}[1]{\href{http://ascl.net/#1}{\nolinkurl{http://ascl.net/#1}}}
\providecommand{\doarXiv}[1]{\href{https://arxiv.org/abs/#1}{\nolinkurl{https://arxiv.org/abs/#1}}}

\bibitem[{{Abbott} {et~al.}(2019){Abbott}, {Abbott}, {Abbott}, {Abraham},
  {Acernese}, {Ackley}, {Adams}, {Adhikari}, {Adya}, {Affeldt}, {Agathos},
  {Agatsuma}, {Aggarwal}, {Aguiar}, {Aiello}, {Ain}, {Ajith}, {Allen},
  {Allocca}, {Aloy}, {Altin}, {Amato}, {Ananyeva}, {Anderson}, {Anderson},
  {Angelova}, {Antier}, {Appert}, {Arai}, {Araya}, {Areeda}, {Ar{\`e}ne},
  {Arnaud}, {Arun}, {Ascenzi}, {Ashton}, {Aston}, {Astone}, {Aubin}, {Aufmuth},
  {AultONeal}, {Austin}, {Avendano}, {Avila-Alvarez}, {Babak}, {Bacon},
  {Badaracco}, {Bader}, {Bae}, {Baker}, {Baldaccini}, {Ballardin}, {Ballmer},
  {Banagiri}, {Barayoga}, {Barclay}, {Barish}, {Barker}, {Barkett}, {Barnum},
  {Barone}, {Barr}, {Barsotti}, {Barsuglia}, {Barta}, {Bartlett}, {Bartos},
  {Bassiri}, {Basti}, {Bawaj}, {Bayley}, {Bazzan}, {B{\'e}csy}, {Bejger},
  {Belahcene}, {Bell}, {Beniwal}, {Berger}, {Bergmann}, {Bernuzzi}, {Bero},
  {Berry}, {Bersanetti}, {Bertolini}, {Betzwieser}, {Bhandare}, {Bidler},
  {Bilenko}, {Bilgili}, {Billingsley}, {Birch}, {Birney}, {Birnholtz},
  {Biscans}, {Biscoveanu}, {Bisht}, {Bitossi}, {Bizouard}, {Blackburn},
  {Blair}, {Blair}, {Blair}, {Bloemen}, {Bode}, {Boer}, {Boetzel}, {Bogaert},
  {Bondu}, {Bonilla}, {Bonnand}, {Booker}, {Boom}, {Booth}, {Bork}, {Boschi},
  {Bose}, {Bossie}, {Bossilkov}, {Bosveld}, {Bouffanais}, {Bozzi},
  {Bradaschia}, {Brady}, {Bramley}, {Branchesi}, {Brau}, {Briant}, {Briggs},
  {Brighenti}, {Brillet}, {Brinkmann}, {Brisson}, {Brockill}, {Brooks},
  {Brown}, {Brunett}, {Buikema}, {Bulik}, {Bulten}, {Buonanno}, {Buscicchio},
  {Buskulic}, {Buy}, {Byer}, {Cabero}, {Cadonati}, {Cagnoli}, {Cahillane},
  {Calder{\'o}n Bustillo}, {Callister}, {Calloni}, {Camp}, {Campbell},
  {Canepa}, {Cannon}, {Cao}, {Cao}, {Capocasa}, {Carbognani}, {Caride},
  {Carney}, {Carullo}, {Casanueva Diaz}, {Casentini}, {Caudill},
  {Cavagli{\`a}}, {Cavalier}, {Cavalieri}, {Cella}, {Cerd{\'a}-Dur{\'a}n},
  {Cerretani}, {Cesarini}, {Chaibi}, {Chakravarti}, {Chamberlin}, {Chan},
  {Chao}, {Charlton}, {Chase}, {Chassande-Mottin}, {Chatterjee}, {Chaturvedi},
  {Chatziioannou}, {Cheeseboro}, {Chen}, {Chen}, {Chen}, {Cheng}, {Cheong},
  {Chia}, {Chincarini}, {Chiummo}, {Cho}, {Cho}, {Cho}, {Christensen}, {Chu},
  {Chua}, {Chung}, {Chung}, {Ciani}, {Ciobanu}, {Ciolfi}, {Cipriano}, {Cirone},
  {Clara}, {Clark}, {Clearwater}, {Cleva}, {Cocchieri}, {Coccia}, {Cohadon},
  {Cohen}, {Colgan}, {Colleoni}, {Collette}, {Collins}, {Cominsky},
  {Constancio}, {Conti}, {Cooper}, {Corban}, {Corbitt}, {Cordero-Carri{\'o}n},
  {Corley}, {Cornish}, {Corsi}, {Cortese}, {Costa}, {Cotesta}, {Coughlin},
  {Coughlin}, {Coulon}, {Countryman}, {Couvares}, {Covas}, {Cowan}, {Coward},
  {Cowart}, {Coyne}, {Coyne}, {Creighton}, {Creighton}, {Cripe}, {Croquette},
  {Crowder}, {Cullen}, {Cumming}, {Cunningham}, {Cuoco}, {Dal Canton},
  {D{\'a}lya}, {Danilishin}, {D'Antonio}, {Danzmann}, {Dasgupta}, {Da Silva
  Costa}, {Datrier}, {Dattilo}, {Dave}, {Davier}, {Davis}, {Daw}, {DeBra},
  {Deenadayalan}, {Degallaix}, {De Laurentis}, {Del{\'e}glise}, {Del Pozzo},
  {DeMarchi}, {Demos}, {Dent}, {De Pietri}, {Derby}, {De Rosa}, {De Rossi},
  {DeSalvo}, {de Varona}, {Dhurandhar}, {D{\'\i}az}, {Dietrich}, {Di Fiore},
  {Di Giovanni}, {Di Girolamo}, {Di Lieto}, {Ding}, {Di Pace}, {Di Palma}, {Di
  Renzo}, {Dmitriev}, {Doctor}, {Donovan}, {Dooley}, {Doravari}, {Dorrington},
  {Downes}, {Drago}, {Driggers}, {Du}, {Ducoin}, {Dupej}, {Dwyer}, {Easter},
  {Edo}, {Edwards}, {Effler}, {Ehrens}, {Eichholz}, {Eikenberry}, {Eisenmann},
  {Eisenstein}, {Essick}, {Estelles}, {Estevez}, {Etienne}, {Etzel}, {Evans},
  {Evans}, {Fafone}, {Fair}, {Fairhurst}, {Fan}, {Farinon}, {Farr}, {Farr},
  {Fauchon-Jones}, {Favata}, {Fays}, {Fazio}, {Fee}, {Feicht}, {Fejer}, {Feng},
  {Fernandez-Galiana}, {Ferrante}, {Ferreira}, {Ferreira}, {Ferrini},
  {Fidecaro}, {Fiori}, {Fiorucci}, {Fishbach}, {Fisher}, {Fishner},
  {Fitz-Axen}, {Flaminio}, {Fletcher}, {Flynn}, {Fong}, {Font}, {Forsyth},
  {Fournier}, {Frasca}, {Frasconi}, {Frei}, {Freise}, {Frey}, {Frey},
  {Fritschel}, {Frolov}, {Fulda}, {Fyffe}, {Gabbard}, {Gadre}, {Gaebel},
  {Gair}, {Gammaitoni}, {Ganija}, {Gaonkar}, {Garcia},
  {Garc{\'\i}a-Quir{\'o}s}, {Garufi}, {Gateley}, {Gaudio}, {Gaur}, {Gayathri},
  {Gemme}, {Genin}, {Gennai}, {George}, {George}, {Gergely}, {Germain},
  {Ghonge}, {Ghosh}, {Ghosh}, {Ghosh}, {Giacomazzo}, {Giaime}, {Giardina},
  {Giazotto}, {Gill}, {Giordano}, {Glover}, {Godwin}, {Goetz}, {Goetz},
  {Goncharov}, {Gonz{\'a}lez}, {Gonzalez Castro}, {Gopakumar}, {Gorodetsky},
  {Gossan}, {Gosselin}, {Gouaty}, {Grado}, {Graef}, {Granata}, {Grant}, {Gras},
  {Grassia}, {Gray}, {Gray}, {Greco}, {Green}, {Green}, {Gretarsson}, {Groot},
  {Grote}, {Grunewald}, {Gruning}, {Guidi}, {Gulati}, {Guo}, {Gupta}, {Gupta},
  {Gustafson}, {Gustafson}, {Haegel}, {Halim}, {Hall}, {Hall}, {Hamilton},
  {Hammond}, {Haney}, {Hanke}, {Hanks}, {Hanna}, {Hannam}, {Hannuksela},
  {Hanson}, {Hardwick}, {Haris}, {Harms}, {Harry}, {Harry}, {Haster},
  {Haughian}, {Hayes}, {Healy}, {Heidmann}, {Heintze}, {Heitmann}, {Hello},
  {Hemming}, {Hendry}, {Heng}, {Hennig}, {Heptonstall}, {Hernandez Vivanco},
  {Heurs}, {Hild}, {Hinderer}, {Hoak}, {Hochheim}, {Hofman}, {Holgado},
  {Holland}, {Holt}, {Holz}, {Hopkins}, {Horst}, {Hough}, {Howell}, {Hoy},
  {Hreibi}, {Huerta}, {Huet}, {Hughey}, {Hulko}, {Husa}, {Huttner},
  {Huynh-Dinh}, {Idzkowski}, {Iess}, {Ingram}, {Inta}, {Intini}, {Irwin},
  {Isa}, {Isac}, {Isi}, {Iyer}, {Izumi}, {Jacqmin}, {Jadhav}, {Jani},
  {Janthalur}, {Jaranowski}, {Jenkins}, {Jiang}, {Johnson}, {Jones}, {Jones},
  {Jones}, {Jonker}, {Ju}, {Junker}, {Kalaghatgi}, {Kalogera}, {Kamai},
  {Kandhasamy}, {Kang}, {Kanner}, {Kapadia}, {Karki}, {Karvinen}, {Kashyap},
  {Kasprzack}, {Katsanevas}, {Katsavounidis}, {Katzman}, {Kaufer}, {Kawabe},
  {Keerthana}, {K{\'e}f{\'e}lian}, {Keitel}, {Kennedy}, {Key}, {Khalili},
  {Khan}, {Khan}, {Khan}, {Khan}, {Khazanov}, {Khursheed}, {Kijbunchoo}, {Kim},
  {Kim}, {Kim}, {Kim}, {Kim}, {Kim}, {Kimball}, {King}, {King},
  {Kinley-Hanlon}, {Kirchhoff}, {Kissel}, {Kleybolte}, {Klika}, {Klimenko},
  {Knowles}, {Koch}, {Koehlenbeck}, {Koekoek}, {Koley}, {Kondrashov}, {Kontos},
  {Koper}, {Korobko}, {Korth}, {Kowalska}, {Kozak}, {Kringel}, {Krishnendu},
  {Kr{\'o}lak}, {Kuehn}, {Kumar}, {Kumar}, {Kumar}, {Kumar}, {Kuo}, {Kutynia},
  {Kwang}, {Lackey}, {Lai}, {Lam}, {Landry}, {Lane}, {Lang}, {Lange}, {Lantz},
  {Lanza}, {Lartaux-Vollard}, {Lasky}, {Laxen}, {Lazzarini}, {Lazzaro},
  {Leaci}, {Leavey}, {Lecoeuche}, {Lee}, {Lee}, {Lee}, {Lee}, {Lee}, {Lee},
  {Lehmann}, {Lenon}, {Leroy}, {Letendre}, {Levin}, {Li}, {Li}, {Li}, {Li},
  {Lin}, {Linde}, {Linker}, {Littenberg}, {Liu}, {Liu}, {Lo}, {Lockerbie},
  {London}, {Longo}, {Lorenzini}, {Loriette}, {Lormand}, {Losurdo}, {Lough},
  {Lousto}, {Lovelace}, {Lower}, {L{\"u}ck}, {Lumaca}, {Lundgren}, {Lynch},
  {Ma}, {Macas}, {Macfoy}, {MacInnis}, {Macleod}, {Macquet},
  {Maga{\~n}a-Sandoval}, {Maga{\~n}a Zertuche}, {Magee}, {Majorana},
  {Maksimovic}, {Malik}, {Man}, {Mandic}, {Mangano}, {Mansell}, {Manske},
  {Mantovani}, {Mapelli}, {Marchesoni}, {Marion}, {M{\'a}rka}, {M{\'a}rka},
  {Markakis}, {Markosyan}, {Markowitz}, {Maros}, {Marquina}, {Marsat},
  {Martelli}, {Martin}, {Martin}, {Martynov}, {Mason}, {Massera}, {Masserot},
  {Massinger}, {Masso-Reid}, {Mastrogiovanni}, {Matas}, {Matichard}, {Matone},
  {Mavalvala}, {Mazumder}, {McCann}, {McCarthy}, {McClelland}, {McCormick},
  {McCuller}, {McGuire}, {McIver}, {McManus}, {McRae}, {McWilliams}, {Meacher},
  {Meadors}, {Mehmet}, {Mehta}, {Meidam}, {Melatos}, {Mendell}, {Mercer},
  {Mereni}, {Merilh}, {Merzougui}, {Meshkov}, {Messenger}, {Messick},
  {Metzdorff}, {Meyers}, {Miao}, {Michel}, {Middleton}, {Mikhailov}, {Milano},
  {Miller}, {Miller}, {Millhouse}, {Mills}, {Milovich-Goff}, {Minazzoli},
  {Minenkov}, {Mishkin}, {Mishra}, {Mistry}, {Mitra}, {Mitrofanov},
  {Mitselmakher}, {Mittleman}, {Mo}, {Moffa}, {Mogushi}, {Mohapatra},
  {Montani}, {Moore}, {Moraru}, {Moreno}, {Morisaki}, {Mours}, {Mow-Lowry},
  {Mukherjee}, {Mukherjee}, {Mukherjee}, {Mukund}, {Mullavey}, {Munch},
  {Mu{\~n}iz}, {Muratore}, {Murray}, {Nagar}, {Nardecchia}, {Naticchioni},
  {Nayak}, {Neilson}, {Nelemans}, {Nelson}, {Nery}, {Neunzert}, {Ng}, {Ng},
  {Nguyen}, {Nichols}, {Nissanke}, {Nocera}, {North}, {Nuttall},
  {Obergaulinger}, {Oberling}, {O'Brien}, {O'Dea}, {Ogin}, {Oh}, {Oh}, {Ohme},
  {Ohta}, {Okada}, {Oliver}, {Oppermann}, {Oram}, {O'Reilly}, {Ormiston},
  {Ortega}, {O'Shaughnessy}, {Ossokine}, {Ottaway}, {Overmier}, {Owen}, {Pace},
  {Pagano}, {Page}, {Pai}, {Pai}, {Palamos}, {Palashov}, {Palomba},
  {Pal-Singh}, {Pan}, {Pang}, {Pang}, {Pankow}, {Pannarale}, {Pant},
  {Paoletti}, {Paoli}, {Parida}, {Parker}, {Pascucci}, {Pasqualetti},
  {Passaquieti}, {Passuello}, {Patil}, {Patricelli}, {Pearlstone}, {Pedersen},
  {Pedraza}, {Pedurand}, {Pele}, {Penn}, {Perez}, {Perreca}, {Pfeiffer},
  {Phelps}, {Phukon}, {Piccinni}, {Pichot}, {Piergiovanni}, {Pillant},
  {Pinard}, {Pirello}, {Pitkin}, {Poggiani}, {Pong}, {Ponrathnam}, {Popolizio},
  {Porter}, {Powell}, {Prajapati}, {Prasad}, {Prasai}, {Prasanna}, {Pratten},
  {Prestegard}, {Privitera}, {Prodi}, {Prokhorov}, {Puncken}, {Punturo},
  {Puppo}, {P{\"u}rrer}, {Qi}, {Quetschke}, {Quinonez}, {Quintero},
  {Quitzow-James}, {Raab}, {Radkins}, {Radulescu}, {Raffai}, {Raja}, {Rajan},
  {Rajbhandari}, {Rakhmanov}, {Ramirez}, {Ramos-Buades}, {Rana}, {Rao},
  {Rapagnani}, {Raymond}, {Razzano}, {Read}, {Regimbau}, {Rei}, {Reid},
  {Reitze}, {Ren}, {Ricci}, {Richardson}, {Richardson}, {Ricker}, {Riles},
  {Rizzo}, {Robertson}, {Robie}, {Robinet}, {Rocchi}, {Rolland}, {Rollins},
  {Roma}, {Romanelli}, {Romano}, {Romel}, {Romie}, {Rose}, {Rosi{\'n}ska},
  {Rosofsky}, {Ross}, {Rowan}, {R{\"u}diger}, {Ruggi}, {Rutins}, {Ryan},
  {Sachdev}, {Sadecki}, {Sakellariadou}, {Salconi}, {Saleem}, {Samajdar},
  {Sammut}, {Sanchez}, {Sanchez}, {Sanchis-Gual}, {Sandberg}, {Sanders},
  {Santiago}, {Sarin}, {Sassolas}, {Sathyaprakash}, {Saulson}, {Sauter},
  {Savage}, {Schale}, {Scheel}, {Scheuer}, {Schmidt}, {Schnabel}, {Schofield},
  {Sch{\"o}nbeck}, {Schreiber}, {Schulte}, {Schutz}, {Schwalbe}, {Scott},
  {Scott}, {Seidel}, {Sellers}, {Sengupta}, {Sennett}, {Sentenac}, {Sequino},
  {Sergeev}, {Setyawati}, {Shaddock}, {Shaffer}, {Shahriar}, {Shaner}, {Shao},
  {Sharma}, {Shawhan}, {Shen}, {Shink}, {Shoemaker}, {Shoemaker},
  {ShyamSundar}, {Siellez}, {Sieniawska}, {Sigg}, {Silva}, {Singer}, {Singh},
  {Singhal}, {Sintes}, {Sitmukhambetov}, {Skliris}, {Slagmolen},
  {Slaven-Blair}, {Smith}, {Smith}, {Somala}, {Son}, {Sorazu}, {Sorrentino},
  {Souradeep}, {Sowell}, {Spencer}, {Spera}, {Srivastava}, {Srivastava},
  {Staats}, {Stachie}, {Standke}, {Steer}, {Steinke}, {Steinlechner},
  {Steinlechner}, {Steinmeyer}, {Stevenson}, {Stocks}, {Stone}, {Stops},
  {Strain}, {Stratta}, {Strigin}, {Strunk}, {Sturani}, {Stuver}, {Sudhir},
  {Summerscales}, {Sun}, {Sunil}, {Suresh}, {Sutton}, {Swinkels},
  {Szczepa{\'n}czyk}, {Tacca}, {Tait}, {Talbot}, {Talukder}, {Tanner},
  {T{\'a}pai}, {Taracchini}, {Tasson}, {Taylor}, {Thies}, {Thomas}, {Thomas},
  {Thondapu}, {Thorne}, {Thrane}, {Tiwari}, {Tiwari}, {Tiwari}, {Toland},
  {Tonelli}, {Tornasi}, {Torres-Forn{\'e}}, {Torrie}, {T{\"o}yr{\"a}},
  {Travasso}, {Traylor}, {Tringali}, {Trovato}, {Trozzo}, {Trudeau}, {Tsang},
  {Tse}, {Tso}, {Tsukada}, {Tsuna}, {Tuyenbayev}, {Ueno}, {Ugolini},
  {Unnikrishnan}, {Urban}, {Usman}, {Vahlbruch}, {Vajente}, {Valdes}, {van
  Bakel}, {van Beuzekom}, {van den Brand}, {Van Den Broeck}, {Vander-Hyde},
  {van der Schaaf}, {van Heijningen}, {van Veggel}, {Vardaro}, {Varma}, {Vass},
  {Vas{\'u}th}, {Vecchio}, {Vedovato}, {Veitch}, {Veitch}, {Venkateswara},
  {Venugopalan}, {Verkindt}, {Vetrano}, {Vicer{\'e}}, {Viets}, {Vine}, {Vinet},
  {Vitale}, {Vo}, {Vocca}, {Vorvick}, {Vyatchanin}, {Wade}, {Wade}, {Wade},
  {Walet}, {Walker}, {Wallace}, {Walsh}, {Wang}, {Wang}, {Wang}, {Wang},
  {Wang}, {Ward}, {Warden}, {Warner}, {Was}, {Watchi}, {Weaver}, {Wei},
  {Weinert}, {Weinstein}, {Weiss}, {Wellmann}, {Wen}, {Wessel}, {We{\ss}els},
  {Westhouse}, {Wette}, {Whelan}, {Whiting}, {Whittle}, {Wilken}, {Williams},
  {Williamson}, {Willis}, {Willke}, {Wimmer}, {Winkler}, {Wipf}, {Wittel},
  {Woan}, {Woehler}, {Wofford}, {Worden}, {Wright}, {Wu}, {Wysocki}, {Xiao},
  {Yamamoto}, {Yancey}, {Yang}, {Yap}, {Yazback}, {Yeeles}, {Yu}, {Yu}, {Yuen},
  {Yvert}, {Zadro{\.z}ny}, {Zanolin}, {Zelenova}, {Zendri}, {Zevin}, {Zhang},
  {Zhang}, {Zhang}, {Zhao}, {Zhou}, {Zhou}, {Zhu}, {Zimmerman}, {Zlochower},
  {Zucker}, {Zweizig}, {LIGO Scientific Collaboration}, \& {Virgo
  Collaboration}}]{gwtc2b}
{Abbott}, B.~P., {Abbott}, R., {Abbott}, T.~D., {et~al.} 2019, \apjl, 882, L24,
  \dodoi{10.3847/2041-8213/ab3800}

\bibitem[{{Abbott} {et~al.}(2020){Abbott}, {Abbott}, {Abraham}, {Acernese},
  {Ackley}, {Adams}, {Adams}, {Adhikari}, {Adya}, {Affeldt}, {Agathos},
  {Agatsuma}, {Aggarwal}, {Aguiar}, {Aiello}, {Ain}, {Ajith}, {Akcay}, {Allen},
  {Allocca}, {Altin}, {Amato}, {Anand}, {Ananyeva}, {Anderson}, {Anderson},
  {Angelova}, {Ansoldi}, {Antelis}, {Antier}, {Appert}, {Arai}, {Araya},
  {Areeda}, {Ar{\`e}ne}, {Arnaud}, {Aronson}, {Arun}, {Asali}, {Ascenzi},
  {Ashton}, {Aston}, {Astone}, {Aubin}, {Aufmuth}, {AultONeal}, {Austin},
  {Avendano}, {Babak}, {Badaracco}, {Bader}, {Bae}, {Baer}, {Bagnasco},
  {Baird}, {Ball}, {Ballardin}, {Ballmer}, {Bals}, {Balsamo}, {Baltus},
  {Banagiri}, {Bankar}, {Bankar}, {Barayoga}, {Barbieri}, {Barish}, {Barker},
  {Barneo}, {Barnum}, {Barone}, {Barr}, {Barsotti}, {Barsuglia}, {Barta},
  {Bartlett}, {Bartos}, {Bassiri}, {Basti}, {Bawaj}, {Bayley}, {Bazzan},
  {Becher}, {B{\'e}csy}, {Bedakihale}, {Bejger}, {Belahcene}, {Beniwal},
  {Benjamin}, {Bennett}, {Bentley}, {Bergamin}, {Berger}, {Bergmann},
  {Bernuzzi}, {Berry}, {Bersanetti}, {Bertolini}, {Betzwieser}, {Bhandare},
  {Bhandari}, {Bhattacharjee}, {Bidler}, {Bilenko}, {Billingsley}, {Birney},
  {Birnholtz}, {Biscans}, {Bischi}, {Biscoveanu}, {Bisht}, {Bitossi},
  {Bizouard}, {Blackburn}, {Blackman}, {Blair}, {Blair}, {Blair}, {Blanch},
  {Bobba}, {Bode}, {Boer}, {Boetzel}, {Bogaert}, {Boldrini}, {Bondu},
  {Bonnand}, {Bonilla}, {Booker}, {Boom}, {Bork}, {Boschi}, {Bose},
  {Bossilkov}, {Boudart}, {Bouffanais}, {Bozzi}, {Bradaschia}, {Brady},
  {Bramley}, {Branchesi}, {Brau}, {Breschi}, {Briant}, {Briggs}, {Brighenti},
  {Brillet}, {Brinkmann}, {Brockill}, {Brooks}, {Brooks}, {Brown}, {Brunett},
  {Bruno}, {Bruntz}, {Buikema}, {Bulik}, {Bulten}, {Buonanno}, {Buscicchio},
  {Buskulic}, {Byer}, {Cabero}, {Cadonati}, {Caesar}, {Cagnoli}, {Cahillane},
  {Calder{\'o}n Bustillo}, {Callaghan}, {Callister}, {Calloni}, {Camp},
  {Canepa}, {Cannon}, {Cao}, {Cao}, {Carapella}, {Carbognani}, {Carney},
  {Carpinelli}, {Carullo}, {Carver}, {Casanueva Diaz}, {Casentini}, {Caudill},
  {Cavagli{\`a}}, {Cavalier}, {Cavalieri}, {Cella}, {Cerd{\'a}-Dur{\'a}n},
  {Cesarini}, {Chaibi}, {Chakravarti}, {Chan}, {Chan}, {Chandra}, {Chanial},
  {Chao}, {Charlton}, {Chase}, {Chassande-Mottin}, {Chatterjee},
  {Chattopadhyay}, {Chaturvedi}, {Chatziioannou}, {Chen}, {Chen}, {Chen},
  {Chen}, {Cheng}, {Cheong}, {Chia}, {Chiadini}, {Chierici}, {Chincarini},
  {Chiummo}, {Cho}, {Cho}, {Cho}, {Choate}, {Christensen}, {Chu}, {Chua},
  {Chung}, {Chung}, {Ciani}, {Ciecielag}, {Cie{\'s}lar}, {Cifaldi}, {Ciobanu},
  {Ciolfi}, {Cipriano}, {Cirone}, {Clara}, {Clark}, {Clark}, {Clarke},
  {Clearwater}, {Clesse}, {Cleva}, {Coccia}, {Cohadon}, {Cohen}, {Colleoni},
  {Collette}, {Collins}, {Colpi}, {Constancio}, {Conti}, {Cooper}, {Corban},
  {Corbitt}, {Cordero-Carri{\'o}n}, {Corezzi}, {Corley}, {Cornish}, {Corre},
  {Corsi}, {Cortese}, {Costa}, {Cotesta}, {Coughlin}, {Coughlin}, {Coulon},
  {Countryman}, {Cousins}, {Couvares}, {Covas}, {Coward}, {Cowart}, {Coyne},
  {Coyne}, {Creighton}, {Creighton}, {Croquette}, {Crowder}, {Cudell},
  {Cullen}, {Cumming}, {Cummings}, {Cunningham}, {Cuoco}, {Curylo}, {Dal
  Canton}, {D{\'a}lya}, {Dana}, {DaneshgaranBajastani}, {D'Angelo}, {Danila},
  {Danilishin}, {D'Antonio}, {Danzmann}, {Darsow-Fromm}, {Dasgupta}, {Datrier},
  {Dattilo}, {Dave}, {Davier}, {Davies}, {Davis}, {Daw}, {Dean}, {DeBra},
  {Deenadayalan}, {Degallaix}, {De Laurentis}, {Del{\'e}glise}, {Del Favero},
  {De Lillo}, {De Lillo}, {Del Pozzo}, {DeMarchi}, {De Matteis}, {D'Emilio},
  {Demos}, {Denker}, {Dent}, {Depasse}, {De Pietri}, {De Rosa}, {De Rossi},
  {DeSalvo}, {de Varona}, {Dhurandhar}, {D{\'\i}az}, {Diaz-Ortiz}, {Didio},
  {Dietrich}, {Di Fiore}, {DiFronzo}, {Di Giorgio}, {Di Giovanni}, {Di
  Giovanni}, {Di Girolamo}, {Di Lieto}, {Ding}, {Di Pace}, {Di Palma}, {Di
  Renzo}, {Divakarla}, {Dmitriev}, {Doctor}, {D'Onofrio}, {Donovan}, {Dooley},
  {Doravari}, {Dorrington}, {Downes}, {Drago}, {Driggers}, {Du}, {Ducoin},
  {Dupej}, {Durante}, {D'Urso}, {Duverne}, {Dwyer}, {Easter}, {Eddolls},
  {Edelman}, {Edo}, {Edy}, {Effler}, {Eichholz}, {Eikenberry}, {Eisenmann},
  {Eisenstein}, {Ejlli}, {Errico}, {Essick}, {Estell{\'e}s}, {Estevez},
  {Etienne}, {Etzel}, {Evans}, {Evans}, {Ewing}, {Fafone}, {Fair}, {Fairhurst},
  {Fan}, {Farah}, {Farinon}, {Farr}, {Farr}, {Fauchon-Jones}, {Favata}, {Fays},
  {Fazio}, {Feicht}, {Fejer}, {Feng}, {Fenyvesi}, {Ferguson},
  {Fernandez-Galiana}, {Ferrante}, {Ferreira}, {Fidecaro}, {Figura}, {Fiori},
  {Fiorucci}, {Fishbach}, {Fisher}, {Fishner}, {Fittipaldi}, {Fitz-Axen},
  {Fiumara}, {Flaminio}, {Floden}, {Flynn}, {Fong}, {Font}, {Forsyth},
  {Fournier}, {Frasca}, {Frasconi}, {Frei}, {Freise}, {Frey}, {Frey},
  {Fritschel}, {Frolov}, {Fronz{\'e}}, {Fulda}, {Fyffe}, {Gabbard}, {Gadre},
  {Gaebel}, {Gair}, {Gais}, {Galaudage}, {Gamba}, {Ganapathy}, {Ganguly},
  {Gaonkar}, {Garaventa}, {Garc{\'\i}a-Quir{\'o}s}, {Garufi}, {Gateley},
  {Gaudio}, {Gayathri}, {Gemme}, {Gennai}, {George}, {George}, {George},
  {Gergely}, {Ghonge}, {Ghosh}, {Ghosh}, {Ghosh}, {Giacomazzo}, {Giacoppo},
  {Giaime}, {Giardina}, {Gibson}, {Gier}, {Gill}, {Giri}, {Glanzer}, {Gleckl},
  {Godwin}, {Goetz}, {Goetz}, {Gohlke}, {Goncharov}, {Gonz{\'a}lez},
  {Gopakumar}, {Gossan}, {Gosselin}, {Gouaty}, {Grace}, {Grado}, {Granata},
  {Granata}, {Grant}, {Gras}, {Grassia}, {Gray}, {Gray}, {Greco}, {Green},
  {Green}, {Gretarsson}, {Griggs}, {Grignani}, {Grimaldi}, {Grimes}, {Grimm},
  {Grote}, {Grunewald}, {Gruning}, {Guerrero}, {Guidi}, {Guimaraes},
  {Guix{\'e}}, {Gulati}, {Guo}, {Gupta}, {Gupta}, {Gupta}, {Gustafson},
  {Gustafson}, {Guzman}, {Haegel}, {Halim}, {Hall}, {Hamilton}, {Hammond},
  {Haney}, {Hanke}, {Hanks}, {Hanna}, {Hannam}, {Hannuksela}, {Hannuksela},
  {Hansen}, {Hansen}, {Hanson}, {Harder}, {Hardwick}, {Haris}, {Harms},
  {Harry}, {Harry}, {Hartwig}, {Hasskew}, {Haster}, {Haughian}, {Hayes},
  {Healy}, {Heidmann}, {Heintze}, {Heinze}, {Heinzel}, {Heitmann}, {Hellman},
  {Hello}, {Helmling-Cornell}, {Hemming}, {Hendry}, {Heng}, {Hennes}, {Hennig},
  {Hennig}, {Hernandez Vivanco}, {Heurs}, {Hild}, {Hill}, {Hines}, {Hochheim},
  {Hofgard}, {Hofman}, {Hohmann}, {Holgado}, {Holland}, {Hollows}, {Holmes},
  {Holt}, {Holz}, {Hopkins}, {Horst}, {Hough}, {Howell}, {Hoy}, {Hoyland},
  {Huang}, {H{\"u}bner}, {Huddart}, {Huerta}, {Hughey}, {Hui}, {Husa},
  {Huttner}, {Hutzler}, {Huxford}, {Huynh-Dinh}, {Idzkowski}, {Iess},
  {Imperato}, {Inchauspe}, {Ingram}, {Intini}, {Isi}, {Iyer},
  {JaberianHamedan}, {Jacqmin}, {Jadhav}, {Jadhav}, {James}, {Jani},
  {Janssens}, {Janthalur}, {Jaranowski}, {Jariwala}, {Jaume}, {Jenkins},
  {Jeunon}, {Jiang}, {Johns}, {Johnson-McDaniel}, {Jones}, {Jones}, {Jones},
  {Jones}, {Jones}, {Jonker}, {Ju}, {Junker}, {Kalaghatgi}, {Kalogera},
  {Kamai}, {Kandhasamy}, {Kang}, {Kanner}, {Kapadia}, {Kapasi}, {Karathanasis},
  {Karki}, {Kashyap}, {Kasprzack}, {Kastaun}, {Katsanevas}, {Katsavounidis},
  {Katzman}, {Kawabe}, {K{\'e}f{\'e}lian}, {Keitel}, {Key}, {Khadka},
  {Khalili}, {Khan}, {Khan}, {Khazanov}, {Khetan}, {Khursheed}, {Kijbunchoo},
  {Kim}, {Kim}, {Kim}, {Kim}, {Kim}, {Kim}, {Kimball}, {King}, {Kinley-Hanlon},
  {Kirchhoff}, {Kissel}, {Kleybolte}, {Klimenko}, {Knowles}, {Knyazev}, {Koch},
  {Koehlenbeck}, {Koekoek}, {Koley}, {Kolstein}, {Komori}, {Kondrashov},
  {Kontos}, {Koper}, {Korobko}, {Korth}, {Kovalam}, {Kozak}, {Kr{\"a}mer},
  {Kringel}, {Krishnendu}, {Kr{\'o}lak}, {Kuehn}, {Kumar}, {Kumar}, {Kumar},
  {Kumar}, {Kuns}, {Kwang}, {Lackey}, {Laghi}, {Lalande}, {Lam}, {Lamberts},
  {Landry}, {Lane}, {Lang}, {Lange}, {Lantz}, {Lanza}, {La Rosa},
  {Lartaux-Vollard}, {Lasky}, {Laxen}, {Lazzarini}, {Lazzaro}, {Leaci},
  {Leavey}, {Lecoeuche}, {Lee}, {Lee}, {Lee}, {Lee}, {Lehmann}, {Leon},
  {Leroy}, {Letendre}, {Levin}, {Li}, {Li}, {Li}, {Li}, {Li}, {Linde},
  {Linker}, {Linley}, {Littenberg}, {Liu}, {Liu}, {Llorens-Monteagudo}, {Lo},
  {Lockwood}, {London}, {Longo}, {Lorenzini}, {Loriette}, {Lormand}, {Losurdo},
  {Lough}, {Lousto}, {Lovelace}, {L{\"u}ck}, {Lumaca}, {Lundgren}, {Ma},
  {Macas}, {MacInnis}, {Macleod}, {MacMillan}, {Macquet}, {Maga{\~n}a
  Hernandez}, {Maga{\~n}a-Sandoval}, {Magazz{\`u}}, {Magee}, {Majorana},
  {Maksimovic}, {Maliakal}, {Malik}, {Man}, {Mandic}, {Mangano}, {Mansell},
  {Manske}, {Mantovani}, {Mapelli}, {Marchesoni}, {Marion}, {M{\'a}rka},
  {M{\'a}rka}, {Markakis}, {Markosyan}, {Markowitz}, {Maros}, {Marquina},
  {Marsat}, {Martelli}, {Martin}, {Martin}, {Martinez}, {Martinez}, {Martynov},
  {Masalehdan}, {Mason}, {Massera}, {Masserot}, {Massinger}, {Masso-Reid},
  {Mastrogiovanni}, {Matas}, {Mateu-Lucena}, {Matichard}, {Matiushechkina},
  {Mavalvala}, {Maynard}, {McCann}, {McCarthy}, {McClelland}, {McCormick},
  {McCuller}, {McGuire}, {McIsaac}, {McIver}, {McManus}, {McRae}, {McWilliams},
  {Meacher}, {Meadors}, {Mehmet}, {Mehta}, {Melatos}, {Melchor}, {Mendell},
  {Menendez-Vazquez}, {Mercer}, {Mereni}, {Merfeld}, {Merilh}, {Merritt},
  {Merzougui}, {Meshkov}, {Messenger}, {Messick}, {Metzdorff}, {Meyers},
  {Meylahn}, {Mhaske}, {Miani}, {Miao}, {Michaloliakos}, {Michel}, {Middleton},
  {Milano}, {Miller}, {Millhouse}, {Mills}, {Milotti}, {Milovich-Goff},
  {Minazzoli}, {Minenkov}, {Mir}, {Mishkin}, {Mishra}, {Mistry}, {Mitra},
  {Mitrofanov}, {Mitselmakher}, {Mittleman}, {Mo}, {Mogushi}, {Mohapatra},
  {Mohite}, {Molina}, {Molina-Ruiz}, {Mondin}, {Montani}, {Moore}, {Moraru},
  {Morawski}, {Moreno}, {Morisaki}, {Mours}, {Mow-Lowry}, {Mozzon},
  {Muciaccia}, {Mukherjee}, {Mukherjee}, {Mukherjee}, {Mukherjee}, {Mukund},
  {Mullavey}, {Munch}, {Mu{\~n}iz}, {Murray}, {Nadji}, {Nagar}, {Nardecchia},
  {Naticchioni}, {Nayak}, {Neil}, {Neilson}, {Nelemans}, {Nelson}, {Nery},
  {Neunzert}, {Nitz}, {Ng}, {Ng}, {Nguyen}, {Nguyen}, {Nguyen}, {Nichols},
  {Nissanke}, {Nocera}, {Noh}, {North}, {Nothard}, {Nuttall}, {Oberling},
  {O'Brien}, {O'Dell}, {Oganesyan}, {Ogin}, {Oh}, {Oh}, {Ohme}, {Ohta},
  {Okada}, {Olivetto}, {Oppermann}, {Oram}, {O'Reilly}, {Ormiston}, {Ortega},
  {O'Shaughnessy}, {Ossokine}, {Osthelder}, {Ottaway}, {Overmier}, {Owen},
  {Pace}, {Pagano}, {Page}, {Pagliaroli}, {Pai}, {Pai}, {Palamos}, {Palashov},
  {Palomba}, {Pan}, {Panda}, {Pang}, {Pankow}, {Pannarale}, {Pant}, {Paoletti},
  {Paoli}, {Paolone}, {Parker}, {Pascucci}, {Pasqualetti}, {Passaquieti},
  {Passuello}, {Patel}, {Patricelli}, {Payne}, {Pechsiri}, {Pedraza},
  {Pegoraro}, {Pele}, {Penn}, {Perego}, {Perez}, {P{\'e}rigois}, {Perreca},
  {Perri{\`e}s}, {Petermann}, {Petterson}, {Pfeiffer}, {Pham}, {Phukon},
  {Piccinni}, {Pichot}, {Piendibene}, {Piergiovanni}, {Pierini}, {Pierro},
  {Pillant}, {Pilo}, {Pinard}, {Pinto}, {Piotrzkowski}, {Pirello}, {Pitkin},
  {Placidi}, {Plastino}, {Pluchar}, {Poggiani}, {Polini}, {Pong}, {Ponrathnam},
  {Popolizio}, {Porter}, {Poverman}, {Powell}, {Pracchia}, {Prajapati},
  {Prasai}, {Prasanna}, {Pratten}, {Prestegard}, {Principe}, {Prodi},
  {Prokhorov}, {Prosposito}, {Prudenzi}, {Puecher}, {Punturo}, {Puosi},
  {Puppo}, {P{\"u}rrer}, {Qi}, {Quetschke}, {Quinonez}, {Quitzow-James},
  {Raab}, {Raaijmakers}, {Radkins}, {Radulesco}, {Raffai}, {Rafferty}, {Rail},
  {Raja}, {Rajan}, {Rajbhandari}, {Rakhmanov}, {Ramirez}, {Ramirez},
  {Ramos-Buades}, {Rana}, {Rao}, {Rapagnani}, {Rapol}, {Ratto}, {Raymond},
  {Razzano}, {Read}, {Regimbau}, {Rei}, {Reid}, {Reitze}, {Rettegno}, {Ricci},
  {Richardson}, {Richardson}, {Richardson}, {Ricker}, {Riemenschneider},
  {Riles}, {Rizzo}, {Robertson}, {Robinet}, {Rocchi}, {Rocha}, {Rodriguez},
  {Rodriguez-Soto}, {Rolland}, {Rollins}, {Roma}, {Romanelli}, {Romano},
  {Romel}, {Romero}, {Romero-Shaw}, {Romie}, {Ronchini}, {Rose}, {Rose},
  {Rose}, {Rosell}, {Rosi{\'n}ska}, {Rosofsky}, {Ross}, {Rowan}, {Rowlinson},
  {Roy}, {Roy}, {Ruggi}, {Ryan}, {Sachdev}, {Sadecki}, {Sadiq},
  {Sakellariadou}, {Salafia}, {Salconi}, {Saleem}, {Samajdar}, {Sanchez},
  {Sanchez}, {Sanchez}, {Sanchis-Gual}, {Sanders}, {Sandles}, {Santiago},
  {Santos}, {Saravanan}, {Sarin}, {Sassolas}, {Sathyaprakash}, {Sauter},
  {Savage}, {Savant}, {Sawant}, {Sayah}, {Schaetzl}, {Schale}, {Scheel},
  {Scheuer}, {Schindler-Tyka}, {Schmidt}, {Schnabel}, {Schofield},
  {Sch{\"o}nbeck}, {Schreiber}, {Schulte}, {Schutz}, {Schwarm}, {Schwartz},
  {Scott}, {Scott}, {Seglar-Arroyo}, {Seidel}, {Sellers}, {Sengupta},
  {Sennett}, {Sentenac}, {Sequino}, {Sergeev}, {Setyawati}, {Shaffer},
  {Shahriar}, {Sharifi}, {Sharma}, {Sharma}, {Shawhan}, {Shen}, {Shikauchi},
  {Shink}, {Shoemaker}, {Shoemaker}, {Shukla}, {ShyamSundar}, {Sieniawska},
  {Sigg}, {Singer}, {Singh}, {Singh}, {Singha}, {Singhal}, {Sintes}, {Sipala},
  {Skliris}, {Slagmolen}, {Slaven-Blair}, {Smetana}, {Smith}, {Smith},
  {Somala}, {Son}, {Soni}, {Soni}, {Sorazu}, {Sordini}, {Sorrentino},
  {Sorrentino}, {Soulard}, {Souradeep}, {Sowell}, {Spencer}, {Spera},
  {Srivastava}, {Srivastava}, {Staats}, {Stachie}, {Steer}, {Steinhoff},
  {Steinke}, {Steinlechner}, {Steinlechner}, {Steinmeyer}, {Stevenson},
  {Stolle-McAllister}, {Stops}, {Stover}, {Strain}, {Stratta}, {Strunk},
  {Sturani}, {Stuver}, {S{\"u}dbeck}, {Sudhagar}, {Sudhir}, {Suh},
  {Summerscales}, {Sun}, {Sun}, {Sunil}, {Sur}, {Suresh}, {Sutton}, {Swinkels},
  {Szczepa{\'n}czyk}, {Tacca}, {Tait}, {Talbot}, {Tanasijczuk}, {Tanner},
  {Tao}, {Tapia}, {Tapia San Martin}, {Tasson}, {Taylor}, {Tenorio},
  {Terkowski}, {Thirugnanasambandam}, {Thomas}, {Thomas}, {Thomas}, {Thompson},
  {Thondapu}, {Thorne}, {Thrane}, {Tiwari}, {Tiwari}, {Tiwari}, {Toland},
  {Tolley}, {Tonelli}, {Tornasi}, {Torres-Forn{\'e}}, {Torrie}, {Melo},
  {T{\"o}yr{\"a}}, {Tran}, {Trapananti}, {Travasso}, {Traylor}, {Tringali},
  {Tripathee}, {Trovato}, {Trudeau}, {Tsai}, {Tsang}, {Tse}, {Tso}, {Tsukada},
  {Tsuna}, {Tsutsui}, {Turconi}, {Ubhi}, {Udall}, {Ueno}, {Ugolini},
  {Unnikrishnan}, {Urban}, {Usman}, {Utina}, {Vahlbruch}, {Vajente}, {Vajpeyi},
  {Valdes}, {Valentini}, {Valsan}, {van Bakel}, {van Beuzekom}, {van den
  Brand}, {Van Den Broeck}, {Vander-Hyde}, {van der Schaaf}, {van Heijningen},
  {Vardaro}, {Vargas}, {Varma}, {Vass}, {Vas{\'u}th}, {Vecchio}, {Vedovato},
  {Veitch}, {Veitch}, {Venkateswara}, {Venneberg}, {Venugopalan}, {Verkindt},
  {Verma}, {Veske}, {Vetrano}, {Vicer{\'e}}, {Viets}, {Vijaykumar},
  {Villa-Ortega}, {Vinet}, {Vitale}, {Vo}, {Vocca}, {Vorvick}, {Vyatchanin},
  {Wade}, {Wade}, {Wade}, {Walet}, {Walker}, {Wallace}, {Wallace}, {Walsh},
  {Wang}, {Wang}, {Wang}, {Wang}, {Ward}, {Warner}, {Was}, {Washington},
  {Watchi}, {Weaver}, {Wei}, {Weinert}, {Weinstein}, {Weiss}, {Wellmann},
  {Wen}, {We{\ss}els}, {Westhouse}, {Wette}, {Whelan}, {White}, {White},
  {Whiting}, {Whittle}, {Wilken}, {Williams}, {Williams}, {Williamson},
  {Willis}, {Willke}, {Wilson}, {Wimmer}, {Winkler}, {Wipf}, {Woan}, {Woehler},
  {Wofford}, {Wong}, {Wrangel}, {Wright}, {Wu}, {Wysocki}, {Xiao}, {Yamamoto},
  {Yang}, {Yang}, {Yang}, {Yap}, {Yeeles}, {Yoon}, {Yu}, {Yu}, {Yuen},
  {Zadro{\.z}ny}, {Zanolin}, {Zelenova}, {Zendri}, {Zevin}, {Zhang}, {Zhang},
  {Zhang}, {Zhang}, {Zhao}, {Zhao}, {Zheng}, {Zhou}, {Zhou}, {Zhu},
  {Zimmerman}, {Zlochower}, {Zucker}, \& {Zweizig}}]{gwtc2}
{Abbott}, R., {Abbott}, T.~D., {Abraham}, S., {et~al.} 2020, arXiv e-prints,
  arXiv:2010.14527.
\newblock \doarXiv{2010.14527}

\bibitem[{{Amaro-Seoane} \& {Chen}(2016)}]{seoane16}
{Amaro-Seoane}, P., \& {Chen}, X. 2016, mn, 458, 3075,
  \dodoi{10.1093/mnras/stw503}

\bibitem[{Amaro-Seoane {et~al.}(2017)}]{amaro17lisa}
Amaro-Seoane, P., {et~al.} 2017.
\newblock \doarXiv{1702.00786}

\bibitem[{{Anagnostou} {et~al.}(2020{\natexlab{a}}){Anagnostou}, {Trenti}, \&
  {Melatos}}]{anagnostou20a}
{Anagnostou}, O., {Trenti}, M., \& {Melatos}, A. 2020{\natexlab{a}}, \pasa, 37,
  e044, \dodoi{10.1017/pasa.2020.35}

\bibitem[{{Anagnostou} {et~al.}(2020{\natexlab{b}}){Anagnostou}, {Trenti}, \&
  {Melatos}}]{anagnostou20b}
---. 2020{\natexlab{b}}, arXiv e-prints, arXiv:2010.06161.
\newblock \doarXiv{2010.06161}

\bibitem[{{Antonini} {et~al.}(2019){Antonini}, {Gieles}, \&
  {Gualandris}}]{antonini19}
{Antonini}, F., {Gieles}, M., \& {Gualandris}, A. 2019, \mnras, 486, 5008,
  \dodoi{10.1093/mnras/stz1149}

\bibitem[{{Arca-Sedda}(2016)}]{AS16}
{Arca-Sedda}, M. 2016, mn, 455, 35, \dodoi{10.1093/mnras/stv2265}

\bibitem[{{Arca Sedda}(2020)}]{arcasedda20d}
{Arca Sedda}, M. 2020, \apj, 891, 47, \dodoi{10.3847/1538-4357/ab723b}

\bibitem[{{Arca-Sedda} {et~al.}(2020){Arca-Sedda}, {Amaro-Seoane}, \&
  {Chen}}]{arcasedda20c}
{Arca-Sedda}, M., {Amaro-Seoane}, P., \& {Chen}, X. 2020, arXiv e-prints,
  arXiv:2007.13746.
\newblock \doarXiv{2007.13746}

\bibitem[{{Arca Sedda} \& {Benacquista}(2019)}]{arcasedda19b}
{Arca Sedda}, M., \& {Benacquista}, M. 2019, \mnras, 482, 2991,
  \dodoi{10.1093/mnras/sty2764}

\bibitem[{{Arca-Sedda} {et~al.}(2015){Arca-Sedda}, {Capuzzo-Dolcetta},
  {Antonini}, \& {Seth}}]{ASCD15He}
{Arca-Sedda}, M., {Capuzzo-Dolcetta}, R., {Antonini}, F., \& {Seth}, A. 2015,
  apj, 806, 220, \dodoi{10.1088/0004-637X/806/2/220}

\bibitem[{{Arca Sedda} {et~al.}(2020{\natexlab{a}}){Arca Sedda}, {Mapelli},
  {Spera}, {Benacquista}, \& {Giacobbo}}]{arcasedda20b}
{Arca Sedda}, M., {Mapelli}, M., {Spera}, M., {Benacquista}, M., \& {Giacobbo},
  N. 2020{\natexlab{a}}, \apj, 894, 133, \dodoi{10.3847/1538-4357/ab88b2}

\bibitem[{{Arca Sedda} {et~al.}(2020{\natexlab{b}}){Arca Sedda}, {Berry},
  {Jani}, {Amaro-Seoane}, {Auclair}, {Baird}, {Baker}, {Berti}, {Breivik},
  {Burrows}, {Caprini}, {Chen}, {Doneva}, {Ezquiaga}, {Saavik Ford}, {Katz},
  {Kolkowitz}, {McKernan}, {Mueller}, {Nardini}, {Pikovski}, {Rajendran},
  {Sesana}, {Shao}, {Tamanini}, {Vartanyan}, {Warburton}, {Witek}, {Wong}, \&
  {Zevin}}]{arcasedda20cqg}
{Arca Sedda}, M., {Berry}, C. P.~L., {Jani}, K., {et~al.} 2020{\natexlab{b}},
  Classical and Quantum Gravity, 37, 215011, \dodoi{10.1088/1361-6382/abb5c1}

\bibitem[{{Banerjee}(2017)}]{banerjee16}
{Banerjee}, S. 2017, mn, 467, 524, \dodoi{10.1093/mnras/stw3392}

\bibitem[{{Banerjee}(2018)}]{banerjee18}
---. 2018, mn, 473, 909, \dodoi{10.1093/mnras/stx2347}

\bibitem[{{Banerjee}(2020)}]{banerjee20}
---. 2020, \mnras, \dodoi{10.1093/mnras/staa2392}

\bibitem[{{Bavera} {et~al.}(2020){Bavera}, {Fragos}, {Qin}, {Zapartas},
  {Neijssel}, {Mandel}, {Batta}, {Gaebel}, {Kimball}, \&
  {Stevenson}}]{bavera20}
{Bavera}, S.~S., {Fragos}, T., {Qin}, Y., {et~al.} 2020, \aap, 635, A97,
  \dodoi{10.1051/0004-6361/201936204}

\bibitem[{{Belczynski}(2020)}]{belczynski21}
{Belczynski}, K. 2020, \apjl, 905, L15, \dodoi{10.3847/2041-8213/abcbf1}

\bibitem[{{Belczynski} \& {Banerjee}(2020)}]{belczynski20c}
{Belczynski}, K., \& {Banerjee}, S. 2020, \aap, 640, L20,
  \dodoi{10.1051/0004-6361/202038427}

\bibitem[{{Belczynski} {et~al.}(2002){Belczynski}, {Kalogera}, \&
  {Bulik}}]{belczynski02}
{Belczynski}, K., {Kalogera}, V., \& {Bulik}, T. 2002, \apj, 572, 407,
  \dodoi{10.1086/340304}

\bibitem[{{Belczynski} {et~al.}(2017{\natexlab{a}}){Belczynski}, {Klencki},
  {Meynet}, {Fryer}, {Brown}, {Chruslinska}, {Gladysz}, {O'Shaughnessy},
  {Bulik}, {Berti}, {Holz}, {Gerosa}, {Giersz}, {Ekstrom}, {Georgy}, {Askar},
  {Lasota}, \& {Wysocki}}]{Belczynski17}
{Belczynski}, K., {Klencki}, J., {Meynet}, G., {et~al.} 2017{\natexlab{a}},
  ArXiv e-prints.
\newblock \doarXiv{1706.07053}

\bibitem[{{Belczynski} {et~al.}(2017{\natexlab{b}}){Belczynski}, {Askar},
  {Arca-Sedda}, {Chruslinska}, {Donnari}, {Giersz}, {Benacquista}, {Spurzem},
  {Jin}, {Wiktorowicz}, \& {Belloni}}]{belczinski17b}
{Belczynski}, K., {Askar}, A., {Arca-Sedda}, M., {et~al.} 2017{\natexlab{b}},
  ArXiv e-prints:1712.00632.
\newblock \doarXiv{1712.00632}

\bibitem[{{Binney} \& {Tremaine}(2008)}]{bt}
{Binney}, J., \& {Tremaine}, S. 2008, {Galactic Dynamics: Second Edition}
  (Princeton University Press)

\bibitem[{{Campanelli} {et~al.}(2007){Campanelli}, {Lousto}, {Zlochower}, \&
  {Merritt}}]{campanelli07}
{Campanelli}, M., {Lousto}, C.~O., {Zlochower}, Y., \& {Merritt}, D. 2007,
  \prl, 98, 231102, \dodoi{10.1103/PhysRevLett.98.231102}

\bibitem[{{Costa} {et~al.}(2021){Costa}, {Bressan}, {Mapelli}, {Marigo},
  {Iorio}, \& {Spera}}]{costa21}
{Costa}, G., {Bressan}, A., {Mapelli}, M., {et~al.} 2021, \mnras, 501, 4514,
  \dodoi{10.1093/mnras/staa3916}

\bibitem[{{Cruz-Osorio} \& {Rezzolla}(2020)}]{osorio20}
{Cruz-Osorio}, A., \& {Rezzolla}, L. 2020, \apj, 894, 147,
  \dodoi{10.3847/1538-4357/ab89aa}

\bibitem[{{de Mink} {et~al.}(2013){de Mink}, {Langer}, {Izzard}, {Sana}, \& {de
  Koter}}]{demink13}
{de Mink}, S.~E., {Langer}, N., {Izzard}, R.~G., {Sana}, H., \& {de Koter}, A.
  2013, \apj, 764, 166, \dodoi{10.1088/0004-637X/764/2/166}

\bibitem[{{Dehnen}(1993)}]{Deh93}
{Dehnen}, W. 1993, mn, 265, 250

\bibitem[{{Di Carlo} {et~al.}(2019){Di Carlo}, {Giacobbo}, {Mapelli},
  {Pasquato}, {Spera}, {Wang}, \& {Haardt}}]{dicarlo19}
{Di Carlo}, U.~N., {Giacobbo}, N., {Mapelli}, M., {et~al.} 2019, \mnras, 487,
  2947, \dodoi{10.1093/mnras/stz1453}

\bibitem[{{Di Carlo} {et~al.}(2020){Di Carlo}, {Mapelli}, {Giacobbo}, {Spera},
  {Bouffanais}, {Rastello}, {Santoliquido}, {Pasquato}, {Ballone}, {Trani},
  {Torniamenti}, \& {Haardt}}]{dicarlo20}
{Di Carlo}, U.~N., {Mapelli}, M., {Giacobbo}, N., {et~al.} 2020, \mnras, 498,
  495, \dodoi{10.1093/mnras/staa2286}

\bibitem[{{Doctor} {et~al.}(2020){Doctor}, {Wysocki}, {O'Shaughnessy}, {Holz},
  \& {Farr}}]{doctor20}
{Doctor}, Z., {Wysocki}, D., {O'Shaughnessy}, R., {Holz}, D.~E., \& {Farr}, B.
  2020, \apj, 893, 35, \dodoi{10.3847/1538-4357/ab7fac}

\bibitem[{{Farmer} {et~al.}(2019){Farmer}, {Renzo}, {de Mink}, {Marchant}, \&
  {Justham}}]{farmer19}
{Farmer}, R., {Renzo}, M., {de Mink}, S.~E., {Marchant}, P., \& {Justham}, S.
  2019, \apj, 887, 53, \dodoi{10.3847/1538-4357/ab518b}

\bibitem[{{Fowler} \& {Hoyle}(1964)}]{fowler64}
{Fowler}, W.~A., \& {Hoyle}, F. 1964, \apjs, 9, 201, \dodoi{10.1086/190103}

\bibitem[{{Fragione} {et~al.}(2020){Fragione}, {Loeb}, \& {Rasio}}]{fragione20}
{Fragione}, G., {Loeb}, A., \& {Rasio}, F.~A. 2020, \apjl, 902, L26,
  \dodoi{10.3847/2041-8213/abbc0a}

\bibitem[{{Georgiev} {et~al.}(2016){Georgiev}, {B{\"o}ker}, {Leigh},
  {L{\"u}tzgendorf}, \& {Neumayer}}]{georgiev16}
{Georgiev}, I.~Y., {B{\"o}ker}, T., {Leigh}, N., {L{\"u}tzgendorf}, N., \&
  {Neumayer}, N. 2016, mn, 457, 2122, \dodoi{10.1093/mnras/stw093}

\bibitem[{{Gerosa} \& {Berti}(2017)}]{gerosa17}
{Gerosa}, D., \& {Berti}, E. 2017, \prd, 95, 124046,
  \dodoi{10.1103/PhysRevD.95.124046}

\bibitem[{{Gieles}(2009)}]{gieles09}
{Gieles}, M. 2009, \mnras, 394, 2113, \dodoi{10.1111/j.1365-2966.2009.14473.x}

\bibitem[{{Giersz} {et~al.}(2019){Giersz}, {Askar}, {Wang}, {Hypki}, {Leveque},
  \& {Spurzem}}]{giersz19}
{Giersz}, M., {Askar}, A., {Wang}, L., {et~al.} 2019, \mnras, 487, 2412,
  \dodoi{10.1093/mnras/stz1460}

\bibitem[{{Giersz} {et~al.}(2015){Giersz}, {Leigh}, {Hypki}, {L{\"u}tzgendorf},
  \& {Askar}}]{giersz15}
{Giersz}, M., {Leigh}, N., {Hypki}, A., {L{\"u}tzgendorf}, N., \& {Askar}, A.
  2015, mn, 454, 3150, \dodoi{10.1093/mnras/stv2162}

\bibitem[{{Glebbeek} {et~al.}(2009){Glebbeek}, {Gaburov}, {de Mink}, {Pols}, \&
  {Portegies Zwart}}]{glebbeek09}
{Glebbeek}, E., {Gaburov}, E., {de Mink}, S.~E., {Pols}, O.~R., \& {Portegies
  Zwart}, S.~F. 2009, aa, 497, 255, \dodoi{10.1051/0004-6361/200810425}

\bibitem[{{Gonz{\'a}lez} {et~al.}(2007){Gonz{\'a}lez}, {Sperhake},
  {Br{\"u}gmann}, {Hannam}, \& {Husa}}]{gonzalez07}
{Gonz{\'a}lez}, J.~A., {Sperhake}, U., {Br{\"u}gmann}, B., {Hannam}, M., \&
  {Husa}, S. 2007, \prl, 98, 091101, \dodoi{10.1103/PhysRevLett.98.091101}

\bibitem[{{Graham} {et~al.}(2020){Graham}, {Ford}, {McKernan}, {Ross}, {Stern},
  {Burdge}, {Coughlin}, {Djorgovski}, {Drake}, {Duev}, {Kasliwal}, {Mahabal},
  {van Velzen}, {Belecki}, {Bellm}, {Burruss}, {Cenko}, {Cunningham}, {Helou},
  {Kulkarni}, {Masci}, {Prince}, {Reiley}, {Rodriguez}, {Rusholme}, {Smith}, \&
  {Soumagnac}}]{graham20}
{Graham}, M.~J., {Ford}, K.~E.~S., {McKernan}, B., {et~al.} 2020, \prl, 124,
  251102, \dodoi{10.1103/PhysRevLett.124.251102}

\bibitem[{{Harris}(2010)}]{harris10}
{Harris}, W.~E. 2010, arXiv e-prints, arXiv:1012.3224.
\newblock \doarXiv{1012.3224}

\bibitem[{{Hobbs} {et~al.}(2005){Hobbs}, {Lorimer}, {Lyne}, \&
  {Kramer}}]{hobbs}
{Hobbs}, G., {Lorimer}, D.~R., {Lyne}, A.~G., \& {Kramer}, M. 2005, mn, 360,
  974, \dodoi{10.1111/j.1365-2966.2005.09087.x}

\bibitem[{{Holley-Bockelmann} {et~al.}(2008){Holley-Bockelmann},
  {G{\"u}ltekin}, {Shoemaker}, \& {Yunes}}]{bockelmann08}
{Holley-Bockelmann}, K., {G{\"u}ltekin}, K., {Shoemaker}, D., \& {Yunes}, N.
  2008, \apj, 686, 829, \dodoi{10.1086/591218}

\bibitem[{{Hurley} {et~al.}(2000){Hurley}, {Pols}, \& {Tout}}]{hurley00}
{Hurley}, J.~R., {Pols}, O.~R., \& {Tout}, C.~A. 2000, mn, 315, 543,
  \dodoi{10.1046/j.1365-8711.2000.03426.x}

\bibitem[{{Hurley} {et~al.}(2002){Hurley}, {Tout}, \& {Pols}}]{hurley02}
{Hurley}, J.~R., {Tout}, C.~A., \& {Pols}, O.~R. 2002, \mnras, 329, 897,
  \dodoi{10.1046/j.1365-8711.2002.05038.x}

\bibitem[{{Jim{\'e}nez-Forteza} {et~al.}(2017){Jim{\'e}nez-Forteza}, {Keitel},
  {Husa}, {Hannam}, {Khan}, \& {P{\"u}rrer}}]{jimenez17}
{Jim{\'e}nez-Forteza}, X., {Keitel}, D., {Husa}, S., {et~al.} 2017, \prd, 95,
  064024, \dodoi{10.1103/PhysRevD.95.064024}

\bibitem[{{King}(1966)}]{King}
{King}, I.~R. 1966, astj, 71, 64, \dodoi{10.1086/109857}

\bibitem[{{Kinugawa} {et~al.}(2021){Kinugawa}, {Nakamura}, \&
  {Nakano}}]{kinugawa21}
{Kinugawa}, T., {Nakamura}, T., \& {Nakano}, H. 2021, \mnras, 501, L49,
  \dodoi{10.1093/mnrasl/slaa191}

\bibitem[{{Kocsis} {et~al.}(2012){Kocsis}, {Ray}, \& {Portegies
  Zwart}}]{kocsis12}
{Kocsis}, B., {Ray}, A., \& {Portegies Zwart}, S. 2012, apj, 752, 67,
  \dodoi{10.1088/0004-637X/752/1/67}

\bibitem[{{Kowalska} {et~al.}(2011){Kowalska}, {Bulik}, {Belczynski},
  {Dominik}, \& {Gondek-Rosinska}}]{kowalska11}
{Kowalska}, I., {Bulik}, T., {Belczynski}, K., {Dominik}, M., \&
  {Gondek-Rosinska}, D. 2011, \aap, 527, A70,
  \dodoi{10.1051/0004-6361/201015777}

\bibitem[{{Kremer} {et~al.}(2020){Kremer}, {Spera}, {Becker}, {Chatterjee}, {Di
  Carlo}, {Fragione}, {Rodriguez}, {Ye}, \& {Rasio}}]{kremer20}
{Kremer}, K., {Spera}, M., {Becker}, D., {et~al.} 2020, arXiv e-prints,
  arXiv:2006.10771.
\newblock \doarXiv{2006.10771}

\bibitem[{{Kroupa} {et~al.}(2013){Kroupa}, {Weidner}, {Pflamm-Altenburg},
  {Thies}, {Dabringhausen}, {Marks}, \& {Maschberger}}]{kroupa13}
{Kroupa}, P., {Weidner}, C., {Pflamm-Altenburg}, J., {et~al.} 2013, {The
  Stellar and Sub-Stellar Initial Mass Function of Simple and Composite
  Populations}, ed. T.~D. {Oswalt} \& G.~{Gilmore}, Vol.~5, 115,
  \dodoi{10.1007/978-94-007-5612-0_4}

\bibitem[{{Law-Smith} {et~al.}(2019){Law-Smith}, {Guillochon}, \&
  {Ramirez-Ruiz}}]{law19}
{Law-Smith}, J., {Guillochon}, J., \& {Ramirez-Ruiz}, E. 2019, \apjl, 882, L25,
  \dodoi{10.3847/2041-8213/ab379a}

\bibitem[{{Liu} \& {Bromm}(2020)}]{liu20}
{Liu}, B., \& {Bromm}, V. 2020, \apjl, 903, L40,
  \dodoi{10.3847/2041-8213/abc552}

\bibitem[{{Lousto} \& {Zlochower}(2008)}]{lousto08}
{Lousto}, C.~O., \& {Zlochower}, Y. 2008, \prd, 77, 044028,
  \dodoi{10.1103/PhysRevD.77.044028}

\bibitem[{{Lousto} {et~al.}(2012){Lousto}, {Zlochower}, {Dotti}, \&
  {Volonteri}}]{lousto12}
{Lousto}, C.~O., {Zlochower}, Y., {Dotti}, M., \& {Volonteri}, M. 2012, \prd,
  85, 084015, \dodoi{10.1103/PhysRevD.85.084015}

\bibitem[{{MacLeod} \& {Ramirez-Ruiz}(2015)}]{macleod15}
{MacLeod}, M., \& {Ramirez-Ruiz}, E. 2015, \apjl, 798, L19,
  \dodoi{10.1088/2041-8205/798/1/L19}

\bibitem[{{Madau} \& {Fragos}(2017)}]{madau17}
{Madau}, P., \& {Fragos}, T. 2017, \apj, 840, 39,
  \dodoi{10.3847/1538-4357/aa6af9}

\bibitem[{{Mapelli}(2016)}]{mapelli16}
{Mapelli}, M. 2016, mn, 459, 3432, \dodoi{10.1093/mnras/stw869}

\bibitem[{{Mapelli} {et~al.}(2020){Mapelli}, {Santoliquido}, {Bouffanais},
  {Arca Sedda}, {Giacobbo}, {Artale}, \& {Ballone}}]{mapelli20}
{Mapelli}, M., {Santoliquido}, F., {Bouffanais}, Y., {et~al.} 2020, arXiv
  e-prints, arXiv:2007.15022.
\newblock \doarXiv{2007.15022}

\bibitem[{{McKernan} {et~al.}(2018){McKernan}, {Ford}, {Bellovary}, {Leigh},
  {Haiman}, {Kocsis}, {Lyra}, {Mac Low}, {Metzger}, {O'Dowd}, {Endlich}, \&
  {Rosen}}]{mckernan18}
{McKernan}, B., {Ford}, K.~E.~S., {Bellovary}, J., {et~al.} 2018, \apj, 866,
  66, \dodoi{10.3847/1538-4357/aadae5}

\bibitem[{{Metzger} \& {Stone}(2016)}]{metzger16}
{Metzger}, B.~D., \& {Stone}, N.~C. 2016, mn, 461, 948,
  \dodoi{10.1093/mnras/stw1394}

\bibitem[{{Mikkola} \& {Aarseth}(1998)}]{mikkola98}
{Mikkola}, S., \& {Aarseth}, S.~J. 1998, \na, 3, 309,
  \dodoi{10.1016/S1384-1076(98)00018-9}

\bibitem[{{Mikkola} \& {Merritt}(2008)}]{mikkola08}
{Mikkola}, S., \& {Merritt}, D. 2008, astj, 135, 2398,
  \dodoi{10.1088/0004-6256/135/6/2398}

\bibitem[{{Nguyen} {et~al.}(2014){Nguyen}, {Seth}, {Reines}, {den Brok},
  {Sand}, \& {McLeod}}]{ngu14}
{Nguyen}, D.~D., {Seth}, A.~C., {Reines}, A.~E., {et~al.} 2014, apj, 794, 34,
  \dodoi{10.1088/0004-637X/794/1/34}

\bibitem[{{Nishizawa} {et~al.}(2016){Nishizawa}, {Berti}, {Klein}, \&
  {Sesana}}]{nishizawa16}
{Nishizawa}, A., {Berti}, E., {Klein}, A., \& {Sesana}, A. 2016, \prd, 94,
  064020, \dodoi{10.1103/PhysRevD.94.064020}

\bibitem[{{O'Leary} {et~al.}(2009){O'Leary}, {Kocsis}, \& {Loeb}}]{oleary09}
{O'Leary}, R.~M., {Kocsis}, B., \& {Loeb}, A. 2009, \mnras, 395, 2127,
  \dodoi{10.1111/j.1365-2966.2009.14653.x}

\bibitem[{Peters(1964)}]{peters64}
Peters, P.~C. 1964, Phys. Rev., 136, B1224, \dodoi{10.1103/PhysRev.136.B1224}

\bibitem[{{Planck Collaboration} {et~al.}(2016){Planck Collaboration}, {Ade},
  {Aghanim}, {Arnaud}, {Ashdown}, {Aumont}, {Baccigalupi}, {Banday},
  {Barreiro}, {Bartlett}, {Bartolo}, {Battaner}, {Battye}, {Benabed},
  {Beno{\^\i}t}, {Benoit-L{\'e}vy}, {Bernard}, {Bersanelli}, {Bielewicz},
  {Bock}, {Bonaldi}, {Bonavera}, {Bond}, {Borrill}, {Bouchet}, {Boulanger},
  {Bucher}, {Burigana}, {Butler}, {Calabrese}, {Cardoso}, {Catalano},
  {Challinor}, {Chamballu}, {Chary}, {Chiang}, {Chluba}, {Christensen},
  {Church}, {Clements}, {Colombi}, {Colombo}, {Combet}, {Coulais}, {Crill},
  {Curto}, {Cuttaia}, {Danese}, {Davies}, {Davis}, {de Bernardis}, {de Rosa},
  {de Zotti}, {Delabrouille}, {D{\'e}sert}, {Di Valentino}, {Dickinson},
  {Diego}, {Dolag}, {Dole}, {Donzelli}, {Dor{\'e}}, {Douspis}, {Ducout},
  {Dunkley}, {Dupac}, {Efstathiou}, {Elsner}, {En{\ss}lin}, {Eriksen},
  {Farhang}, {Fergusson}, {Finelli}, {Forni}, {Frailis}, {Fraisse},
  {Franceschi}, {Frejsel}, {Galeotta}, {Galli}, {Ganga}, {Gauthier}, {Gerbino},
  {Ghosh}, {Giard}, {Giraud-H{\'e}raud}, {Giusarma}, {Gjerl{\o}w},
  {Gonz{\'a}lez-Nuevo}, {G{\'o}rski}, {Gratton}, {Gregorio}, {Gruppuso},
  {Gudmundsson}, {Hamann}, {Hansen}, {Hanson}, {Harrison}, {Helou},
  {Henrot-Versill{\'e}}, {Hern{\'a}ndez-Monteagudo}, {Herranz}, {Hildebrand t},
  {Hivon}, {Hobson}, {Holmes}, {Hornstrup}, {Hovest}, {Huang}, {Huffenberger},
  {Hurier}, {Jaffe}, {Jaffe}, {Jones}, {Juvela}, {Keih{\"a}nen}, {Keskitalo},
  {Kisner}, {Kneissl}, {Knoche}, {Knox}, {Kunz}, {Kurki-Suonio}, {Lagache},
  {L{\"a}hteenm{\"a}ki}, {Lamarre}, {Lasenby}, {Lattanzi}, {Lawrence}, {Leahy},
  {Leonardi}, {Lesgourgues}, {Levrier}, {Lewis}, {Liguori}, {Lilje},
  {Linden-V{\o}rnle}, {L{\'o}pez-Caniego}, {Lubin}, {Mac{\'\i}as-P{\'e}rez},
  {Maggio}, {Maino}, {Mandolesi}, {Mangilli}, {Marchini}, {Maris}, {Martin},
  {Martinelli}, {Mart{\'\i}nez-Gonz{\'a}lez}, {Masi}, {Matarrese}, {McGehee},
  {Meinhold}, {Melchiorri}, {Melin}, {Mendes}, {Mennella}, {Migliaccio},
  {Millea}, {Mitra}, {Miville-Desch{\^e}nes}, {Moneti}, {Montier}, {Morgante},
  {Mortlock}, {Moss}, {Munshi}, {Murphy}, {Naselsky}, {Nati}, {Natoli},
  {Netterfield}, {N{\o}rgaard-Nielsen}, {Noviello}, {Novikov}, {Novikov},
  {Oxborrow}, {Paci}, {Pagano}, {Pajot}, {Paladini}, {Paoletti}, {Partridge},
  {Pasian}, {Patanchon}, {Pearson}, {Perdereau}, {Perotto}, {Perrotta},
  {Pettorino}, {Piacentini}, {Piat}, {Pierpaoli}, {Pietrobon}, {Plaszczynski},
  {Pointecouteau}, {Polenta}, {Popa}, {Pratt}, {Pr{\'e}zeau}, {Prunet},
  {Puget}, {Rachen}, {Reach}, {Rebolo}, {Reinecke}, {Remazeilles}, {Renault},
  {Renzi}, {Ristorcelli}, {Rocha}, {Rosset}, {Rossetti}, {Roudier},
  {Rouill{\'e} d'Orfeuil}, {Rowan-Robinson}, {Rubi{\~n}o-Mart{\'\i}n},
  {Rusholme}, {Said}, {Salvatelli}, {Salvati}, {Sandri}, {Santos},
  {Savelainen}, {Savini}, {Scott}, {Seiffert}, {Serra}, {Shellard}, {Spencer},
  {Spinelli}, {Stolyarov}, {Stompor}, {Sudiwala}, {Sunyaev}, {Sutton},
  {Suur-Uski}, {Sygnet}, {Tauber}, {Terenzi}, {Toffolatti}, {Tomasi},
  {Tristram}, {Trombetti}, {Tucci}, {Tuovinen}, {T{\"u}rler}, {Umana},
  {Valenziano}, {Valiviita}, {Van Tent}, {Vielva}, {Villa}, {Wade}, {Wandelt},
  {Wehus}, {White}, {White}, {Wilkinson}, {Yvon}, {Zacchei}, \&
  {Zonca}}]{planck15}
{Planck Collaboration}, {Ade}, P.~A.~R., {Aghanim}, N., {et~al.} 2016, \aap,
  594, A13, \dodoi{10.1051/0004-6361/201525830}

\bibitem[{{Portegies Zwart} \& {McMillan}(2002)}]{zwart02}
{Portegies Zwart}, S.~F., \& {McMillan}, S.~L.~W. 2002, apj, 576, 899,
  \dodoi{10.1086/341798}

\bibitem[{{Portegies Zwart} {et~al.}(2010){Portegies Zwart}, {McMillan}, \&
  {Gieles}}]{pz10}
{Portegies Zwart}, S.~F., {McMillan}, S. L.~W., \& {Gieles}, M. 2010, \araa,
  48, 431, \dodoi{10.1146/annurev-astro-081309-130834}

\bibitem[{{Qin} {et~al.}(2018){Qin}, {Fragos}, {Meynet}, {Andrews},
  {S{\o}rensen}, \& {Song}}]{Qin18}
{Qin}, Y., {Fragos}, T., {Meynet}, G., {et~al.} 2018, \aap, 616, A28,
  \dodoi{10.1051/0004-6361/201832839}

\bibitem[{{Qin} {et~al.}(2019){Qin}, {Marchant}, {Fragos}, {Meynet}, \&
  {Kalogera}}]{Qin19}
{Qin}, Y., {Marchant}, P., {Fragos}, T., {Meynet}, G., \& {Kalogera}, V. 2019,
  \apjl, 870, L18, \dodoi{10.3847/2041-8213/aaf97b}

\bibitem[{{Renzo} {et~al.}(2020){Renzo}, {Cantiello}, {Metzger}, \&
  {Jiang}}]{renzo20}
{Renzo}, M., {Cantiello}, M., {Metzger}, B.~D., \& {Jiang}, Y.~F. 2020, \apjl,
  904, L13, \dodoi{10.3847/2041-8213/abc6a6}

\bibitem[{{Rizzuto} {et~al.}(2021){Rizzuto}, {Naab}, {Spurzem}, {Giersz},
  {Ostriker}, {Stone}, {Wang}, {Berczik}, \& {Rampp}}]{rizzuto20}
{Rizzuto}, F.~P., {Naab}, T., {Spurzem}, R., {et~al.} 2021, \mnras, 501, 5257,
  \dodoi{10.1093/mnras/staa3634}

\bibitem[{{Robson} {et~al.}(2019){Robson}, {Cornish}, \& {Liu}}]{robson19}
{Robson}, T., {Cornish}, N.~J., \& {Liu}, C. 2019, Classical and Quantum
  Gravity, 36, 105011, \dodoi{10.1088/1361-6382/ab1101}

\bibitem[{{Rodriguez} {et~al.}(2019){Rodriguez}, {Zevin}, {Amaro-Seoane},
  {Chatterjee}, {Kremer}, {Rasio}, \& {Ye}}]{rodriguez19}
{Rodriguez}, C.~L., {Zevin}, M., {Amaro-Seoane}, P., {et~al.} 2019, \prd, 100,
  043027, \dodoi{10.1103/PhysRevD.100.043027}

\bibitem[{{Romero-Shaw} {et~al.}(2020){Romero-Shaw}, {Lasky}, {Thrane}, \&
  {Calderon Bustillo}}]{romero20}
{Romero-Shaw}, I.~M., {Lasky}, P.~D., {Thrane}, E., \& {Calderon Bustillo}, J.
  2020, arXiv e-prints, arXiv:2009.04771.
\newblock \doarXiv{2009.04771}

\bibitem[{{Schr{\o}der} {et~al.}(2020){Schr{\o}der}, {MacLeod}, {Loeb},
  {Vigna-G{\'o}mez}, \& {Mandel}}]{schroder20}
{Schr{\o}der}, S.~L., {MacLeod}, M., {Loeb}, A., {Vigna-G{\'o}mez}, A., \&
  {Mandel}, I. 2020, \apj, 892, 13, \dodoi{10.3847/1538-4357/ab7014}

\bibitem[{{Secunda} {et~al.}(2020){Secunda}, {Bellovary}, {Mac Low}, {Ford},
  {McKernan}, {Leigh}, {Lyra}, {S{\'a}ndor}, \& {Adorno}}]{secunda20}
{Secunda}, A., {Bellovary}, J., {Mac Low}, M.-M., {et~al.} 2020, \apj, 903,
  133, \dodoi{10.3847/1538-4357/abbc1d}

\bibitem[{{Shiokawa} {et~al.}(2015){Shiokawa}, {Krolik}, {Cheng}, {Piran}, \&
  {Noble}}]{shiokawa15}
{Shiokawa}, H., {Krolik}, J.~H., {Cheng}, R.~M., {Piran}, T., \& {Noble}, S.~C.
  2015, \apj, 804, 85, \dodoi{10.1088/0004-637X/804/2/85}

\bibitem[{{Spera} \& {Mapelli}(2017)}]{spera17}
{Spera}, M., \& {Mapelli}, M. 2017, mn, 470, 4739,
  \dodoi{10.1093/mnras/stx1576}

\bibitem[{{Stevenson} {et~al.}(2019){Stevenson}, {Sampson}, {Powell},
  {Vigna-G{\'o}mez}, {Neijssel}, {Sz{\'e}csi}, \& {Mandel}}]{stevenson19}
{Stevenson}, S., {Sampson}, M., {Powell}, J., {et~al.} 2019, \apj, 882, 121,
  \dodoi{10.3847/1538-4357/ab3981}

\bibitem[{{Tagawa} {et~al.}(2020){Tagawa}, {Kocsis}, {Haiman}, {Bartos},
  {Omukai}, \& {Samsing}}]{tagawa20}
{Tagawa}, H., {Kocsis}, B., {Haiman}, Z., {et~al.} 2020, arXiv e-prints,
  arXiv:2012.00011.
\newblock \doarXiv{2012.00011}

\bibitem[{{Tanikawa} {et~al.}(2021){Tanikawa}, {Kinugawa}, {Yoshida},
  {Hijikawa}, \& {Umeda}}]{tanikawa21}
{Tanikawa}, A., {Kinugawa}, T., {Yoshida}, T., {Hijikawa}, K., \& {Umeda}, H.
  2021, \mnras, \dodoi{10.1093/mnras/stab1421}

\bibitem[{{The LIGO Scientific Collaboration} {et~al.}(2020){The LIGO
  Scientific Collaboration}, {the Virgo Collaboration}, \& et~al}]{gw190521b}
{The LIGO Scientific Collaboration}, {the Virgo Collaboration}, \& et~al. 2020,
  arXiv e-prints, arXiv:2009.01190.
\newblock \doarXiv{2009.01190}

\bibitem[{{Valsecchi} {et~al.}(2010){Valsecchi}, {Glebbeek}, {Farr}, {Fragos},
  {Willems}, {Orosz}, {Liu}, \& {Kalogera}}]{valsecchi10}
{Valsecchi}, F., {Glebbeek}, E., {Farr}, W.~M., {et~al.} 2010, \nat, 468, 77,
  \dodoi{10.1038/nature09463}

\bibitem[{{Vink} {et~al.}(2021){Vink}, {Higgins}, {Sander}, \&
  {Sabhahit}}]{vink21}
{Vink}, J.~S., {Higgins}, E.~R., {Sander}, A. A.~C., \& {Sabhahit}, G.~N. 2021,
  \mnras, 504, 146, \dodoi{10.1093/mnras/stab842}

\bibitem[{{Wang} {et~al.}(2015){Wang}, {Spurzem}, {Aarseth}, {Nitadori},
  {Berczik}, {Kouwenhoven}, \& {Naab}}]{wang15}
{Wang}, L., {Spurzem}, R., {Aarseth}, S., {et~al.} 2015, \mnras, 450, 4070,
  \dodoi{10.1093/mnras/stv817}

\bibitem[{{Wang} {et~al.}(2016){Wang}, {Spurzem}, {Aarseth}, {Giersz}, {Askar},
  {Berczik}, {Naab}, {Schadow}, \& {Kouwenhoven}}]{wang16}
---. 2016, mn, 458, 1450, \dodoi{10.1093/mnras/stw274}

\bibitem[{{Wong} {et~al.}(2012){Wong}, {Valsecchi}, {Fragos}, \&
  {Kalogera}}]{wong12}
{Wong}, T.-W., {Valsecchi}, F., {Fragos}, T., \& {Kalogera}, V. 2012, \apj,
  747, 111, \dodoi{10.1088/0004-637X/747/2/111}

\bibitem[{{Woosley}(2017)}]{woosley17}
{Woosley}, S.~E. 2017, \apj, 836, 244, \dodoi{10.3847/1538-4357/836/2/244}

\bibitem[{{Woosley} {et~al.}(2007){Woosley}, {Blinnikov}, \&
  {Heger}}]{woosley07}
{Woosley}, S.~E., {Blinnikov}, S., \& {Heger}, A. 2007, \nat, 450, 390,
  \dodoi{10.1038/nature06333}

\bibitem[{{Woosley} \& {Heger}(2021)}]{woosley21}
{Woosley}, S.~E., \& {Heger}, A. 2021, arXiv e-prints, arXiv:2103.07933.
\newblock \doarXiv{2103.07933}

\bibitem[{{Yang} {et~al.}(2019){Yang}, {Bartos}, {Gayathri}, {Ford}, {Haiman},
  {Klimenko}, {Kocsis}, {M{\'a}rka}, {M{\'a}rka}, {McKernan}, \&
  {O'Shaughnessy}}]{yang19}
{Yang}, Y., {Bartos}, I., {Gayathri}, V., {et~al.} 2019, \prl, 123, 181101,
  \dodoi{10.1103/PhysRevLett.123.181101}

\bibitem[{{Zevin} {et~al.}(2020){Zevin}, {Bavera}, {Berry}, {Kalogera},
  {Fragos}, {Marchant}, {Rodriguez}, {Antonini}, {Holz}, \&
  {Pankow}}]{zevin20b}
{Zevin}, M., {Bavera}, S.~S., {Berry}, C. P.~L., {et~al.} 2020, arXiv e-prints,
  arXiv:2011.10057.
\newblock \doarXiv{2011.10057}

\end{thebibliography}
\bibliographystyle{aasjournal}

\end{document}